\documentclass[preprint2]{aastex62}
\usepackage{hyperref}
\usepackage{graphicx}   
\usepackage{amsmath}    
\usepackage{amssymb}    
\newcommand{\MK}[1]{$M_{K_{\text{S}}}$#1}
\newcommand{\Ks}[1]{$K_{\text{S}}$#1}
\newcommand{\teff}[1]{$T_{\text{eff}}$#1}
\newcommand{\logg}[1]{$\log{g}$#1}
\newcommand{\gaia}[1]{{\it GAIA}#1}
\newcommand{\tess}[1]{{\it TESS}#1}

\newcommand{\pipeline}[1]{\texttt{ORION}#1}
\newcommand{\vespa}[1]{\texttt{vespa}#1}

\shorttitle{\tess{} candidates around low mass stars}
\shortauthors{R. Cloutier}

\begin{document}
\title{The independent discovery of planet candidates around low mass stars and astrophysical false positives from the first two \tess{} sectors}

\correspondingauthor{Ryan Cloutier}
\email{cloutier@astro.utoronto.ca}

\author{Ryan Cloutier}
\affiliation{Department of Astronomy \& Astrophysics, University of Toronto \\
  50 St. George Street, Toronto, Ontario, Canada, M5S 3H4}
\affiliation{Centre for Planetary Sciences, Department of Physical \& Environmental Sciences, University of Toronto \\
  1265 Military Trail, Toronto, Ontario, Canada, M1C 1A4}
\affiliation{Institut de recherche sur les exoplan\`etes, D\'epartement de physique, Universit\'e de Montr\'eal \\
  2900 boul. \'Edouard-Montpetit, Montr\'eal, Quebec, Canada, H3T 1J4}

\begin{abstract}
  Continuous data releases throughout the \tess{} primary mission will provide unique opportunities for
  the exoplanet community at large to contribute to maximizing \tess{'}s scientific return via the discovery and validation
  of transiting planets. This paper introduces our independent pipeline for the detection of periodic transit events along
  with the results of its inaugural application to the recently released 2 minute light curves of low mass stars
  from the first two \tess{} sectors. The stellar parameters within our
  sample are refined using precise parallax measurements from the \gaia{} DR2 which reduces the number
  of low mass stars in our sample relative to those listed in the TESS Input Catalog.
  In lieu of the follow-up observations required to confirm or refute the planetary nature of
  transit-like signals, a validation of transit-like events flagged by our pipeline is performed statistically.
  The resulting vetted catalog contains seven probable blended eclipsing binaries, eight known TOIs, plus eight
  new planet candidates smaller than 4 Earth radii.
  This work demonstrates the ability of our pipeline to detect sub-Neptune-sized planet candidates
  which to-date represent some of the most attractive targets for
  future atmospheric characterization via transmission or thermal emission spectroscopy and for radial velocity
  efforts aimed at the completion
  of the \tess{} level one requirement to deliver 50 planets smaller than 4 Earth radii with measured masses.
\end{abstract}

\section{Introduction}
With our current observational capabilities, nearby transiting planets offer the best targets
to characterize exoplanetary systems in detail. By their proximity many of
these planets are amenable to follow-up observations to, for example, refine their radii and orbital
ephemerides from the ground \citep{stefansson17,cooke18} and from space \citep{broeg13,gaidos17},
measure planetary masses via precision radial velocities \citep{cloutier18b}, and study
their atmospheric compositions, dynamics, and thermal structures \citep{louie18,kempton18}. NASA's
\emph{Transiting Exoplanet Survey Satellite} \citep[\tess{;}][]{ricker15}, which launched on April 18
2018, is a purpose-built survey observatory and currently offers the best opportunity to discover
nearby transiting planets smaller than Neptune around stars within $\sim 1000$ pc \citep{stassun17}.
Indeed \tess{} has already produced a number of new confirmed planet detections
\citep{esposito18,gandolfi18,huang18a,trifonov18,vanderspek18} 
in addition to its set of \tess{} Objects of Interest or \tess{} `alerts'.

\tess{} features four refractive lens that provide a combined wide field-of-view of
$24^{\circ} \times 96^{\circ}$ (i.e. $\sim 2300$ square degrees) for a single sector. The primary
\tess{} mission splits the sky into 26 equal sectors (13 per hemisphere) anchored on the ecliptic
poles and extending towards the ecliptic plane where fields at the lowest ecliptic latitudes
($\sim 63$\% of the sky) will be continuously monitored for $\sim 27$ days. Conversely, fields
centered at the galactic poles ($\sim 2$\% of the sky) will be continuously monitored for
$\sim 350$ days and overlap with continuous viewing zone of the \emph{James Webb Space Telescope}
(\emph{JWST}). In total  
\tess{} will survey $\sim 85$\% of the entire sky over its two year-long survey targeting
2-4$\times 10^5$ predominantly bright dwarf stars listed in the TESS Input Catalog
\citep[TIC;][]{stassun17} with 2 minute cadence.
Full Frame Images for all visible objects within each field will also be released with a 30 minute
cadence. From these data products \tess{}
is expected to discover thousands of new transiting exoplanets 
\citep{sullivan15,ballard18,barclay18} plus potentially thousands more from a variety of proposed
extended missions \citep{bouma17,huang18b}. The launch of \tess{}
and its recent large data release marks the beginning of a new era of exoplanetary survey
science that will carry on the legacy of the infamous \emph{Kepler} space telescope which was
decommissioned on November 16 2018, after nearly a decade of transformative exoplanet observations
and thousands of planet discoveries.

\tess{} is also unique to past space-based transiting exoplanet survey observatories
(e.g. \emph{Kepler}, \emph{CoRoT}) in that its bandpass extends further redward into the near-IR:
600-1000 nm. This enables \tess{} to access more cool M dwarf stars at high signal-to-noise
than previous missions. Systems of sub-Neptune-sized planets are common around M dwarfs
\citep{dressing13,morton14,dressing15a} and are required in order to provide a global view of
outcomes of the planet formation process across the Initial Mass Function.
Given their lower luminosities relative
to Sun-like stars, detecting close-in planets around low mass stars probes a subset of exoplanets
with systematically lower equilibrium temperatures including temperate planets orbiting
within the habitable zone \citep{kasting93,kopparapu13}. Given their relative
abundance within the solar neighbourhood \citep{winters15}, planet masses around nearby low mass
stars may be
readily characterized with radial velocities to build up a statistically significant view of the
mass-radius relationship for small planets \citep{weiss14,rogers15,wolfgang16,chen17}. The small sizes
of low mass stars also works to
increase observational signatures of transiting planets thus making their planetary systems of
particular interest for the atmospheric characterization of terrestrial to super-Earth-sized planets
whose scale heights are expected to be inherently small \citep[$\lesssim$][]{millerricci09}
and thus difficult to detect even with state-of-art instrumentation on-board the up-coming
\emph{JWST} \citep{morley17}.

The recent public data release from the first two \tess{} sectors, processed and validated by the
TESS Science Processing Operations Center \citep{jenkins16,twicken18,li18}, provides an opportunity
for members of the extended exoplanet community to pursue a variety of unique science cases. This
includes the search for new transiting planets using transit detection algorithms that are independent
of those used by the TESS Science Team and on distinct subsets of stars targeted by \tess{.}
In this study, focus on low mass dwarf stars from the TIC and use \gaia{} parallaxes to infer precise
stellar parameters and the refine the sample probable M dwarf TICs. We then search for transiting
exoplanets around these low mass dwarfs in the high cadence \tess{} light curves using our
custom-built transit detection pipeline described herein.

In Sect.~\ref{sect:stars} we present the derivation of our input target list of low mass TICs.
In Sect.~\ref{sect:pipe} we present the details of our transit detection pipeline \pipeline{.} In
Sect.~\ref{sect:search} we present our pipeline results and our supplementary efforts to
classify flagged transit-like events via human vetting and statistical validation before
culminating our final list of planet candidates and astrophysical false positives.
We conclude with a discussion in Sect.~\ref{sect:disc}.

\section{Stellar Sample} \label{sect:stars}
\subsection{Initial stellar sample}
Our initial stellar sample is retrieved from version 7 of the TESS Input Catalog (TIC-7) which
is accessed via the Barbara A. Mikulski Archive for Space Telescope (MAST)
Portal\footnote{\url{https://mast.stsci.edu/portal/Mashup/Clients/Mast/Portal.html}}. Among other
parameters, the TIC-7 table contains estimates of each star's physical parameters (i.e. effective
temperatures \teff{,} surface gravities \logg{,} radii $R_s$, masses $M_s$, etc.),
astrometry (either from the \emph{Tycho}-\gaia{} astrometric solution; \citealt{gaia16,brown16}
or from \emph{Hipparcos}), $G$-band magnitude from the \gaia{} data release 1, and 2MASS photometry
\citep{cutri03}. To identify putative low mass dwarf stars within the TIC-7, we first restrict our sample to
sources flagged as dwarf stars based on their 2MASS colors and the reduced proper motion criterion
from \cite{stassun17}, modified from \cite{collier07}. We further restrict our sample to stars whose
`priority' is $\geq 10^{-3}$ where the TIC priority metric is based on the relative probability of the TIC
of detecting small planetary transits. As such, the priority is dependent on $R_s$, the
expected photometric precision, the number of \tess{} sectors in which the TIC will be visible, and its
contamination ratio: the ratio of contamination to source flux where contamination is computed over ten
\tess{} pixels from the source ($\sim 3.5$ arcmin).

Next we establish our initial sample of low mass stars based on the physical stellar parameters from the
TIC-7 and using the following criteria:

\begin{itemize}
\item \teff{} $\in [2700,4200]$ K,
\item \logg{} $>4$,
\item $R_s < 0.75$ R$_{\odot}$,
\item $M_s < 0.75$ M$_{\odot}$.
\end{itemize}

\begin{figure}
  \centering
  \includegraphics[scale=.9]{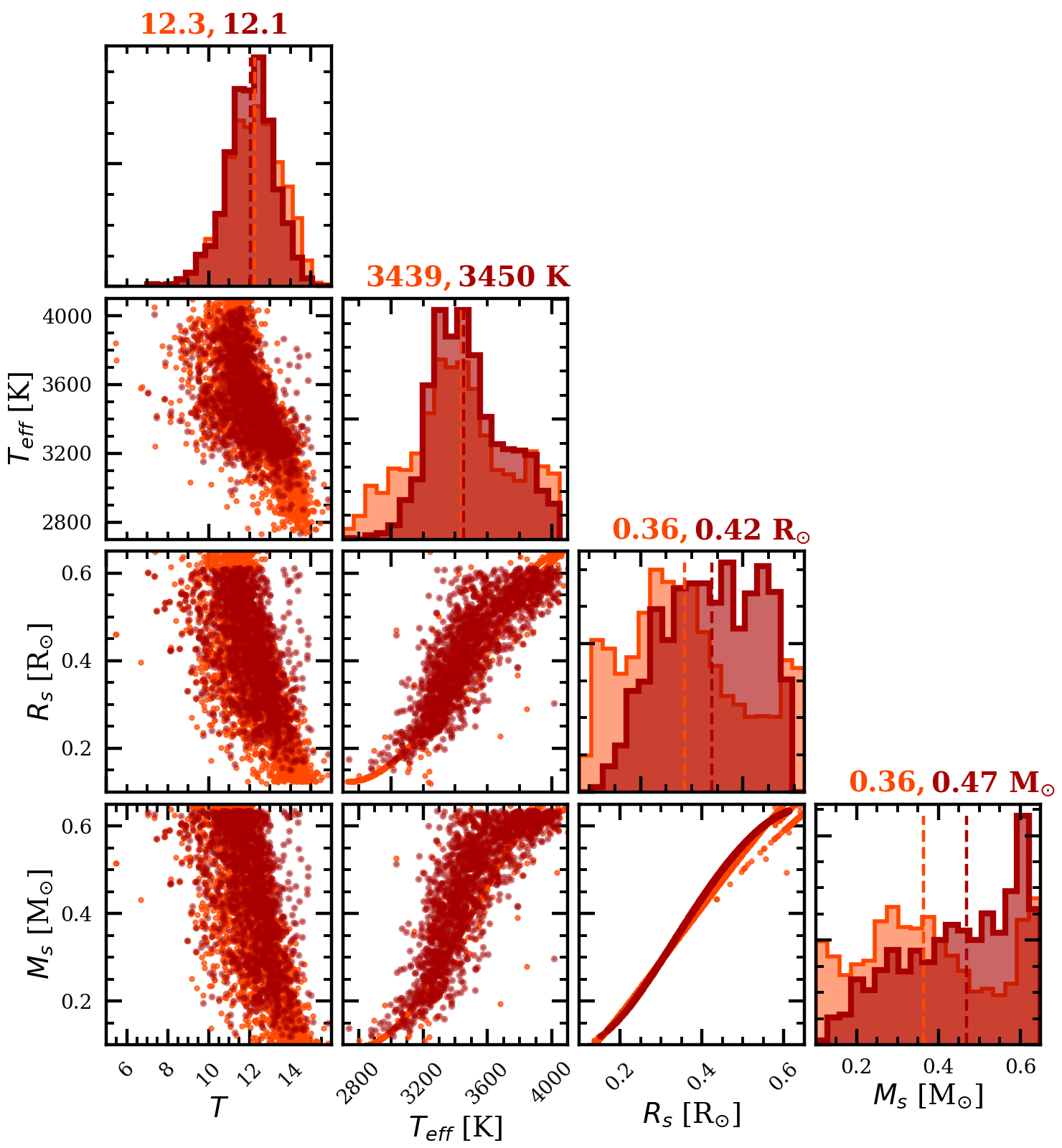}
  \caption{Distributions of \tess{} apparent magnitudes $T$, effective temperatures,
    stellar radii, and stellar masses for our initial (translucent orange markers) and final
    (dark red markers) stellar samples. Our initial sample contains 2849 low mass stars observed in
    sectors 1 and/or 2 and identified by their stellar parameters from the TIC-7. Our final sample contains
    1599 low mass stars with refined stellar parameters based on \gaia{}
    DR2 parallaxes. The median parameter values for each sample are annotated above each histogram.}
  \label{fig:stars}
\end{figure}

\noindent We note that these criteria are not intended to reflect the exact M dwarf parameter ranges of
interest but instead are chosen to be intentionally conservative as to avoid missing any potential
M dwarfs prior to their final classification (for use within this study) based on \teff{} and 
near-IR luminosities  (\MK{} $\in [4.5,10]$; \citealt{delfosse00, benedict16}) that will be refined
in Sect.~\ref{sect:gaia} using \gaia{} parallaxes.
At this stage we find a total of 93090 TICs that obey our criteria. Of these, 2849 TIC are observed in
one or both of \tess{} sectors 1 and 2 and are depicted in Fig.~\ref{fig:stars}.

\subsection{Refined stellar sample based on \gaia{} DR2} \label{sect:gaia}
The stellar parameters used to derive our initial stellar sample were obtained from a variety of sources
as outlined in \cite{stassun17}. Effective temperatures within our sample were predominantly obtained
from the cool dwarf catalog \citep{muirhead18} or alternatively from spectroscopic catalogs or $V$-\Ks{}
colors.
Most stellar masses and radii also come from the cool dwarf catalog or from the \cite{torres10}
spectroscopic relations. Stellar surface gravities follow from measurements of $R_s$ and $M_s$.
The TIC-7 stellar radii are typically known at the level of $\sim 16$\% which often dominates
the error budget of the measured planetary radii from transit observations. Here we aim to produce a
homogeneously-derived set of precise stellar parameters by exploiting the exquisite precision of their
2MASS photometry and stellar parallaxes from the \gaia{} DR2 \cite{gaia18} available for the majority of
stars in our initial sample.

For the accurate and precise characterization of transiting planets we are principally
interested in the measurement of host stellar radii $R_s$. Additionally, the derivation of other fundamental
parameters such as stellar masses and effective temperatures are of importance for a more complete understanding
of the effect that stars can have on their host planetary systems. Here, we re-derive
the stellar parameters of our initial sample by deriving their near-IR absolute magnitudes coupled with
empirically-derived M dwarf radius-luminosity relations \citep{mann15}. We begin by querying the \gaia{} DR2
archive using the star's right ascension and declination ($\alpha,\delta$) with a search radius of 10-60 arcseconds.
Cross-matching the TIC-7 with the \gaia{} DR2 data is necessary to obtain each star's updated parallax from the DR2,
additional \gaia{} photometry (i.e. $G_{\text{BP}}$ and $G_{\text{RP}}$) which was not included in the TIC-7,
and point-estimates of their respective measurement uncertainties that we will approximate as Gaussian distributed.
Our querying procedure utilizes the \texttt{astropy.astroquery python} package \citep{ginsburg17}.
Next we identify source matches according to their predicted photometric colors based on the 2MASS-\gaia{}
color-color relations reported in \cite{evans18}. Explicitly, we use the quadratic polynomial fits from
\cite{evans18} to predict each of the colors
$G-K_{\text{S}}$, $G_{\text{BP}}-K_{\text{S}}$, $G_{\text{RP}}-K_{\text{S}}$ and $G_{\text{BP}}-G_{\text{RP}}$ and then compare
those predictions for potential source matches to the measured TIC colors. Each of the four color-color
relations is accompanied by a characteristic scatter of $0.3692, 0.4839, 0.2744,$ and $0.2144$ magnitudes
respectively. We claim a source match when all of the calculated colors are within $3\sigma$ of their predicted
values. Based on numerous checks of individual known TICs, we determined
that such a tolerance is required to ensure accurate source matches. This dispersion is also expected given
that higher order effects not taken into account by the polynomial fits, can have stark effects on the accuracy
of the photometric predictions.

We proceed with identifying bona-fide low mass TICs within our initial stellar sample by using the \gaia{} data
of matched sources to refine the stellar parameters that were initially used to flag low mass stars. 
We will classify low mass stars within this study based on their absolute \Ks{-}band magnitudes
\citep{delfosse00,mann15,benedict16} and effective temperatures which are derived in the coming sub-sections.
We focus our analysis on the \Ks{-}band in this study due to the reduced effects of dust
extinction at near-IR wavelengths compared to in the visible. The absolute \Ks{-}band magnitude is

\begin{equation}
  M_{K_{\text{S}}} = K_{\text{S}} - \mu - A_{K_{\text{S}}}, \label{eq:MK}
\end{equation}

\noindent where \Ks{} is the source's \Ks{-}band apparent magnitude, $\mu = 5\log_{10}{(d/\text{1 pc})} - 5$
is the distance modulus given the distance to the source $d$, and $A_{K_{\text{S}}}$ is the source extinction in the
\Ks{-}band. Therefore, in order to compute \MK{} for our stellar sample we must first obtain the parameters $d$ and
$A_{K_S}$.

\subsubsection{Stellar distances from \gaia{}}
The \gaia{} DR2 reports precise stellar parallaxes $\varpi$ for the majority of stars in our initial sample
The typical parallax uncertainty for the stars in our sample is $\sim 0.2$\%.
As noted by numerous authors \citep[e.g.][]{bailorjones15,astraatmadja16,luri18}, reliable distances to the
majority of stars in the \gaia{} DR2 cannot be obtained by simply inverting the stellar parallax. Given $\varpi$
values with posterior probability density functions (PDFs) that are presumed Gaussian distributed, and are
therefore fully described by their mean values and $1\sigma$ dispersions, the non-linearity
of the transformation from $\varpi$ to $d$ will result in an asymmetric $d$ posterior PDF whose skewness
is dependent on the absolute $\varpi$ measurement value and its signal-to-noise \citep{luri18}.
By the proximity of the majority of sources in the TIC-7, their parallaxes are measured with high precision such
that the resulting $d$ PDF obtained using the standard formula ($d/$ pc) = ($\varpi$/arcsec, can be well-approximated
as a Gaussian distribution \citep{bailorjones18}. The median relative distance uncertainty therefore corresponds to 
that of the measured parallax (i.e. $\sim 0.2$\%). The maximum a-posteriori (MAP) $d$ value and its $1\sigma$
uncertainty are then propagated to the calculation of $\mu$ which we will ultimately use in Eq.~\ref{eq:MK} to
calculate \MK{} after the extinction coefficient are obtained (see Sect.~\ref{sect:AK}).

There are known systematic effects in the \gaia{} astrometric solution in the form of a non-zero parallax zero-point
that is dependent on the source position, $G$-band magnitude, and possibly color \citep{lindegren18}.
In computing $d$ from $\varpi$ we first apply a simple correction by adding the globally-averaged parallax
zero-point of 29 $\mu$as \citep{lindegren18} to the verbatim stellar parallaxes from the \gaia{} DR2.

\subsubsection{Source extinction estimates} \label{sect:AK}
The source extinction is dependent on the source's location on the sky and particularly on
its proximity to the galactic plane where the dust column density is highest. To estimate the \Ks{-}band extinction
for each source we utilize the \texttt{mwdust} package \citep{bovy16} which queries one of three E(B-V) reddening maps
\citep[i.e.][]{drimmel03,marshall06,green15} based on the applicability of each map to the input source position.
Given the source's galactic coordinates ($l,b$), \gaia{} distance, and uncertainties as input, \texttt{mwdust}
queries the reddening maps and returns the extinction coefficient $A_{\lambda}=R_{\lambda} \text{E(B-V)}$ in the desired
band using the extinction vector scaling $R_{K_{\text{S}}}=0.31$ from \cite{schlafly11}. Uncertainties in $A_{K_{\text{S}}}$ are
derived from the $d$ measurement uncertainty and from inherent uncertainties in the value of $R_{K_{\text{S}}}$
\citep[e.g.]{green18}
which we attempt to account for via the quadrature addition of a 30\% fractional uncertainty on $R_{K_{\text{S}}}$ following
the methodology of \citep{fulton18}.

\subsubsection{Deriving the set of refined stellar radii}
Combining the retrieved values of \Ks{,} $\mu$, and $A_{K_{\text{S}}}$ into Eq.~\ref{eq:MK} returns the
distribution of \MK{} for all of the 2489 stars in our initial sample for which 2MASS photometry and \gaia{}
parallaxes are available.

Calculations of M dwarf stellar radii from their bolometric magnitudes would require \Ks{-}band bolometric
corrections which for cool stars, are known to often suffer from comparatively large inaccuracies 
\citep[\teff{} $\lesssim 4100$ K;][]{berger18}. We therefore adopt the alternative approach from
\cite{berger18} which used the empirically-derived M dwarf radius-luminosity relation (RLR) from
\cite{mann15} to update M dwarf stellar radii in the \emph{Kepler} field using \gaia{} distances.
The fitted RLR uses a quadratic function to map
\MK{} to directly measured $R_s$ from interferometry and parallaxes. Because we are interested in deriving a
self-consistent sample of low mass stellar radii, we restrict our
analysis to TICs with \MK{} values that are applicable to the \cite{mann15} RLR which
is valid for M dwarfs with \MK{} $\in (4.6,9.8)$. This condition will be used to establish our final
sample of low mass dwarf stars following the derivation of \teff{} within our initial sample.
The radii inferred from the RLR have a fractional residual dispersion of 0.0289 $R_s$ which we add
in quadrature to the radius uncertainty propagated from \MK{.} 

\begin{figure}
  \centering
  \includegraphics{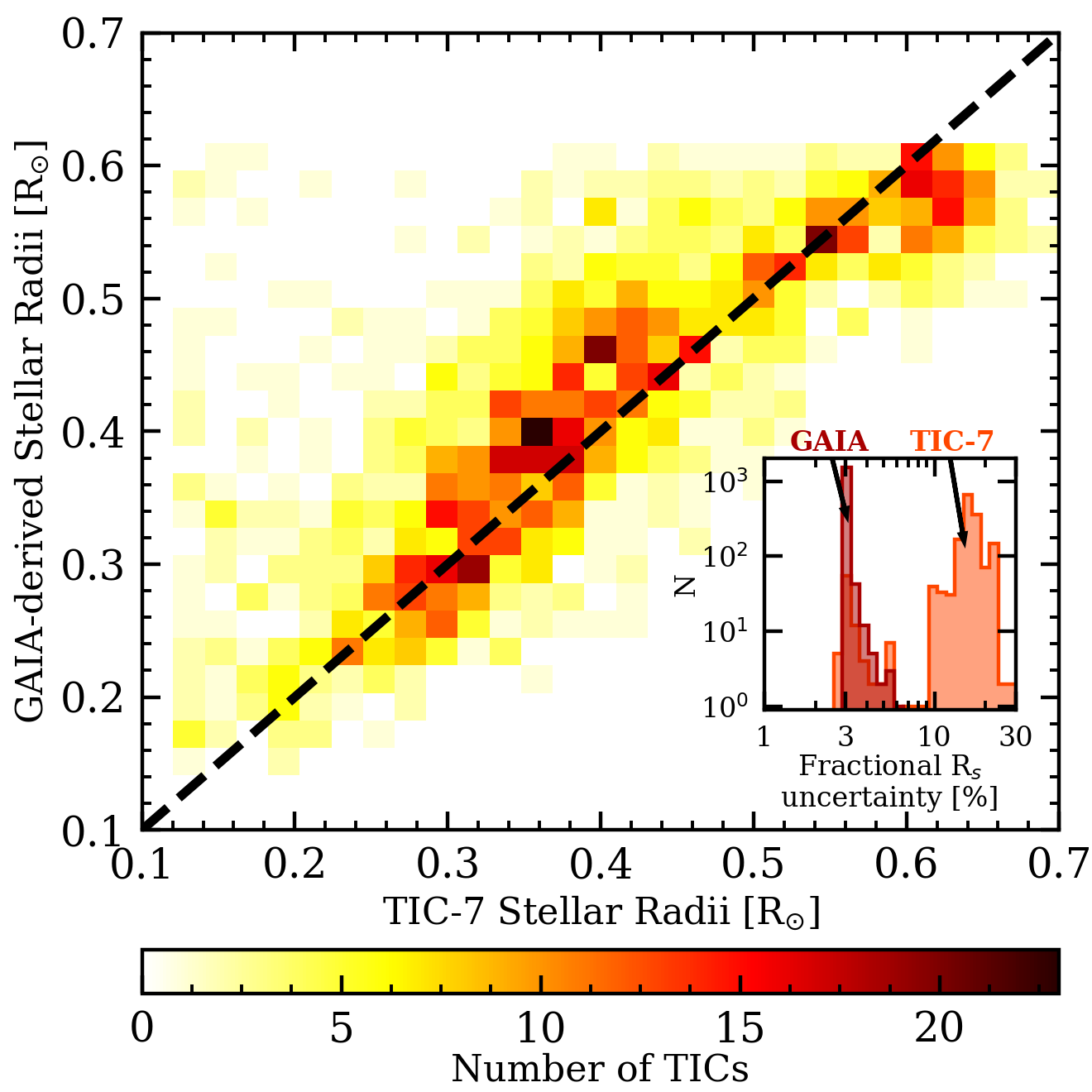}
  \caption{A 2-dimensional histogram comparing the stellar radii in our sample of 1599 low mass stars,
    derived from \gaia{} parallaxes and the M dwarf radius-luminosity relation from \cite{mann15}, to those
    from the TESS Input Catalog (TIC-7) which are compiled from a variety of sources. The subpanel compares the
    distributions of fractional stellar radius uncertainties in each catalog. The refined stellar radii 
    based on \gaia{} parallaxes have a typical precision improvement of $\sim 4-5$ compared to the TIC-7
    parameters.}
  \label{fig:Rs}
\end{figure}

Fig.~\ref{fig:Rs} compares the TIC-7 stellar radii (compiled from various input sources) with those
derived from \gaia{} distances and and the M dwarf RLR from \cite{mann15}. The relation is largely one-to-one
but with a slight translation of the updated $R_s$ distribution to larger radii ($\sim 3.7$\% median increase).
The effect is already known \citep{berger18} and is the result of many sources having their distance
measures increased following the release of the \gaia{} DR2 parallaxes. More notably for the measurement of
transit planet radii is the significant reduction in the fractional radius uncertainty as evidence in the
histogram included in Fig.~\ref{fig:Rs}. The typical fractional radius uncertainty $\sigma_{R_{s}}/R_s$ within
our updated sample is $\sim 4-5$ smaller than in the TIC-7. The median fractional uncertainty on our
\gaia{-}derived stellar radii is $\sim 3$\%.

\subsubsection{Deriving the set of refined stellar effective temperatures}
Similarly to the RLR, \citep{mann15} also parameterized an empirically-derived M dwarf temperature-color-metallicity
relation (TCMR). Our sample contains both 2MASS and \gaia{} photometry so we use these photometric systems to infer
\teff{} for the stars in our newly refined sample. Specifically, we adopt the fitted TCMR from \cite{mann15} which is
cubic in $G_{\text{BP}}$-$G_{\text{RP}}$ and quadratic in $J$-$H$. The latter color is used as a proxy for metallicity
\citep{leggett92,johnson12,mann13,newton14}. The TCMR used here has a residual temperature dispersion of 49 K which we
add in quadrature to the uncertainties in \teff{} propagated from the input photometric uncertainties.

\subsubsection{Deriving the set of refined stellar masses}
We revise the stellar masses using the empirically-derived M dwarf mass-luminosity relation (MLR) from  
\cite{benedict16}. Their fitted relation uses a quartic function to map \MK{} to directly measured $M_s$ from
dynamical analyses of binary star systems and is valid for M dwarfs with \MK{} $\leq 10$. This valid range of 
\MK{} is consistent with the range required for the \cite{mann15} empirical relations. The five fitted
coefficients that parameterize the MLR all have an associated uncertainty which we sample from, along with \MK{}
sampled from a Gaussian distribution, to infer the $M_s$ PDF. Point estimates of each TIC's $M_s$ and uncertainty
comes from the MAP of the PDF and the average of its $16^{\text{th}}$ and $84^{\text{th}}$ percentiles. 

\subsubsection{Final stellar sample}
Using the refined stellar parameters obtained from cross-matching putative low mass stars from the TIC-7 with
the \gaia{} DR2, we now construct our final stellar sample as stars that obey the following criteria:

\begin{itemize}
\item \MK{} $\in (4.6,9.8)$
\item \teff{} $- \sigma_{T_{\text{eff}}} < 4000$ K,
\item \logg{} $+ \sigma_{\log{g}}>4$,
\item $R_s -\sigma_{R_s} < 0.75$ R$_{\odot}$,
\item $M_s - \sigma_{M_s} < 0.75$ M$_{\odot}$.
\end{itemize}

\noindent That is that we retain all TICs whose luminosities, effective temperatures, surface
gravities, radii, and masses that are consistent with those of late-K to M dwarf parameter ranges
\citep{pecaut13} at the $1\sigma$ level.

Our final stellar sample contains 1599 low mass stars with 537, 694, and 368
observed within \tess{} sectors 1, 2, and both respectively.
The final stellar sample is over-plotted in Fig.~\ref{fig:stars}. The distribution of $T$
in our stellar samples spans 7-15 with a median $T=12.1$. The distribution of
effective temperatures extends from 2740-4040 K with a median \teff{} $=3450$ K
whose approximate spectral type is M3V \citep{pecaut13}. The \teff{} distributions from the TIC-7 and
our \gaia{-}derived values are roughly consistent. The stellar radii span 0.15-0.61 R$_{\odot}$
with a median $R_s=0.42$ R$_{\odot}$. The stellar masses span 0.12-0.63 M$_{\odot}$
with a median $M_s=0.47$ M$_{\odot}$. Owing to the increased distances of many TICs in our final sample,
the distributions of $R_s$ and $M_s$ are both translated to slightly larger radii and masses compared
to the values listed in the TIC-7.

\section{Overview of the \pipeline{} transit detection pipeline} \label{sect:pipe}
Here we present our independent transit detection pipeline \pipeline{} that burrows many of the strategies and
vetting procedures from established methods focused on transit detections primarily with \emph{Kepler} and
\emph{K2} (see references herein).
Our pipeline can be thought of as having six sequential steps that
take as input the TIC identifier and stellar parameters to 
produce a set of transiting planet candidates (PCs) with measured orbital periods $P$, times of mid-transit
$T_0$, scaled semi-major axes $a/R_s$, scaled planetary radii $r_p/R_s$, and orbital inclinations $i$.
The six steps within \pipeline{} are 1) to obtain the extracted \tess{} light curves and ancillary data for
of input TIC, 2) to derive an initial systematic model for light curve de-trending, 3) to perform a linear search
for transit-like events, 4) to perform a periodic search for repeating transit-like events, 5) to subject putative
PCs to a set of vetting criteria in an automated way, and 6) to re-model the light curve with a joint
systematics plus transit mode for all vetted transit-like events. These stages are described
in detail in the succeeding sections.


\subsection{\tess{} light curve acquisition}
The execution of \pipeline{} on a TIC begins with downloading the star's publicly available
2 minute \tess{} extracted light curves and target pixel files for all available sectors. 
The \tess{} data is downloaded from the MAST data service\footnote{\url{https://archive.stsci.edu/tess/index.html}}.
Only TICs observed at 2 minute cadence are considered at this time with their extracted light curve made available
following its processing by the TESS Science Processing Operations Center. Efforts to extract 30 minute light
curve data from the \tess{} Full Frame Images and significantly expand the list of \tess{} targets accessible to
\pipeline{} are underway but are reserved for a future study.
Target pixels files are principally used to quickly assess the data quality and will be used to
infer the TIC's point spread function during the statistical validation of putative PCs in Sect.~\ref{sect:vespa}.

For each available sector of data, the chronological vectors of valid observing times $\mathbf{t}$
(i.e. not NaNs), measured fluxes $\mathbf{f}$, and the associated $1\sigma$ flux
uncertainties $\boldsymbol{\sigma}_{f}$ are constructed. Fluxes are obtained from the Simple Aperture Photometry Pre-search
Data Conditioning extraction which includes artifact mitigation \citep{smith12}. These vectors are
attributed to the following fields: \texttt{TIME} [BJD], \texttt{PDCSAP\_FLUX} [e$^-$/s], and
\texttt{PDCSAP\_FLUX\_ERR} [e$^-$/s]. The flux and flux uncertainty vectors are converted into normalized
flux units via division by median($\mathbf{f}$).

\subsection{Initial light curve de-trending} \label{sect:detrend}
Residual systematic effects are clearly visible in the many of the extracted light curves. Due to the
inherent photometric and pointing precision of the first \tess{} sectors, these systematic effects
are often largely attributable to astrophysical noise sources such as flicker \citep{bastien13} in
Sun-like stars but more commonly for low mass stars from
large-scale variability caused by active regions on the rotating stellar surface. As an initial de-trending step
to correct for temporally-correlated noise sources from either systematics or intrinsic stellar phenomena,
a semi-parametric Gaussian process (GP) regression model is fit to the extracted \tess{} photometry.

GP regression models provide a flexible and probabilistic framework to model the temporal
covariances between photometric measurements following the removal of a mean model
$\boldsymbol{\mu}(\theta)$ which is parameterized by the set of observable parameters $\theta$
(e.g. orbital period, time of mid-transit, etc.). The posterior PDFs of the $\theta$ elements 
can be sampled simultaneously with the GP hyperparameters $\Theta$ which parameterize
the residual covariances through the covariance matrix K($\Theta$) and are fit by
optimizing the $\ln$ likelihood function

\begin{multline}
  \ln{\mathcal{L}(\theta,\Theta)} = \\
  -\frac{1}{2} \left[ (\mathbf{f}-\boldsymbol{\mu}(\theta))^{\text{T}}
    \cdot \text{K}(\Theta)^{-1} \cdot (\mathbf{f}-\boldsymbol{\mu}(\theta)) \right. \\
    \left. + \ln{\mathrm{det} \text{K}} + N \ln{2 \pi} \right] \label{eq:lnL}
\end{multline}

\noindent along with appropriately chosen priors on the parameters in $\theta$ and $\Theta$.
Here, $N$ is the number of photometric measurements in the light curve. Similar routines based on GP regression 
have been adopted for \emph{K2} systematic corrections \citep[e.g.][]{aigrain15,crossfield15,aigrain16} 
and can also be used to infer accurate photometric stellar rotation periods \citep{angus18}. For cases in which
the origin of the apparent photometric variations are likely dominated by active regions on rotating spotted
stars, the resulting photometry will vary non-sinusoidally as the active regions 
evolve in size, brightness, and location over the observational baseline. This physically motivates the
use of a quasi-periodic covariance matrix K$_{i,j} = \delta_{i,j} \sigma_{f,i} + k_{i,j}$ where $\delta_{i,j}$
is the Kronecker delta and $k_{i,j}$ is the covariance kernel of the form

\begin{equation}
  k_{i,j} = a_{\text{GP}}^2 \exp{\left[ - \frac{|t_i-t_j|^2}{2\lambda^2} -\Gamma^2
      \sin^2{\left(\frac{\pi|t_i-t_j|}{P_{\text{GP}}} \right)} \right]}.
\end{equation}

\noindent The covariance kernel is parameterized by the four hyperparameters
$\Theta=\{a_{\text{GP}},\lambda,\Gamma,P_{\text{GP}}\}$ where
$a$ is the correlation amplitude, $\lambda$ the exponential timescale, $\Gamma$ the coherence scale,
and $P_{\text{GP}}$ the periodic timescale of the correlations. Moreover,
a quasi-periodic covariance kernel is favorable for cases in which the origin of
apparent photometric variations are dominated by systematics which need not have a strong periodic
component. In this limit, the coherence parameter $\Gamma$ approaches small values such
that the covariance kernel becomes well-approximated by a squared exponential kernel with a single
effective timescale. Note that because 
systematic and astrophysical noise sources within the GP noise model are not distinguished,
the fitted hyperparameter values are unable to be used to interpret the origin of the photometric
variability. Furthermore, during the remainder of this paper the covariance structures modelled by the
GP will be solely referred to as `systematics' despite the possibility that their (partial) origin may
be astrophysical.

The logarithmic hyperparameters are initialized and subsequently optimized in an iterative manner and
are performed on each \tess{} sector independently assuming a null mean function (i.e.
$\boldsymbol{\mu}=\mathbf{0}$).
The periodic GP timescale $P_{\text{GP}}$ is initialized by peaks in the Lomb-Scargle
periodogram \citep[LSP;][]{scargle82} of the extracted light curve whereby in each of the iterations
performed, $P_{\text{GP}}$ is initialized to the $i^{\text{th}}$ most significant peak in the LSP where $i$
is the iteration index $\in [1,10]$. Periods within 5\% of a 1 day are excluded due to the
inherent LSP noise at sampled frequencies close to 1 days$^{-1}$. Following the use of GP regression
modelling for radial velocity (RV) activity mitigation in \cite{dittmann17a}, where the physical source of
activity is largely common between the optical RVs and broadband \tess{} photometry,  
$\ln{\lambda}$ is initialized to $\ln{3P_{\text{GP}}}$. In each iteration $\ln{\Gamma}$ is initialized to
$0$ and $\ln{a_{\text{GP}}}=\ln{\text{max}(|\mathbf{f}_{\text{bin}} - \text{median}(\mathbf{f}_{\text{bin}})|)}$
where $\mathbf{f}_{\text{bin}}$ is the vector of binned photometric points whose temporal bin width is set
such that a single periodic GP timescale is sampled by at least eight measurements.

For each iteration in each \tess{} sector, the uniquely initialized GP hyperparameters are optimized
using the \texttt{scipy.optimize.minimize python} function to minimize the negative
$\ln{\mathcal{L}(\Theta)}$ from Eq.~\ref{eq:lnL}
given the Jacobian of $\ln{\mathcal{L}(\Theta)}$ with respect to the hyperparameters in
$\Theta$. During optimization the $\ln$ GP hyperparameters are bounded by broad uninformative priors
which are explicitly reported in Table~\ref{table:priors}. Broad $\ln$ uniform priors enable
the generalization of the \pipeline{} de-trending method across all of the input \tess{} light curves
which greatly benefit from semi-parametric modelling given the wide range of covariance timescales
exhibited by TICs in photometry.
Given an optimized set of hyperparameters, the resulting GP posterior PDF is a $N$-dimensional
multi-variate Gaussian distribution whose mean function is taken to be a potential systematic correction.
The mean function of the GP from the iteration whose optimized hyperparameters maximize
$\ln{\mathcal{L}(\Theta)}$ is assigned as the initial systematic
correction and is used to de-trend the photometry prior to the search for periodic transit events.

\begin{figure*}
  \centering
  \includegraphics[scale=.96]{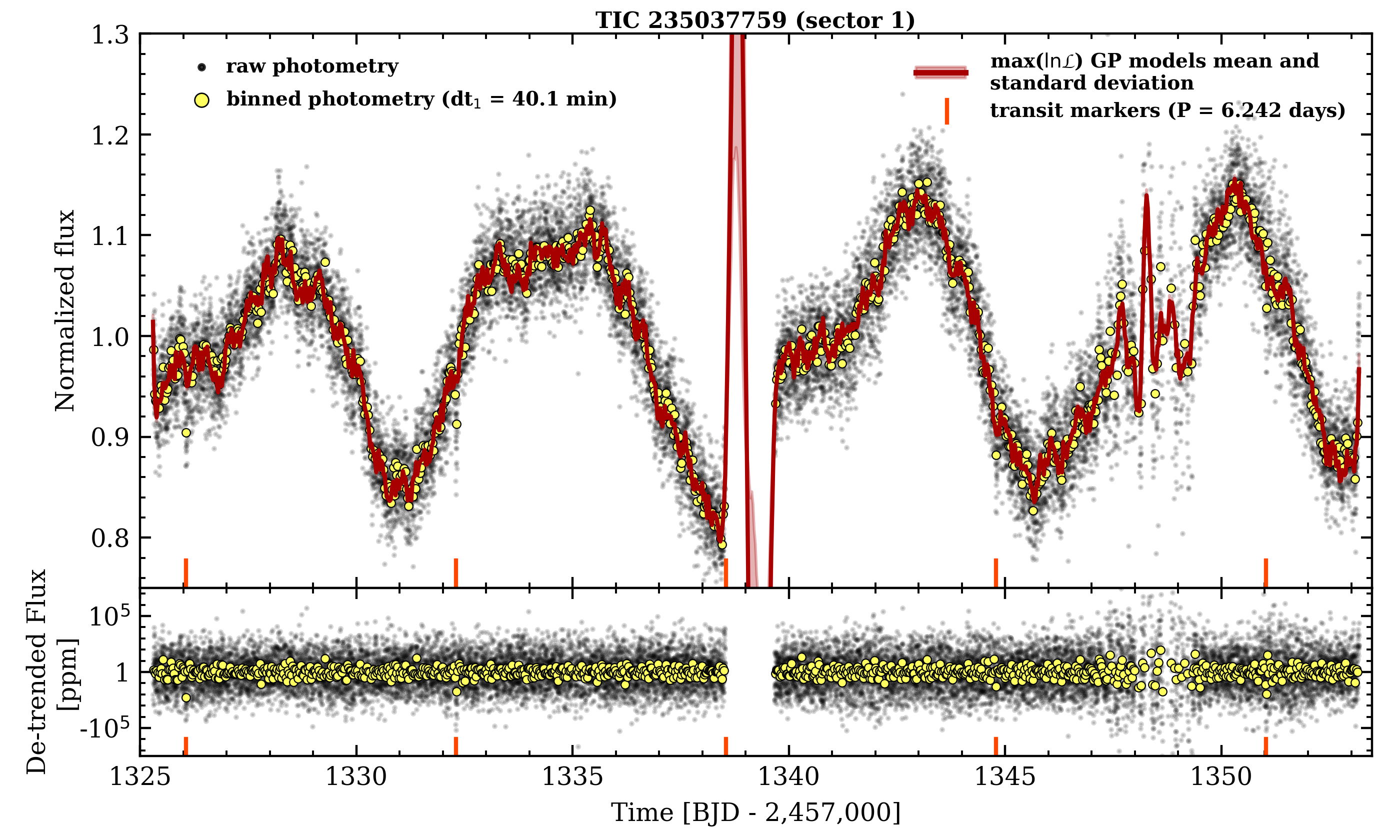}
  \includegraphics[scale=.96]{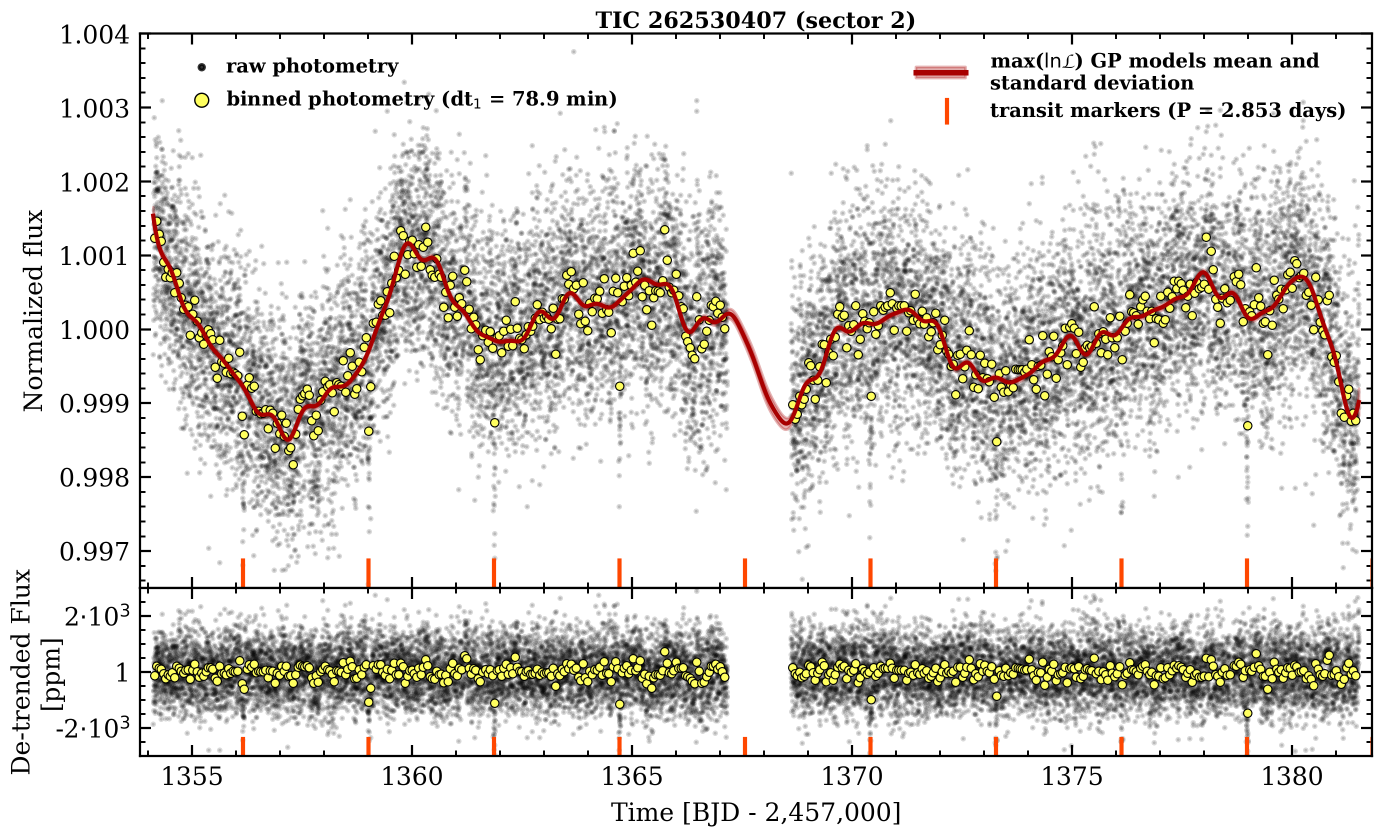}
  \caption{Two demonstrations of the \pipeline{} initial de-trending stage for TIC 235037759 and 262530407.
    \emph{Large panels}: the 2 minute raw and binned \tess{} light curves with temporal
    binning chosen to sample eight GP periodic timescales (40.1 and 78.9 minute bins for TIC 235037759
    and 262530407 respectively). Solid red curves and their surrounding shaded regions
    depict the mean GP model used for de-trending and its $1\sigma$ confidence intervals which are often
    small and difficult to visualize everywhere except during the data transfer gap at the centers of each
    light curve. Vertical ticks along the abscissa axes are indicative of the transit times of planet
    candidates which will ultimately be flagged by \pipeline{} as possible planet candidates at
    6.2 and 2.9 days around TIC 235037759 and 262530407 respectively. \emph{Shallow panels}: the raw and binned light curves after de-trending.}
  \label{fig:detrend}
\end{figure*}

Fig.~\ref{fig:detrend} depicts two examples of the results of this iterative de-trending procedure over
individual \tess{} sectors for TICs 235037759 and 262530407. The accuracy of each mean GP regression model,
with maximum $\ln$ likelihood hyperparameters, is clearly exhibited. The systematics model for TIC 235037759
is required to be much more aggressive than that for TIC 262530407
given the star's large photometric variability with a peak-to-peak
amplitude in the binned light curve of 280,000 ppm and a 85,000 ppm rms. Unlike in the raw
light curve, the de-trended light curve lacks any low frequency variations and exhibits a significantly
reduced rms of 12,000 ppm.

The TIC 262530407 light curve also exhibits photometric variability albeit with a much lower peak-to-peak
amplitude and rms of 1700 and 1000 ppm respectively. After de-trending, the rms is slightly reduced to 
810 ppm. The most important residual feature of the de-trended light curves is that they appear free
of the majority of large-scale systematic effects. This fact will facilitate the linear search for
transit-like events with minimal contamination from residual systematic features. Indeed 
a transiting planet candidate is detected
around each of these systems although the putative PC around TIC 235037759
is ultimately favored by an astrophysical false positive interpretation as presented in
Sect.~\ref{sect:gaiafps}.

One notable limitation to the effectively corrected light curve systematics is the prevalent increase
in rms for TIC 235037759 between BJD - 2,4570,00 $\sim 1347-1349$ during a brief period of loss of
\tess{} pointing precision.  This is a common feature to
many of the TICs observed during \tess{} sector 1. During this time, the extracted light curve is only
partially corrected for the pointing precision loss while the systematic GP model provides only marginal
improvements if any at all.
To ensure that the GP hyperparameters were not being strongly affected by the anomalous systematics structure
during this period, those measurements were masked and the GP hyperparameters re-optimized with the
remaining data. The resulting GP systematics only varies marginally from that shown in
Fig.~\ref{fig:detrend} for TIC 235037759 such that we are confident that the \pipeline{} de-trending is
largely robust to the loss in pointing precision for TICs observed in \tess{} sector 1.

Recall that in the initial de-trending step discussed in this section that the methodology
assumes a null mean function which implies that any transient events such as flares or transits are still
present during the optimization of the de-trending model.
The principal caveat with this methodology is that one cannot guarantee
that the GP model does not (partially) capture any of the in-transit light curve deprecations that
\pipeline{} is searching for. If partially suppressed by the initial GP model, planets will be more
difficult to detect due to the reduced signal-to-noise (S/N) of individual transit events.
Furthermore, transit events
that remain detectable within the de-trended light curve could result in underestimated transit depths
and correspondingly smaller planet sizes. To ensure a self-consistent planet+systematics model
for putative PC from \pipeline{,} the light model is revisited in Sect.~\ref{sect:joint}
with the inclusion of a transiting planet mean function in place of the null mean function used during
the de-trending step.

\subsection{Linear transit search} \label{sect:linearsearch}
Next a linear search for individual transit-like events is conducted on the de-trended light curves over
their full duration. The following methodology is reminiscent of a number of individual transit event
search algorithms
(e.g. Box Least Squares, \texttt{BLS}; \citealt{kovacs02}, Transiting Planet Search, \texttt{TPS};
\citealt{jenkins10,christiansen13,christiansen15,christiansen16}, \texttt{TERRA};
\citealt{petigura13a}, \citealt{foremanmackey15a}). The aim here is
to identify high S/N transit-like events along with their associated mid-transit times $T_0$,
durations $D$, and depths $Z$ which will feed into the search for repeating transit-like
events and ultimately the list of putative transiting PCs.

The linear search for transit-like events begins with stepping through
a two-dimensional grid of $T_0$ and $D$. At each $(T_0,D)$ grid point a simple box model of the form

\begin{equation}
  m(t) =
  \begin{cases}
    1-Z & \quad \text{if } T_0-\frac{D}{2} \leq t \leq T_0+\frac{D}{2} \\
    1 & \quad \text{otherwise,}
  \end{cases}
\end{equation}

\noindent is constructed with fixed $T_0$ and $D$. The box depth $Z$ (or mock transit depth)
is fit by $\ln$ likelihood maximization and saved along with the value of $\ln{\mathcal{L}}$
given the unique set of parameters $\{T_0,D,Z\}$. Computing $\ln{\mathcal{L}}$ with Eq.~\ref{eq:lnL}
implicitly assumes that the flux uncertainties are Gaussian distributed which allows for the
construction of a diagonal covariance matrix K with elements K$_{i,j} = \delta_{i,j} \sigma_{f,i}$.
The linear search along the $T_0$ dimension proceeds by stepping through the observation epochs
$\mathbf{t}$ in 30 minute bins and assigning
$t_{\text{bin},i}$ to $T_0$ $\forall$ $i=1,\dots,N_{\text{bin}}$.
This fixed binning is the first of many \pipeline{} free parameters which are listed in
Table~\ref{table:freeparams} along with their default values and brief explanations of their
effects. Initial \pipeline{} tests on synthetic light curves with injected transit models
determined that finer temporal binning did not result in a significant variation in the number
of detected high S/N transit-like events. This is likely due to 30 minute bins being more comparable
to typical transit durations of the types of planets that can be detected in 27-54 day baselines.
The $D$ dimension is sampled on a much coarser grid
given that the precision of the box model parameters are not yet required to infer planet properties
but only to identify epochs at which transit-like events are likely to have occurred. Explicitly, the
adopted linear search $D$ grid contains three possible transit durations of either 1.2, 2.4, or 4.8 hours.

\begin{figure}
  \centering
  \includegraphics[scale=.92]{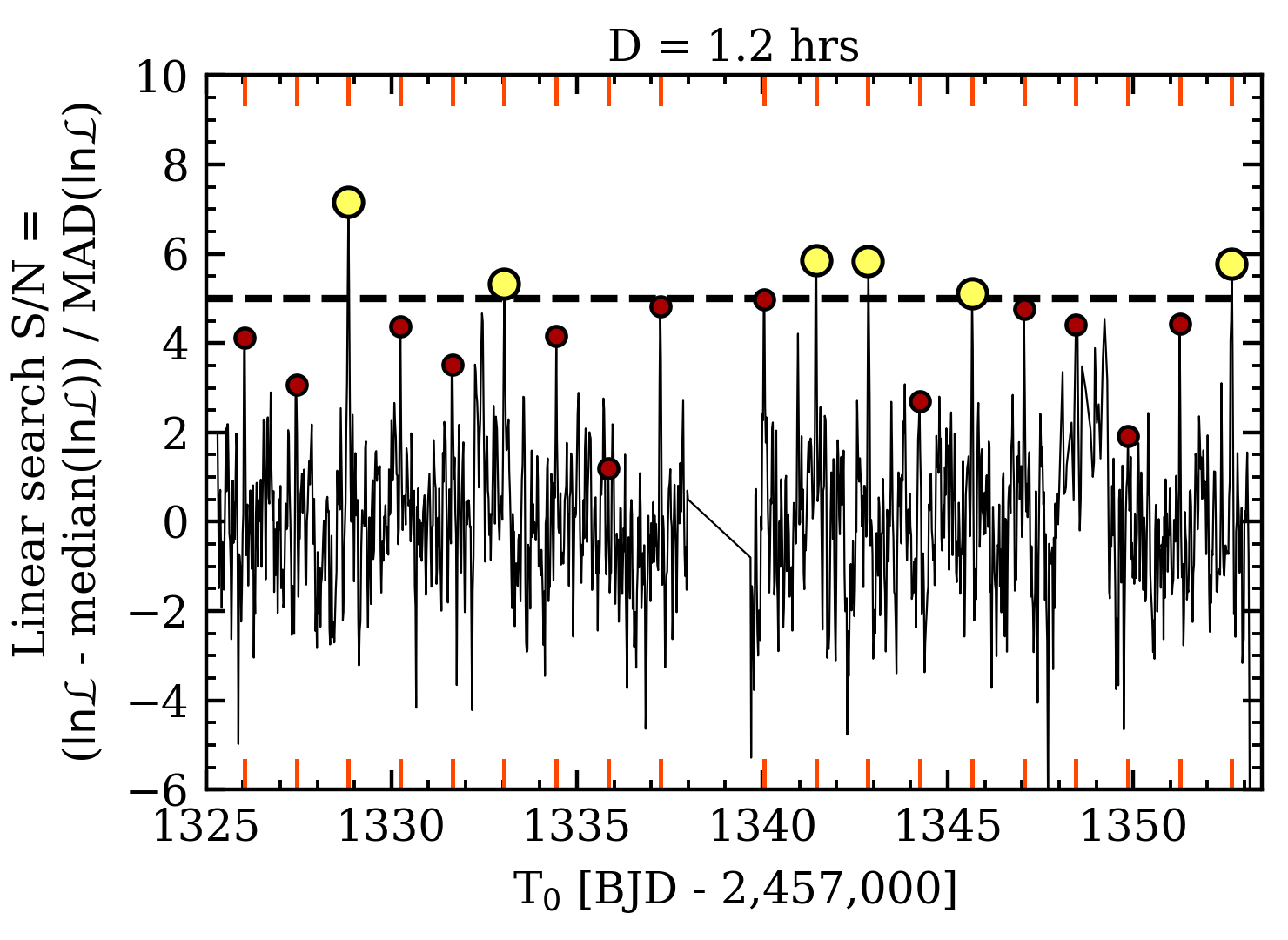}
  \caption{The linear search S/N spectrum versus transit times for TIC 234994474.
    The linear search S/N is calculated from the $\ln$ likelihood of the de-trended light
    curve in 30 minute bins under a box model centered at each transit time and with a fixed
    transit duration of 1.2 hours in this example. Mid-transit times of the PC candidate
    hosted by TIC 234994474 (TOI-134.01, $P=1.4$ days) are depicted by the vertical ticks
    and highlighted on the S/N spectrum with circular markers. The larger yellow markers
    indicate transit times when the S/N spectrum exceeds the imposed \pipeline{}
    detection threshold of S/N$_{\text{thresh}}\geq 5\sigma$.}
  \label{fig:linearsearch}
\end{figure}

This procedure produces a $N_{\text{bin}} \times 3$ matrix of transit times and durations
with each entry having an associated $\ln$ likelihood and optimized transit depth $Z$.
From the $\ln$ likelihoods a S/N spectrum as a function of transit times
is computed for each $D$ value considered. By translating the $\ln$ likelihoods by
their median value and normalizing by their median absolute deviation (MAD),
the aforementioned linear search S/N spectrum versus transit times is calculated.
The linear search S/N spectrum is analogous to the Signal Detection Efficiency in the
\texttt{BLS} algorithm. The conversion from
$\ln$ likelihoods to the ad hoc S/N spectrum centered around zero such aids in its interpretation as
each TIC's spectrum can be searched in absolute terms.
In adopting the median and MAD $\ln$ likelihood values over the mean and standard deviation, the S/N
is less sensitive to contamination by stochastic, short timescale photometric features
and results in a S/N spectrum whose baseline  is dominated by the light curve's
inherent photometric precision. An example linear search S/N spectrum is shown in
Fig.~\ref{fig:linearsearch} for a fixed duration of 1.2 hours. We note that referring to the
linear search S/N spectra as a S/N is a misnomer given that its values over $T_0$ can be negative.
However, we find this language to be a clear descriptor of the quantity's aim and its
interpretability.

The S/N spectra are then searched for high S/N transit-like events analogously to the individual
significance events in the primary Kepler mission which are combined and flagged as Threshold Crossing
Events \citep{jenkins10}.
High S/N transit-like events are flagged as peaks in the linear search S/N spectrum when
S/N$_{\text{thresh}}\geq 5$.
All $n_{\text{T}}$ transit times $\mathbf{T}_0$ with S/N exceeding S/N$_{\text{thresh}}$ for
any value of $D$, are compiled into a set of potential transit-like events.
Because transit times in the linear search are sampled on a fixed grid (i.e.
$\mathbf{t}$ binned to $\Delta t$), each transit time is refined by Gaussian smoothing the light curve
around $\pm 2D$ of $T_0$ using \texttt{scipy.ndimage.filters.gaussian\_filter} and updating $T_0$ to the
central time of the box model minimum before proceeding to search for periodic events that may
be indicative of transiting PCs. In the example shown in Fig.~\ref{fig:linearsearch},
a PC exists with $\sim 1.4$ day orbital period. Six out of the nineteen transit events
that occur within the sector 1 baseline are detected above S/N$_{text{thres}}$. This includes two
consecutive transits between 1341 and 1343 BJD - 2,457,000 which are used in the subsequent
section to infer its possible period equal to the time difference between the two events. 

\subsection{Periodic transit search} \label{sect:periodic}
\begin{figure*}
  \centering
  \includegraphics[scale=1]{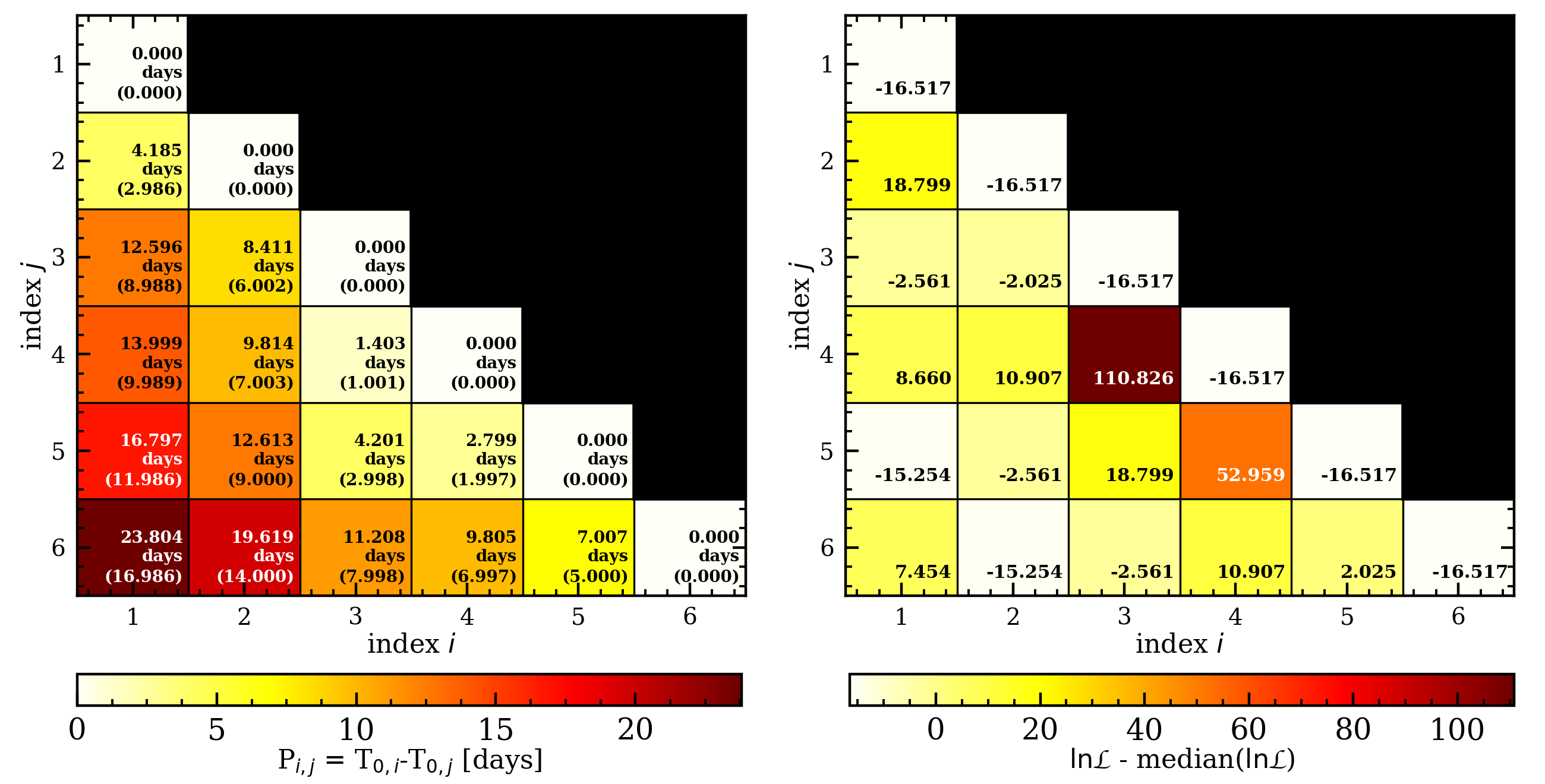}
  \caption{The results of the periodic search for repeating transit-like events in the
    light curve of TIC 234994474 which hosts TOI-134.01 at 1.40131 days.
    \emph{Left panel}: the $n_{\text{T}} \times n_{\text{T}}$ P matrix of
    possible periods of repeating transit-like events from the $n_{\text{T}}=6$ high S/N transit
    times detected in the linear search stage (see Fig.~\ref{fig:linearsearch}).
    P is skew-symmetric such that only the periods $>0$ below the diagonal are valid potential
    periods. Each P element is annotated in
    each grid cell along with its ratio to the true orbital period.
    Because all of the transit times detected during the linear search
    are associated with a transit of the PC, the P matrix elements are all close to
    integer multiples of the true orbital period. \emph{Right panel}: the
    $n_{\text{T}} \times n_{\text{T}}$ matrix of data $\ln$ likelihoods under a box transit
    model with orbital periods from the matrix P, and with mid-transit times, depths, and
    fixed duration (i.e. 1.2 hours) from the linear search stage.
    $\ln$ likelihood values along the diagonal correspond to the null hypothesis: a transit
    model with zero period. The PC period at $\sim 1.4$ days is clearly seen with the largest
    $\ln$ likelihood which then discards all other potential periods as multiples of $P$.}
  \label{fig:periodicsearch}
\end{figure*}

The chronologically sorted set of $n_{\text{T}}$ high S/N transit times in $\mathbf{T}_0$
are used to construct a matrix of differential transit times with
elements P$_{i,j} = T_{0,i} - T_{0,j}$ $\forall$ $i,j=1,\dots,n_{\text{T}}$.
A separate matrix is populated for each unique value of $D$. An example of
P is shown in Fig.~\ref{fig:periodicsearch} for TIC 234994474 for the fixed duration of 1.2 hours. 
The $n_{\text{T}} \times n_{\text{T}}$ matrix P represents potential transit periods of
PCs whose individual transit events
may be separated in time by any of the off-diagonal elements of P or some multiple thereof.
By its construction, the matrix P is skew-symmetric implying that only the positive non-zero matrix elements
below the main diagonal are valid periods for consideration, or $\sum_{i=1}^{n_{\text{T}}-1}i$ periods.
Because this search if for repeating transit-like events, it is required that $n_{\text{T}}>1$. In this
way, more than two transit events are not required in order to detect a repeating putative 
PC. This fact extends the \pipeline{} detection sensitivity to nearly the full observational baseline
or $\sim 27$ days for the majority of TICs.

Because each linear search with a unique $D$ is independent of the others, the P matrices of
differential transit times are considered together. In this way a single master set of periods
is compiled whose elements are referred to as Periods-of-Interest (POIs). Recall that each POI has
an associated time of mid-transit $T_0$, duration $D$, depth $Z$ and $\mathcal{L}$
from the linear search. At this stage,
the computationally tractable box transit model is substituted in favor of a more
physical transit model whose model parameters are initialized using the box model parameters before
optimization. The \cite{mandel02} transit model is used through its implementation within the
\texttt{batman python} package \citep{kreidberg15} to
compute model realizations given the input parameters
$\theta=\{P, T_0,a/R_s, r_p/R_s,i,e,\omega,a_{\text{LDC}},b_{\text{LDC}}\}$ where $e$ is orbital eccentricity,
$\omega$ is the argument of periastron, and $\{a_{\text{LDC}},b_{\text{LDC}}\}$ are the quadratic limb
darkening coefficients. In practice, only
the parameters $\theta=\{P,T_0,a/R_s,r_p/R_s,i\}$ are optimized by assuming circular orbits and fixing
the quadratic limb darkening coefficients in the \tess{} bandpass to the values interpolated
from the \cite{claret17} grid over \teff{,} \logg{,} and assuming solar metallicity. The $\theta$
parameters are optimized using the same routine which was used to optimize the GP hyperparameters during
the initial de-trending stage (see Sect.~\ref{sect:detrend}). With each POI's optimized transit
model, the $\ln$ likelihood of the data is computed for use during the succeeding steps
aimed at identifying repeating transit-like events from the initial set of POIs.

A series of cuts is performed on the set of optimized POIs to identify the most likely independent
periods within the data. The first cut is to remove repeated period multiples.
If POIs with integer period multiples (i.e. $2P_i,3P_i,\dots$) are indeed due to a transiting planet,
then those POIs are likely to be manifestations of the same object. The exact $P$ value that is retained
from a set of apparent period multiples is that
with the largest $\ln$ likelihood. Any arbitrary pair of POIs ($P_i,P_j$) is flagged
as a period multiple if $P_i$ and $n\cdot P_j$ are within $f_{\text{P}}=1$\% for any
$n=2,\dots,n_{\text{transit}}$ where $n_{\text{transit}}$ is the number of transits can occur within the
TIC's observational baseline given $P_i$ and $T_{0,i}$.
One important caveat to the removal of integer multiple periods is that resonant multi-planet systems
are undetectable within the periodic transit search because all but one of the POIs will always be
rejected in favor of its maximum $\ln$ likelihood multiple.

Similarly, because the set of POIs are derived from peaks in the linear search S/N spectrum, 
rational multiples of each POI must also be sampled (i.e. $P_i/2,P_i/3,\dots$).
Consider a S/N spectrum derived from a light curve containing a single transiting planet with
$n_{\text{transit}}>2$ but whose individual transits are only marginally detectable  due to their
amplitude relative to the photometric precision. Consider in this case that only a fraction of
transit events detected during the linear search.
The detection of only some transit events may result in a misidentified
POI that is an integer multiple times greater than the underlying true period if one or more
intermediate transit-events go undetected due to the effects of random noise.
Therefore fractional multiples of each POI must be considered. These new periods are equal to
$P_i/n$ for all integers $n$ resulting in a reduced period greater than or equal to the minimum
orbital period considered by \pipeline{:} 0.5 days. The $\ln$ likelihood of the data under
the box model with reduced period $P_i/n$ and remaining parameters $\{T_{0,i},D_i,Z_i\}$ is calculated
to be compared with the $\mathcal{L}$ value for the model with $P_i$. Here, the latter three parameters
are fixed regardless of the input period. In the search over period
multiples retains the period with the largest $\ln$ likelihood.
The lower period limit of 0.5 days is imposed to limit the number of rational multiples of each POI that
are investigated and because the temporal bin width used during the linear search stage already limits
the sensitivity to short-period planets whose transit durations are comparable to the 30 minute bins used
therein. Limiting the periodic search to orbital periods $< 0.5$ is
unlikely to result in a large number of missed transits owing to the intrinsically low occurrence rate of
ultra-short-period planets \citep[$\lesssim 1$\%;][]{sanchisojeda14,adams16}.

The result of the cuts to the initial set POIs are a set of repeating transit-like events which may or
may not correspond to a transiting planet or some other form or periodic astrophysical source
such as an eclipsing binary. The next steps in \pipeline{} are to vet the surviving POIs for systematic
false positives and eclipsing binaries given their distinctive light curve features that can largely
be vetted in an automated way.

\subsection{Automated planet vetting} \label{sect:vetting}
\subsubsection{Automated vetting based on light curve features} \label{sect:autovetting}
Here, POIs are automatically vetted using a set of eight vetting criteria which investigate
the flagged transit-like features in the de-trended light curve (see Sect.~\ref{sect:periodic}).
This automated vetting stage is intended to identify false or insignificant transit-like events
and thus provides a preliminary list of putative PCs prior to more selective human vetting
and statistical vetting for astrophysical false positives.

The automated vetting criteria are controlled by the set of free parameters
$\{c_i\}$ for $i=1,\dots,10$ which are described below and are included in the
summary Table~\ref{table:freeparams}.
The adopted values of these parameters controls the performance of \pipeline{} in terms of its
detection sensitivity and false positive rate and were derived
from early \pipeline{} executions on both archival \emph{Kepler} and simulated \tess{} light
curves\footnote{\url{https://archive.stsci.edu/tess/ete-6.html}} prior to the first \tess{} data
release. We do not however make any significant claims of their optimality.

\begin{figure}
  \centering
  \includegraphics[scale=.92]{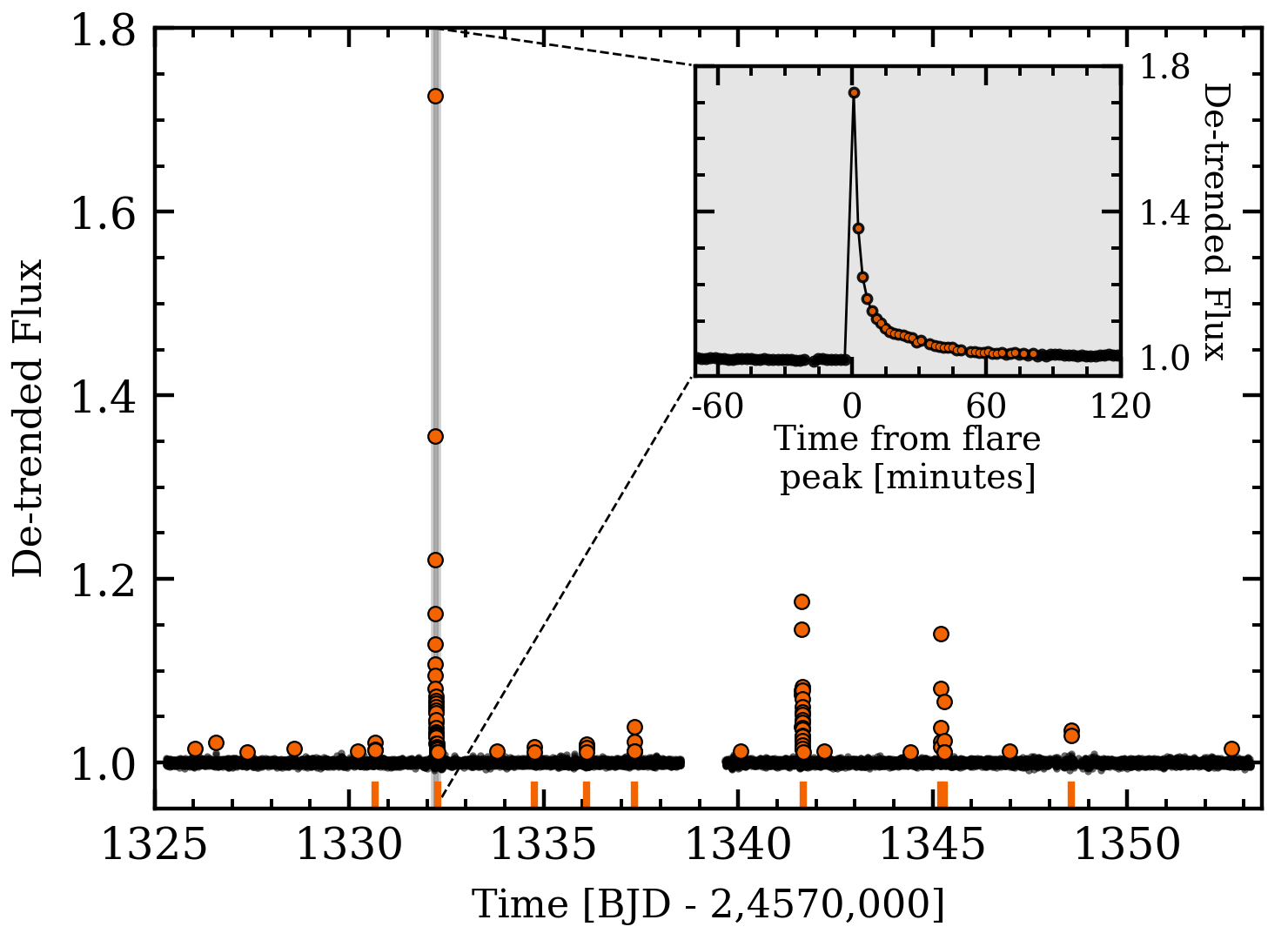}
  \caption{The de-trended light curve of the flare star TIC 25200252. Measurements initially flagged
    as being potentially associated with a flare event are highlighted by the
    large orange markers in the light curve. Only windows containing at least two successive
    bright measurements above this threshold are flagged as flares. TIC 25200252 is found to shows nine flares
    during the \tess{} sector 1 baseline which are marked by the vertical ticks along the abscissa axis.
    This includes two flares in quick succession near BJD-2,457,000 = 1345.2.
    \emph{Subpanel}: zoom-in of the event centered on 1332.212 BJD-2,4570,000 depicting the steep rise
    in flux and exponential decay which are characteristic of stellar flares.}
  \label{fig:flare}
\end{figure}

The eight automated vetting criteria are defined as follows:

\begin{enumerate}
\item It is required that each POI's transit depth $Z$ from its optimized transit model be $>0$.
\item The transit S/N is
  \begin{equation}
    \text{S/N}_{\text{transit}} = \frac{Z}{\sigma_{f,\text{transit}}} \sqrt{n_{\text{transit}}(P,T_0,\mathbf{t})}
  \end{equation}
  where $\sigma_{f,\text{transit}}$ is the photometric precision over the transit duration timescale
  and acts as a proxy for the Combined Differential Photometric Precision
  \citep[CDPP$_{\text{transit}}$;][]{christiansen12}. The number of observed transits is
  $n_{\text{transit}}$ given the vector of observations  $\mathbf{t}$, the POI's orbital
  period $P$, and mid-transit time $T_0$. The S/N$_{\text{transit}}$ is required to be $> c_1=8.4$.  
\item The transit parameters $P$, $a/R_s$, $Z$, and $i$ 
  are used to phase-fold the light curve and compute the transit duration \citep{winn10}
  such that the in-transit points, including those in ingress or egress, can be isolated.
  It is required that the difference in the median in and out-of-transit fluxes
  exceed $c_2=2.4$ median absolute deviations of the out-of-transit flux.
\item If a misidentified POI happens to be a less than the true period, then the phase-folded light curve
  will appear to contain out-of-transit points in-transit. This is combated by requiring that the number of
  in-transit points lying below $Z+\sigma_Z$ (where $\sigma_Z$ is the $1\sigma$ uncertainty on the
  transit depth) accounts for at least the a $c_3=0.7$ fraction of all in-transit points.
\item It is required that the in-transit sampling be approximately symmetric in time by insisting that the
  number of points between $T_1$ and $T_2$ be $\in [50-c_4,50+c_4]$\% of the total
  number of in-transit points (i.e. between $T_1$ and $T_4$). $c_4$ is set to $10$\%.
\item Flare stars such as TIC 25200252 as shown in Fig.~\ref{fig:flare} are found to result in a number
  of misidentified transit-events. Flare events are therefore searched within each light curve by first
  flagging individual flux measurements which are $>c_5=8$ median absolute deviations brighter than the
  median flux baseline. However, by the aforementioned criterion, individual stochastic flux jumps can also
  mimic flares. It is therefore required that any window over which a possible flare event occurs must contain
  $>c_6=2$ successive bright measurements above the $c_5$ threshold in order to identify a flare.
  Flux measurements occurring within a flare window are identified from the $q^{\text{th}}$ percentile
  of the light curve flux distribution where $q$ is the fraction of the observational baseline
  that occurs within a flare's duration. The total flare duration over the light curve is calculated from the
  number of detected flares multiplied by the characteristic M dwarf flare duration $c_7=30$ minutes
  \citep{moffett74,walkowicz11,hawley14}. Transit-like
  events with an identified flare occurring within $c_8=4$ transit durations from $T_0$ of a POI are vetted as
  flares.
\item Visual inspection of a number of TIC light curves observed during sector 1 frequently reveals sharp
  flux losses at the light curve edges. This signature is often falsely attributed to transit-like
  signal but is clearly a systematic effect that is not always well-modelled during de-trending stage.
  Because this edge effect appears to operate over the final $\sim 4-5$ hours of the light curve,
  POIs with mid-transit times within $c_9=4.8$ hours of either the first or final flux measurements
  are automatically flagged as probable false positives.
\item The optimized transit models of the POIs that satisfy all of the seven aforementioned criteria are
  removed from the light curve. This produces a maximally clean light curve whose residuals should only
  arise from random noise in photometry or from inaccuracies in the independent systematic GP and transit
  models. The \texttt{numpy.correlate python} function is used to compute the autocorrelation of the
  residual light curve as a function of time delay as light curves demonstrating large autocorrelations
  due to imperfect systematic models, can often mimic transit-like events which satisfy
  the previous vetting criteria. This criterion is particularly important for the use of \pipeline{} on
  \emph{K2} light curves which often exhibit significant temporal correlations due to the thrusts used for
  re-orientation of the spacecraft and its imperfect correction \citep{vanderburg14}. It is requires that
  the autocorrelation function for delays greater than zero be $\leq c_{10}=0.6$.
\end{enumerate}

\subsubsection{Automated vetting of eclipsing binaries} \label{sect:autoEB}
POIs that obey all eight automated vetting criteria from Sect.~\ref{sect:autovetting}
are passed along to vetting of eclipsing binary (EB) astrophysical false
positives. Six free parameters control the
performance of the astrophysical vetting procedure: $\{c_{\text{EB},i}\}$ for $i=1,\dots,6$.
The EB vetting criteria are adopted from a variety of sources
\citep{batalha10,bryson13,gunther17,crossfield18}
and are used to flag light curve features consistent with being an EB rather than a transiting
planet. The EB vetting criteria are defined as follows:

\begin{enumerate}
\item POIs are required to have $r_p/R_s < c_{\text{EB},1}=0.5$.
\item It is also required that the inferred companion radius $r_p < c_{\text{EB},2}=30$
  R$_{\oplus}$.
\item The observed transit duration $D$ is required to be less than the transit
  duration corresponding to planet with radius $c_{\text{EB},2}=30$ R$_{\oplus}$.
\item Searches for secondary eclipses are conducted by first sampling eclipse duty cycles (i.e. the fraction of the
  orbit during eclipse) from the \cite{shan15} duty cycle PDF (see their Fig.~4). This distribution was
  derived from a synthetic population of M dwarf EBs based on \emph{Kepler} binary statistics and is
  dependent on the population of EB total radii, total masses, orbital periods, and eccentricities. In
  each POI's light curve, phase-folded points occurring within a duty cycle centered on
  phase$=0.5$ are considered for possible contamination by a secondary eclipse.
  With the in-eclipse points defined, the secondary eclipse depth $Z_{\text{occ}}$ and 
  photometric precision during the occultation $\sigma_{\text{occ}}$ are computed. It is required that EBs
  satisfy the following conditions:
  \begin{align}
    \frac{Z_{\text{occ}}}{\sigma_{\text{occ}}} &>c_{\text{EB},4}, \\
    \frac{Z-Z_{\text{occ}}}{\sqrt{\sigma_{f,\text{transit}}^2+\sigma_{\text{occ}}^2}} &>c_{\text{EB},4}
    \label{eq:occ}
  \end{align}
  \noindent Recall that $Z$ and $\sigma_{f,\text{transit}}$ are the transit depth and in-transit photometric
  precision. $c_{\text{EB},4}$ is set to 5 \citep{gunther17} and EBs identified by this criterion are required
  to have $>c_{\text{EB},5}=50$\% duty cycle samples satisfy the above conditions.
\item EBs also exhibit `V'-shaped transits due to the self-luminosity of each companion. To search for
  `V'-shaped transits, the ingress time $T_{12}$ and egress time $T_{34}$ are calculated from the
  optimized transit model and compared to the total transit duration $D$. 
  `V'-shaped transits are required to have $T_{12}+T_{34}$ which are $c_{\text{EB},6}\geq 90$\% of $D$.
  Notably, `V'-shaped transits may also be indicative of planetary transits at large impact parameters
  so `V'-shaped transits are not explicitly discarded but are instead assigned a
  non-definitive disposition based solely on this criterion.
\end{enumerate}

\subsection{Joint systematic plus transiting planet modelling} \label{sect:joint}
The set of transit-like events that satisfy all of the  vetting criteria
presented in Sects.~\ref{sect:autovetting} and~\ref{sect:autoEB} are treated
as PCs in this, the final \pipeline{} stage. At this point the modelling of systematic
light curve effects using a 1-dimensional GP regression model from Sect.~\ref{sect:detrend}
is revisited. The alternative is to simultaneously sample the joint GP plus transit light curve
parameter posterior PDF using Markov Chain Monte-Carlo (MCMC) simulations.
Explicitly, the light curve model is modified by replacing the previously null mean model
$\boldsymbol{\mu}(\mathbf{t})$ with a full transit model containing all putative PCs. 
Overfitting by the systematic model, which can partially fill in planetary
transits, is mitigated by simultaneously modelling systematics and PCs.
The resulting joint systematic+planet model is therefore derived in a self-consistent
manner with more robust solutions for the transiting PC parameters of interest.
MCMC sampling of the transit parameter marginalized posterior PDFs allows us to compute
point estimates of their MAP values and uncertainties for later use.

For systems containing $N_{\text{PC}}$ PCs, $4+5N_{\text{PC}})$ parameter PDFs are sampled
by continuing to insist on circular orbits and fixed limb-darkening coefficients.
Explicitly, the GP
hyperparameters $\Theta = \{\ln{a_{\text{GP}}},\ln{\lambda},\ln{\Gamma},\ln{P_{\text{GP}}}\}$
are fit along with the
transiting planet parameters $\theta = \{P_i,T_{0,i},a_i/R_s,r_{p,i}/R_s,i_i \}$
$\forall$ $i=1,\dots,N_{\text{PC}}$. The GP hyperparameters are initialized to their maximum
likelihood values from the de-trending stage and are continued to be bounded by the broad
uniform priors listed in Table~\ref{table:priors}. Transit parameters are
initialized by their maximum likelihood values assuming fixed GP hyperparameters from
de-trending. The adopted prior PDFs on the transit model parameters are also reported
in Table~\ref{table:priors} for the most common case of fitting transiting PCs with
multiple transits over the observational baseline.
As we will see in Sect.~\ref{sect:ste}, some priors will be modified when sampling transit
parameters used to model only a single transit. 

MCMC sampling is performed with the \texttt{emcee}
ensemble-sampler \citep{foremanmackey13}. One hundred walkers are initialized in small Gaussian balls
centered on each parameter's initial value. Throughout the MCMC, each walker's acceptance
fraction is monitored and warns the user when its mean value over all walkers does not
fall within the desired range of 20-60\% in either of the burn-in or final sampling stages.
The desired duration of each MCMC stage is $\gtrsim 10$ autocorrelation times. Warnings are again produced
if the MCMC chains fail to reach this length.

\begin{figure*}
  \centering
  \includegraphics[scale=.88]{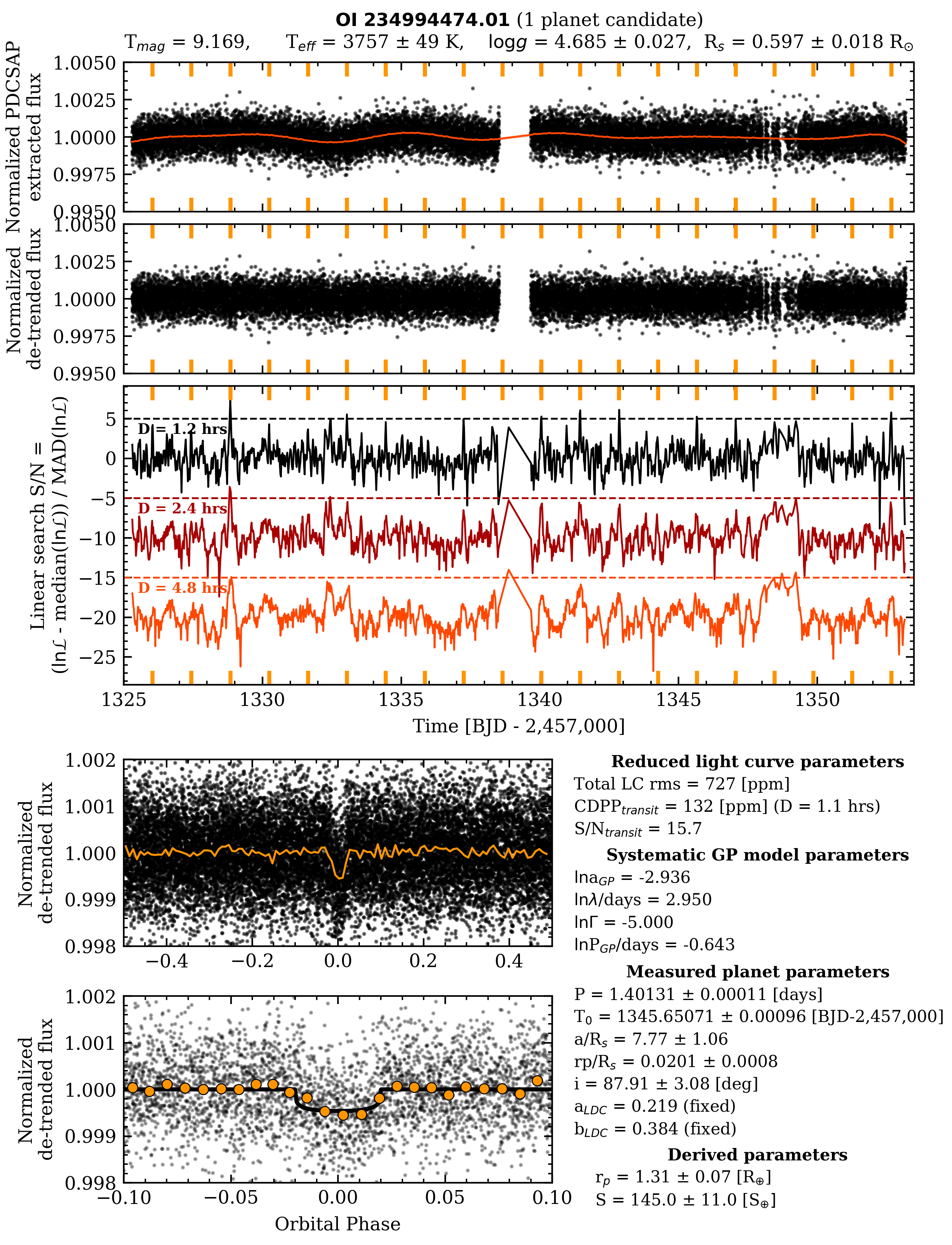}
  \caption{The summary image output from running \pipeline{} on TIC 234994474 and the resulting
    detection of OI 234994474.01 which is consistent with the known TOI-134.01. The \tess{}
    magnitude $T$ and \gaia{-}derived physical stellar parameters 
    are annotated at the top. \emph{Top panel}: the 2 minute extracted light curve from
    the TESS Science Processing Operations Center along with the mean GP systematic model
    (orange line) and the
    times of TOI-134.01 planetary transits indicated by the vertical ticks. \emph{Second panel}: the
    de-trended light curve. \emph{Third panel}: the linear search S/N spectra calculated
    from the likelihood of the data given a box model with fixed mid-transit time $T_0$ and for
    each of the three fixed transit durations $D$ (i.e. 1.2, 2.4, and 4.8 hours).
    Each spectrum is offset for clarity along with the $5\sigma$ S/N threshold.
    \emph{Fourth panel}: the complete and binned ($\Delta t = 0.2D/P$) de-trended light curve
    phase-folded to the MAP orbital period $P$ and mid-transit time $T_0$ of the planet candidate.
    \emph{Bottom panel}: zoom-in on the transit in the de-trended and phase-folded light curve.
    Various diagnostic quantities are reported in the lower right corner along with measured
    and derived transit parameters.}
  \label{fig:summary}
\end{figure*}

The main \pipeline{} output is a list of objects of interest (OIs)
along with samples from the model parameter marginalized
posterior PDFs and point estimates of each parameter's MAP value and uncertainties derived from the
$16^{\text{th}}$ and $84^{\text{th}}$ percentiles of their 1-dimensional marginalized PDF.
The raw light curves and models
sampled at observation times $\mathbf{t}$ are also saved to produce summary images 
such as the example shown in Fig.~\ref{fig:summary} for OI 234994474.01  
whose PC is a known TESS Object of Interest: TOI-134.01, a close-in terrestrial planet being validated
by HARPS and PFS RVs \citep{astudillodefru19}.

\section{\pipeline{} planet search around low mass stars in TESS sectors 1 \& 2} \label{sect:search}
For its inaugural application we apply the \pipeline{} transit detection pipeline to the
2 minute extracted light curves from the first two \tess{} sectors. 
Overall, \pipeline{} produces a set of 121 OIs around 96 of the 1599 low mass TICs in our stellar
sample after automated vetting.
It is expected that many of these postulated PCs will be false positives due to
imperfect corrections of systematic effects or to astrophysical false positives other than eclipsing
binaries as EBs have certain distinct photometric features which are flagged during the automated 
vetting stage. As such, we proceed with manual vetting of all \pipeline{} OIs via human inspection
of the pipeline's output. This step is particularly important for newly developed transit
search algorithms to develop an understanding of common sources of false positives and if or how
they may be corrected in future versions.

\subsection{Manual vetting of \pipeline{} planet candidates} \label{sect:manual}
We conduct a visual inspection of each of the raw, de-trended, and phase-folded light curves for
each OI produced by \pipeline{.} 
In this analysis we flag 97/121 PCs as being residual systematic effects which are
misidentified as transits. Of those, 19 appear to have been directly affected by measurements
obtained between $\sim$ 1347 and 1349 BJD - 2,4570,000 during sector 1 at times when \tess{} briefly lost much
of its pointing precision. This effect is not perfectly corrected for in many of the sector 1 extracted light
curves nor by our own systematic modelling. Most of the remaining OIs flagged as false positives are attributable
to residual systematics mimicking transits. Visual inspection indicates that these OIs are clearly
inconsistent with being a planetary transit.

\subsubsection{Peculiar light curve features}
\begin{figure}
  \centering
  \includegraphics{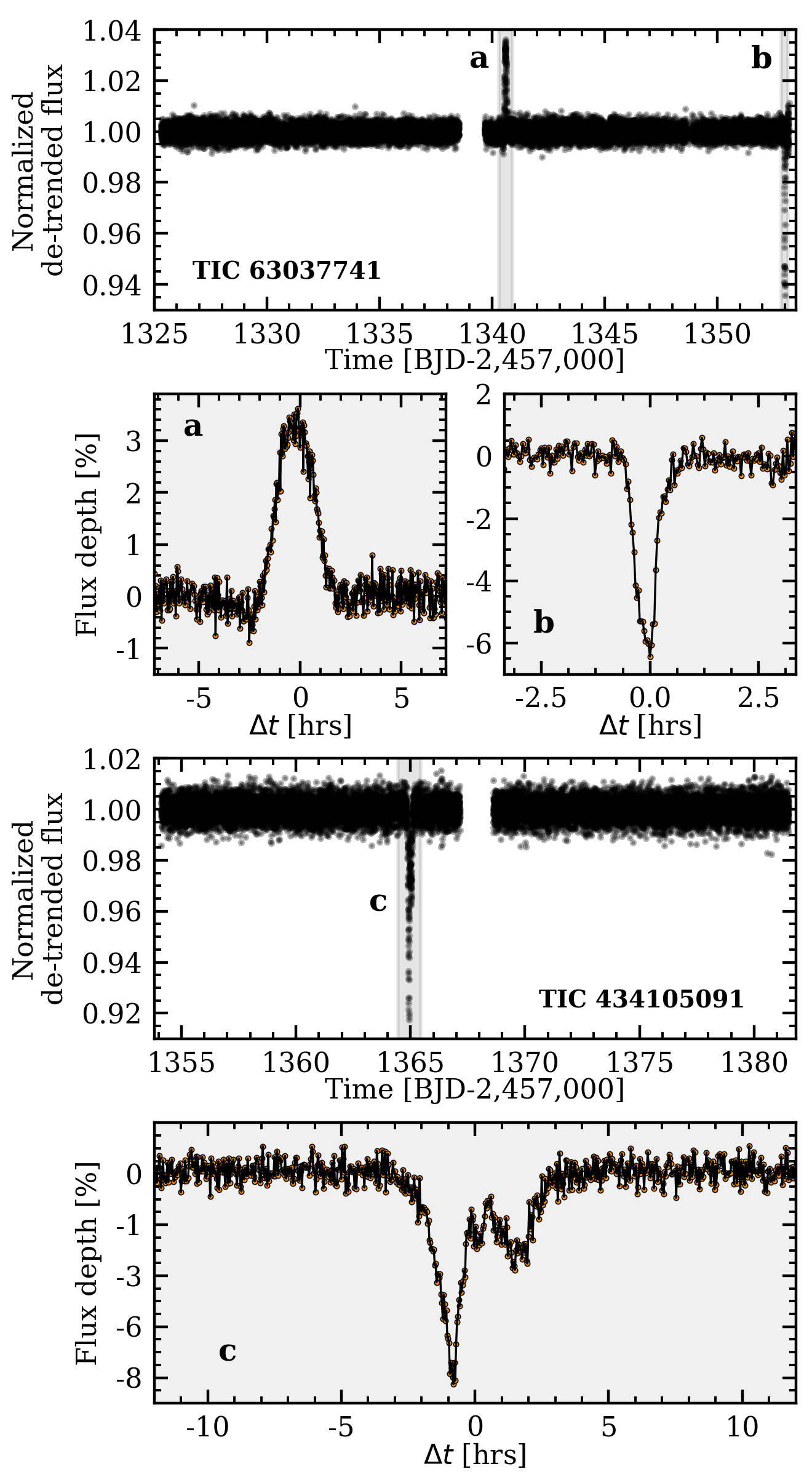}
  \caption{The raw (i.e. pre-de-trended)
    light curves of the TICs 63037741 and 434105091 which exhibit peculiar
    photometric features. \emph{Upper panel}: the full TIC 63037741 light curve
    showing two anomalous features at 1340.5 and 1353 BJD - 2,457,000 respectively. \emph{Panels a and b}:
    zoom-in on these features to reveal their structure. \emph{Fourth panel}: the full TIC 434105091
    light curve with an anomalous feature at 1365 BJD - 2,457,000. \emph{Panel c}: zoom-in on this
    feature depicting 2-3 dips over its full duration. The origins of these features are as of
    yet unknown.}
  \label{fig:wtf}
\end{figure}

The two OIs 63037741.01 and 434105091.01 produced by \pipeline{} exhibit peculiar
light curve features which do not appear to be consistent with planetary transits or residual systematics.
These features are depicted in Fig.~\ref{fig:wtf}. Each of the two TIC 63037741 features and the
TIC 434105091 feature are clearly seen at a high S/N either above or below the surrounding flux continuum.
The TIC 63037741 flux excess feature (`a' in Fig.~\ref{fig:wtf}) is approximately 4\% brighter than the baseline
flux and has a temporal evolution which is inconsistent with being a transient flare given its lack of a steep rise
in flux followed by an exponential decay \citep{hawley14}. The second TIC 63037741 feature (`b') has a 6\%
decrease in flux with a similar structure to feature `a', except that it is inverted and with approximately half
of the timescale. Feature `b' also appears to have
a slightly longer egress time than its ingress which can be a signpost of an extended transiting object if
astrophysical in origin.
Lastly, feature `c' around TIC 434105091 shows 2-3 flux dips over the entire `transit' duration lasting
$\sim 6$ hours. We remain agnostic as to the exact interpretation of these features but their existence is highlighted
here for any interested parties.

\subsubsection{Single transit events} \label{sect:ste}
The interpretations of the remaining 24/121 OIs as transiting planets
are deemed plausible by our manual vetting. These preliminary PC dispositions are based purely on the
resemblance of the light curve features to periodic transit events or in some instances, to single
transits (ST) events that do not show compelling evidence for periodicity over the observational baseline.
We note that \pipeline{} is not optimized nor intended to be sensitive to the detection of ST
events. However, some ST events with moderate to high S/N can be detected but incorrectly classified as
periodic events if at least one other transit time in the linear search S/N spectrum exceeds
S/N$_{\text{thresh}}$, even if that event is resulting from noise (see Sect.~\ref{sect:linearsearch}).
In this case, the ST event is folded to the time difference between the ST and any of the other S/N
events exceeding S/N$_{\text{thresh}}$. If the ST event has a sufficiently high S/N on its own
then the addition of noise by phase-folding to the incorrect
period, but correct $T_0$, may result in a feature that still passes our automated vetting criteria due solely
to the significance of the ST event. The inferred period of such a ST event in \pipeline{} will therefore
always be less than its true period if the feature is indeed the result of a singly transiting planet.

Three OIs resembling ST events are identified during manual vetting around the TICs 49678165, 92444219,
and 415969908. The latter TIC already hosts the known TOI-233.01 at 11.7 days but the proposed ephemeris for
TOI-233.01 is inconsistent with the $T_0$ of our putative ST event. A more complete discussion of
this and the other individual systems is reserved for Sect.~\ref{sect:indiv}.

For each of the three OIs classified as a putative ST, we refine
their transit parameters by re-modelling their transits by isolating the light curve around $10D$ and
using MCMC to sample the transit model parameters with just a single transit \citep{seager03}. In these
simulations the $P$, $T_0$, and $a/R_s$ priors are modified as listed in Table~\ref{table:priors}.
The orbital period of the ST is further restricted to periods greater than the largest time difference
between $T_0$ and both edges of the light curve's baseline. 
The resulting periods are largely uncertain with their posterior PDFs showing extended tails out to long
periods $\gtrsim 100$ days, as is expected for transit models which lack multiple events to constrain $P$.
The revied transit parameters for our three putative ST events, derived from the isolated data and with
the aforementioned updated priors, are used in place of the transit model solutions
produced by \pipeline{} with their likely underestimated periods.

\subsection{Statistical validation of transiting planet candidates} \label{sect:vespa}
We are currently not in a position to distinguish between confirmed planets and various astrophysical false
positive scenarios in an absolute sense. This is because of the lack of follow-up observations in this study
which are ultimately required to validate or disprove the planetary nature of our OIs. 
Despite the lack of such follow-up observations, it is still advisable to attempt to statistically
validate OIs by inferring the relative probabilities of a variety of astrophysical false
positive (AFP) scenarios which can be compared to the planetary interpretation.
Such considerations are further motivated given that the rate of AFPs in
the 2 minute \tess{} light curves is expected to be significant \citep[$\sim 60$\%;][]{sullivan15}.

We attempt to statistically validate our 24 OIs around 22 TICs using the \texttt{PyMultinest}
\citep{buchner14} implementation of the probabilistic transit validation software
\vespa{} \citep{morton12,morton15} which computes the planetary false positive probability (FPP)
for use in establishing the final dispositions of our OIs.
\vespa{} considers six AFP scenarios as potential explanations for transit-like
signals. These include undiluted eclipsing binaries (EB), hierarchical triple EBs (HEB), and blended
background or foreground EBs which are not physically associated with the target (BEB). Each of
these scenarios then have two instances with the first assuming the input orbital period and the
second assuming twice the input orbital period.
We note however that the forthcoming statistical OI interpretations are not treated as abolsutely
definitive in lieu of follow-up observations aimed at distinguishing transiting planets from AFPs.

For \vespa{} input we use the TIC's celestial coordinates ($\alpha$,$\delta$),
stellar parameters \teff{,} \logg{,} and $\varpi$, along
with the star's $JHK_{\text{S}}$ photometry. \vespa{} also requires the photometric band in which the
putative transit is detected but the code cannot properly handle the \tess{} bandpass in its current
version. Fortunately, the central wavelengths of the \tess{} bandpass and the Cousins $I_C$-band
are similar but with the \tess{} band being much wider \citep{sullivan15}. Given the similarity of the
$I_C$ and SDSS $i$-band, and the compatibility of the latter within \vespa{,} we use $T$ and
\Ks{} to derive $i$ using the color relation from \cite{muirhead18}. We also pass to \vespa{} the
OI's orbital period and planet-to-star radius ratio along with 
its de-trended light curve following the removal of all candidate transit models that are not associated
with the OI being statistically validated. The light curves are phase-folded and restricted
to $\pm 3D$ around $T_0$ for comparison to light curve models generated under the transiting planet
and each AFP scenario.

\vespa{} also requires constraints on the maximum angular separation
(\texttt{maxrad}) from the target star that
should be searched for potential blending sources. We limit this separation to be less than the median
full width at half maximum (FWHM) of the target's approximate point spread function (PSF).
The FWHM of the PSF is derived by fitting a 2-dimensional Gaussian profile to the
target image in each target pixel file over time and adopting the median FWHM as \vespa{} input.
The median FWHM value among the 22 TICs is $\sim 37$ arcsec or nearly two \tess{} pixels across.
Over such a large field, it is reasonable to expect that many of the OIs may be favored by either the
BEB or BEB2 models.

Lastly, \vespa{} requires
the maximum permissible depth of a secondary eclipse of an EB (\texttt{secthresh}) to be specified.
Recall that attempts within \pipeline{} were made to automatically vet EBs among our OIs in
Sect.~\ref{sect:autoEB}.
We therefore expect that \vespa{} is unlikely to detect any probable EBs. Nevertheless, the input
\texttt{secthresh} value for each TIC is derived from
the box model depths fitted to each transit time in $\mathbf{T}_0$ during the linear search stage
(see Sect.~\ref{sect:linearsearch}). After masking measurements occurring within the PC's
transit window and extrapolating the fitted depths to the PC's transit duration,
we adopt the 95$^{\text{th}}$ percentile of the depth distribution depths as the value of \texttt{secthresh}
\citep{crossfield18}. The median \texttt{secthresh} is $\sim 2100$ ppm. The input \texttt{maxrad} and
\texttt{secthresh} for each OI are reported in Table~\ref{table:vespa} along with the results of
our \vespa{} calculations.

The true power of \vespa{} is realized when additional follow-up observations such as 
contrast curves from AO-assisted imaging or photometric follow-up
are used to inform the interpretation of transiting PCs.
Given the lack of such data in this study, we adopt conservative limits on the interpretation of the
resulting \vespa{} probabilities. Similarly, we also do not claim to validate planets with ultra low FPP
\citep[$<0.01$; e.g.][]{montet15,crossfield18,livingston18} as we will caution in Sect.~\ref{sect:gaiafps}
that \vespa{} results should not be taken absolutely in the absence of follow-up
observations. Our limiting values on interpreting FPPs are as follows: OIs with FPP$<0.1$
are classified as PCs. Similarly, OIs are classified as AFPs when
FPP $\geq 0.9$ and have their dispositions assigned to the specific AFP model with the highest
probability. OIs with intermediate FPPs are classified as putative planet candidates (pPC).

The statistical validation calculations with \vespa{} result in 13/24 of our
OIs being classified as PCs plus 3/24 as pPCs.
The OIs 49678165.01, 415969908.02, and 92444219.01 correspond to the three
ST events detected during the manual vetting stage. We reclassify these objects as STs and a pST
respectively. Seven of the remaining eight OIs are favored by either BEB
model with 4/7 BEBs and 3/7 BEB2s. The
MCMC during the \vespa{} calculation of the lone remaining OI 235037759.01 failed to converge
leaving its disposition as of yet undefined. We will
show in the following subsection that despite the failure of the FPP calculation, the nature of OI
235037759.01 is likely to be an AFP. Although we are unable to distinguish between the different
AFP scenarios. The derived rate of AFPs from this small sample of OIs is $\sim 33\pm 12$\% which
is somewhat lower than expected AFP rate of 60\% from the \tess{} simulations by \citep{sullivan15}.

\subsection{Querying \gaia{} sources to supplement statistical validation calculations} \label{sect:gaiafps}
\begin{figure*}
  \centering
  \includegraphics{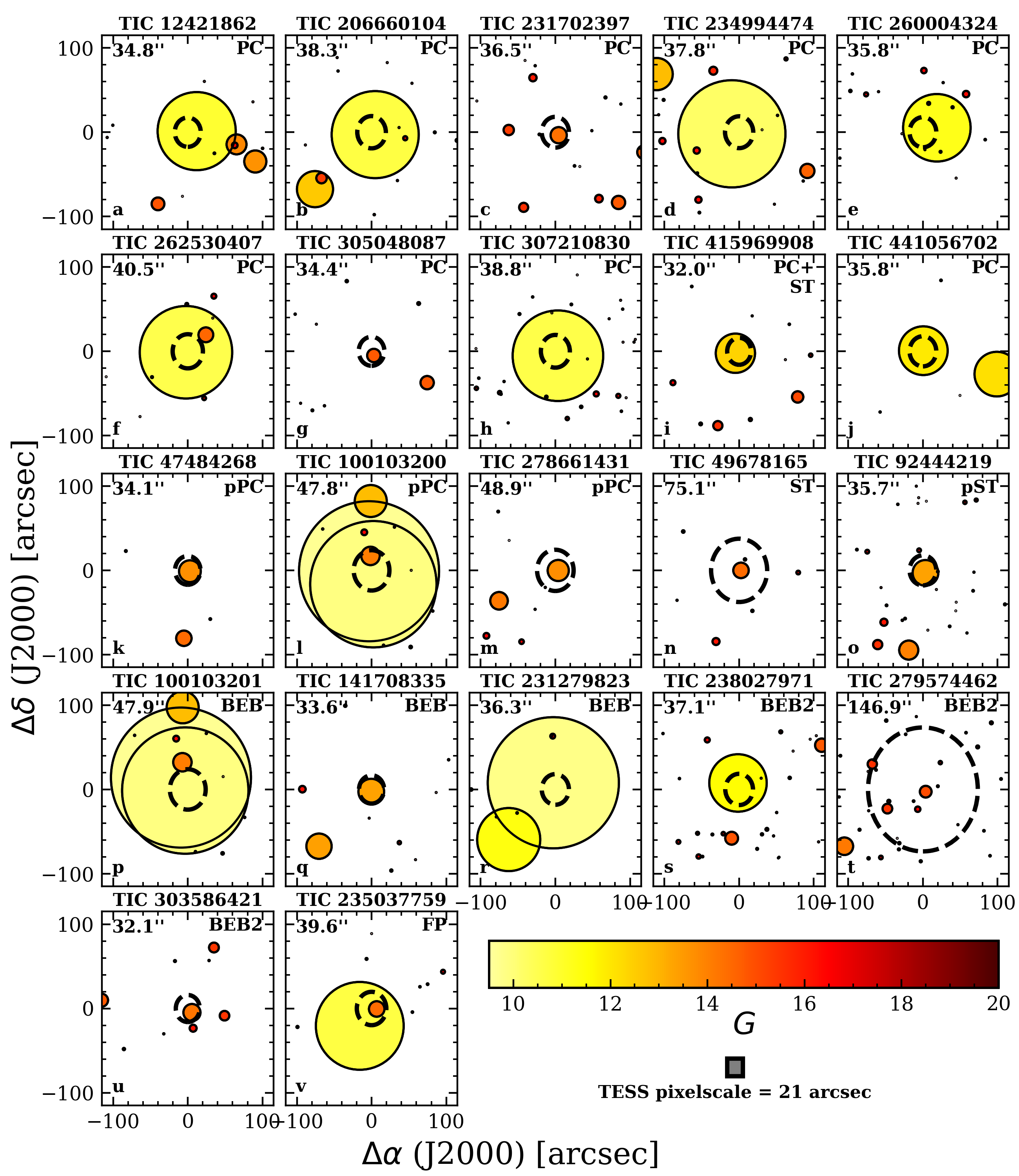}
  \caption{Star maps containing sources from the \gaia{} DR2 in the vicinity of each TIC identified
    as hosting an object of interest during the manual vetting stage. The panels are ordered by the
    dispositions which are annotated in the upper right of each panel (see Sect.~\ref{sect:vespa} for
    definitions). The fitted FWHM of each targeted TIC's PSF is annotated in the upper left
    of its panel in arcsec as well as being depicted by the dashed black circle centered on the
    panel's origin. The colorbar is indicative of $G$-band magnitudes while marker sizes are proportional
    to the source flux in that band. For reference, the size of a single \tess{} detector pixel is shown
    in the lower right of the figure.}
  \label{fig:gaiafps}
\end{figure*}

\vespa{} calculations are based on synthetic stellar populations from the
\texttt{TRILEGAL} galaxy model \citep{girardi05}. These synthetic results can be supplemented
by querying the \gaia{} DR2 in the vicinity of each TIC to investigate the number density and brightness
distribution of nearby sources on the sky. In this way, we hope to find supporting empirical evidence for
any of the BEB interpretations of OIs with high FPPs as those inferences should be expected if nearby bright
sources fall within or near the PSF of the targeted TIC. The resulting maps of \gaia{} sources around
the 22 TICs with OIs in our sample are shown in Fig.~\ref{fig:gaiafps}. Querying the \gaia{} DR2 is
performed identically to the method used in Sect.~\ref{sect:gaia} to match TICs with the \gaia{} DR2 catalog
although here we conduct our searches with a fixed radius of 105 arcseconds or $\sim 10$ \tess{} pixels
in diameter.

From Fig.~\ref{fig:gaiafps} it is clear that many of the statistically favored interpretations as either 
some form of planet candidate or a BEB are consistent with the lack or prevalence of nearby bright sources
to the targeted TIC respectively.
All panels with planet candidate OIs in Fig.~\ref{fig:gaiafps}, with the exception of panel `l', show no
or minor sources of comparable brightness within or very close to the target PSF edges as to significantly
contaminate the measured TIC photometry and consequently result in a probable FP.
Similarly, all AFP panels other than `q'
and `s' do have at least one neighbouring source of comparable brightness that may be responsible
for the favorability of an AFP scenario by \vespa{.}
This includes the TIC 235037759 whose \vespa{} calculation failed.
Two bright sources are clearly seen to contribute to the flux within the target's PSF thus supporting the
probable interpretation of the OI 235037759.01 transit-like event as being caused by an AFP.

\begin{figure}
  \centering
  \includegraphics{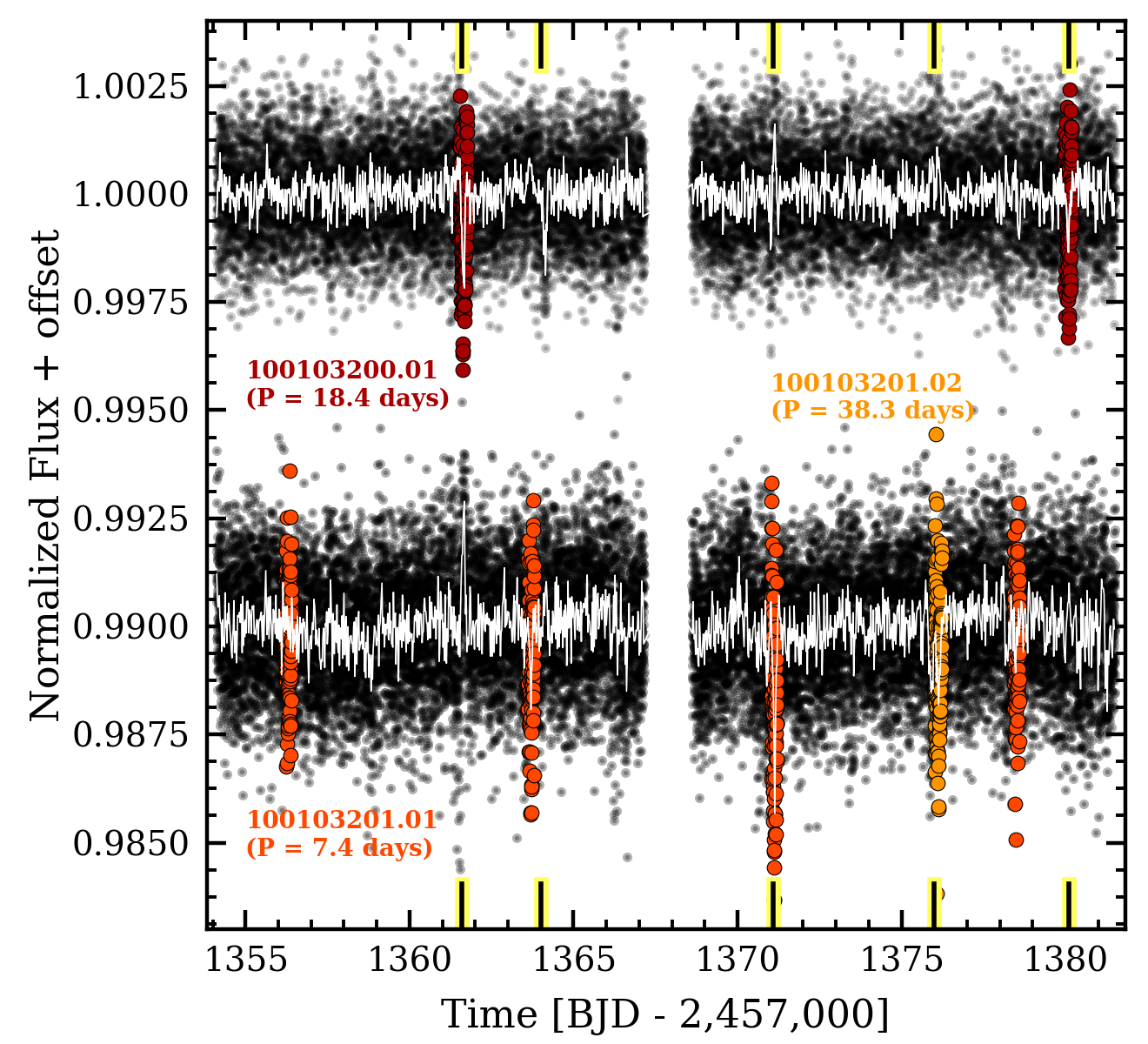}
  \caption{The de-trended light curves of TICs 100103200 (top) and 100103201 (bottom).
    The in-transit points for the three
    OIs in these systems are highlighted by the colored points with each OI having a unique color.
    The OI identifications and periods are annotated next to one of its transit-like events.
    The white curves represent the light curves binned to 30 minutes. The binned light curves reveal
    $\sim 5$ transit-like events (indicated by the vertical ticks along the abscissa axes) which appear to be
    due to transit-events in one TIC's light curve but also having clear manifestations in the other
    TIC's light curve despite the ephemerides of all three OIs being out of phase and non-commensurate.}
  \label{fig:tic300}
\end{figure}

We note however that some discrepancies between the distributions of \gaia{} sources and our \vespa{}
interpretations still persist. Particularly with regards to the TICs 100103200 and 100103201 (panels `l'
and `p' in Fig.~\ref{fig:gaiafps}) which strongly favor the PC and BEB models respectively despite being
located within 1 \tess{} pixel of one another on the sky, having very
similar brightnesses (i.e. $J=7.50, 7.66$), and being located at effectively identical distances (i.e.
$d=16.745$ pc). Such properties are reminiscent of an M dwarf binary system.
Perhaps naively, we might expect this architecture to favor almost any AFP scenario for
both TICs including blends, an EB, or an HEB. Indeed the apparent flux dips which appear qualitatively
consistent with a transiting planet around either one of the TICs, is also seen to have a clear manifestation
in the light curve of the other as evidenced in Fig.~\ref{fig:tic300}. This is almost certainly caused by the
overlap of each target's PSF.
However \vespa{} results indicate FPPs that differ by over two orders of magnitude between the two TICs.
Perhaps it is feasible, although seemingly unlikely, for TIC 100103200 to host a detectable transiting
PC while being blended with the nearby TIC 100103201 whose transit-like events are strongly favored by the
BEB scenario. Indeed the transit times of each TIC's transit-like events flagged by \pipeline{} appear out of
phase as they do not align nor do they overlap in Fig.~\ref{fig:tic300} implying that transit-like events detected
around each TIC by \pipeline{} do not affect the transit-like events in the light curve of the other. Even if
there are regions of the out-of-transit light curve that are mutually affected.
We are therefore left with the questionable interpretation
of these OIs by the aforementioned conflicts. We opt to use the \vespa{} results to conclude final
dispositions of these OIs but we also exercise caution by modifying the statistical disposition of
OI 10010300.01 from a PC to a pPC.

\subsection{Population of planet candidates}
\begin{figure*}
  \centering
  \includegraphics[scale=1]{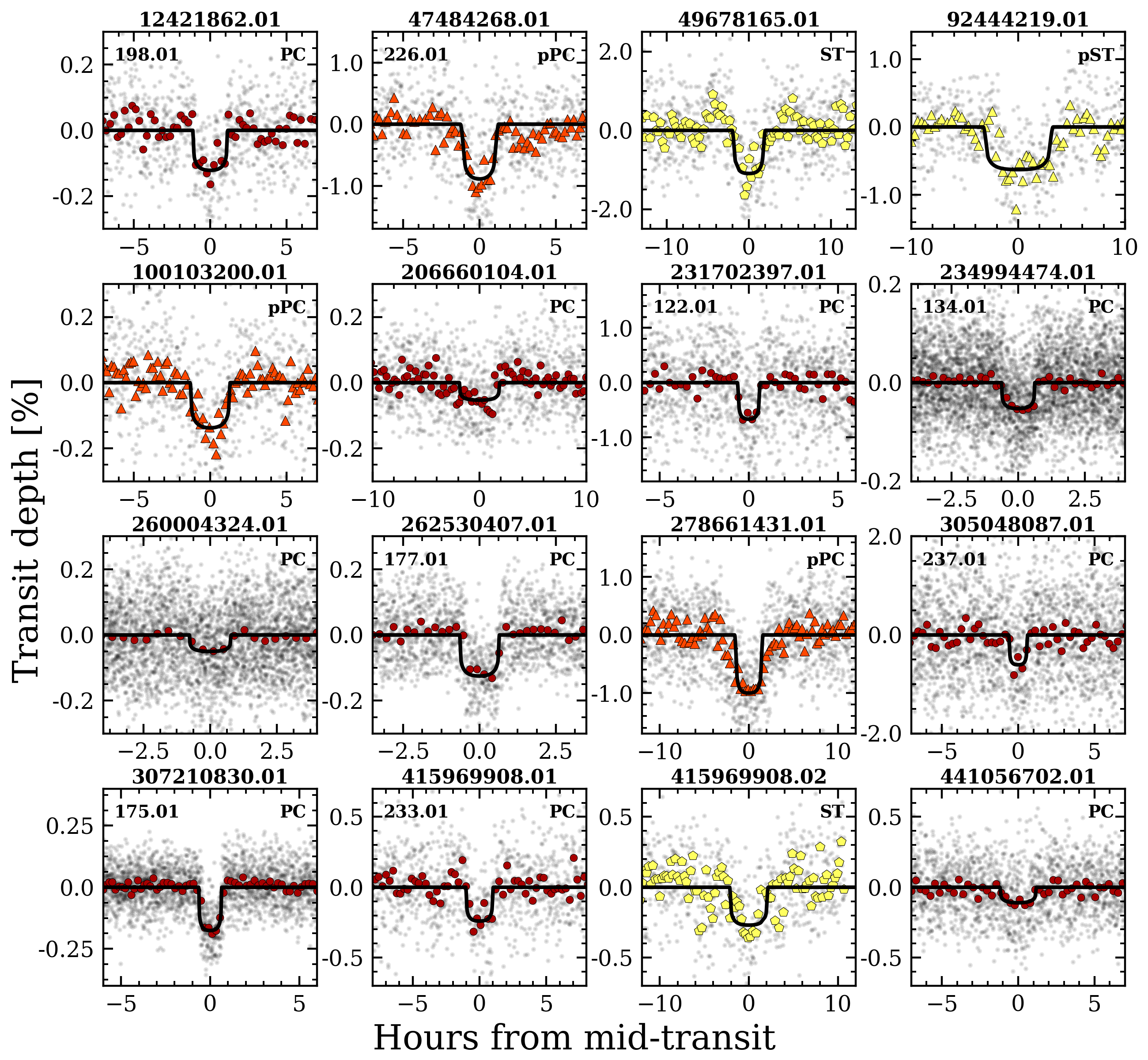}
  \caption{Phase-folded transit light curves for our set of 16 planet candidates. The temporal binning and axes
    ranges for each candidate are chosen to optimize visual clarity. The marker colors for each candidate's binned
    light curve is indicative of its disposition which also is annotated in the upper
    right of its panel. Candidates detected by \pipeline{} which are also \tess{} Objects of Interest 
    have their TOI ID annotated in the upper left.}
  \label{fig:LCs}
\end{figure*}

After manual vetting and statistical validation we are left with sixteen candidate planets. These include
ten PCs, two STs, three pPCs, and one pST. Half of our candidates are `new' having not yet been released
as TOIs\footnote{As of December 19, 2018.}.
Point estimates of observable and derived planetary parameters for these candidates are reported in
Table~\ref{table:planets}.  Fig.~\ref{fig:LCs} also depicts their phase-folded light curves along with
the transit models computed using the MAP parameter values from Table~\ref{table:planets}.

\begin{figure*}
  \centering
  \includegraphics[scale=1]{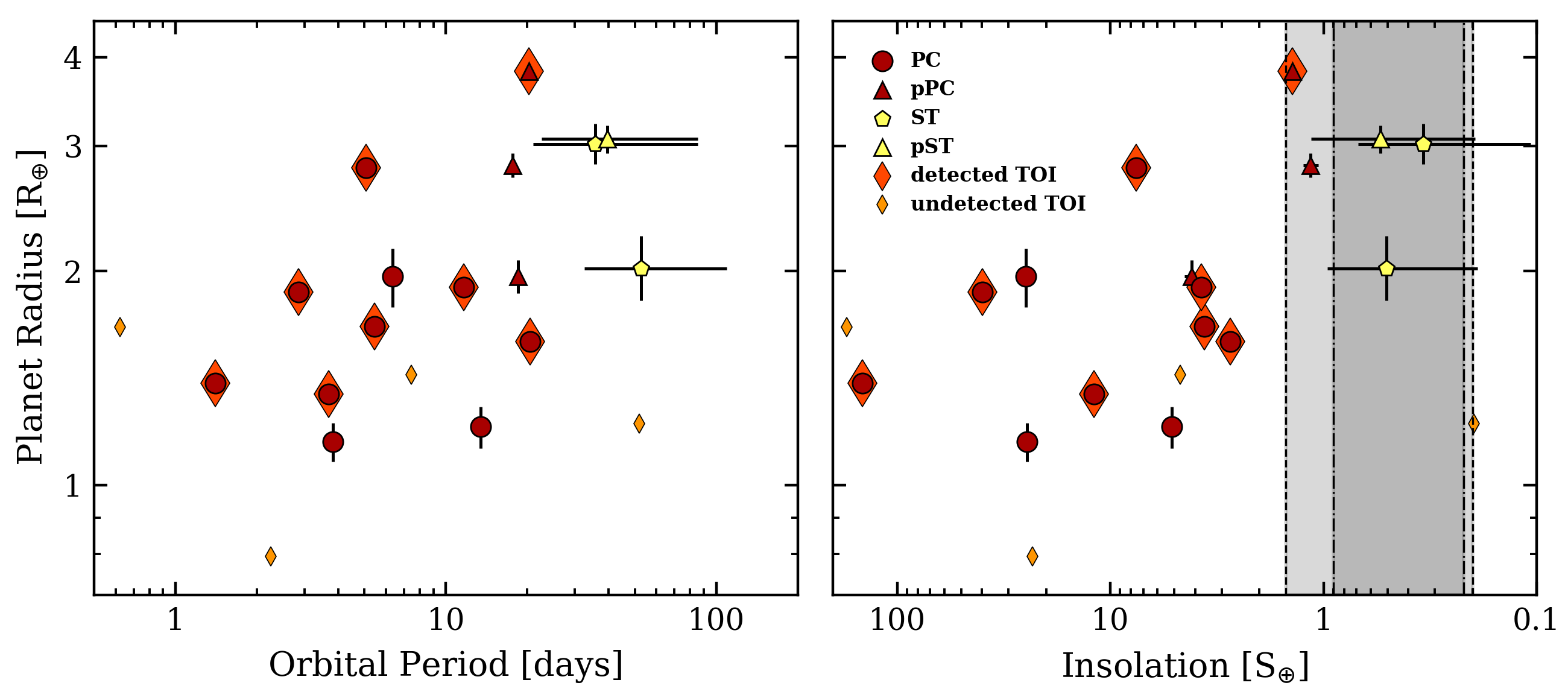}
  \caption{The resulting planet candidates from running \pipeline{} on the 2 minute extracted light curves
    from the first two \tess{} sectors in the period/radius and insolation/radius parameter spaces.
    The legend labels are planet candidates (PC), putative planet candidates (pPC), 
    single transit events (ST), and putative single transit events (pST).
    TOIs which are also detected by \pipeline{} are highlighted with
    orange diamonds surrounding the associated candidate's marker. TOIs which remain undetected by \pipeline{}
    are depicted as small orange diamonds. 
    The outer shaded region in the insolation panel marks the `recent-Venus' and `early-Mars'
    limits of the habitable zone around low mass stars from \cite{kopparapu13}. The inner shaded region
    marks the more conservative `water-loss' and `maximum-greenhouse' habitable zone limits.}
  \label{fig:planets}
\end{figure*}

Fig.~\ref{fig:planets} depicts our candidate population and compares it to the twelve TOIs whose TICs are
included in our stellar sample. Our candidates have orbital periods from 1.4-20 days for
PCs and from 35-50 days for ST events. Although ST periods exhibit large uncertainties of order the MAP
period value. Our candidates have radii between 1.1-3.8 R$_{\oplus}$ potentially making them targets of interest
for the completion of the \tess{} level one science requirement of delivering 50 planets with $r_p < 4$
R$_{\oplus}$ with measured masses. In Sect.~\ref{sect:rv} we will discuss the prospects that our
candidates have  for contributing to the realization of the \tess{} level one science requirement.
We do not detect any hot sub-Neptunes in the photoevaporation desert
\citep{lundkvist16} nor any small planets ($\lesssim 1.5$ R$_{\oplus}$) on orbits longer than $\sim 20$
days. This is largely attributable to our poor detection sensitivity in that regime due to the limited
\tess{} baselines of just 27 days.

Fig.~\ref{fig:planets} also depicts our candidates as a function of insolation. The majority of candidates
(11/16) experience incident insolation levels $S$ in excess of twice that of the Earth. However five candidates,
including all three ST events and two pPCs, are likely more temperate and experience insolations
$\lesssim 1.5$ S$_{\oplus}$.
This $S$ limit marks the `recent-Venus' inner edge of the low mass star habitable zone \citep[HZ;][]{kopparapu13}.
Our STs also lie within the more conservative HZ definition bounded by the `water-loss' inner
edge, where an increase in insolation results in the photolysis of stratospheric water vapor causing the atmosphere
to experience rapid hydrogen escape, and the `maximum-greenhouse' outer edge where an increase in CO$_2$ no longer
results in a net surface heating due to the increased albedo. Our five temperate candidates may represent attractive
targets for the characterization of HZ exoplanets around nearby low mass stars. We will address the prospect of
atmospheric characterization of these planets in Sect.~\ref{sect:atmospheres}.

\begin{figure*}
  \centering
  \includegraphics[scale=1]{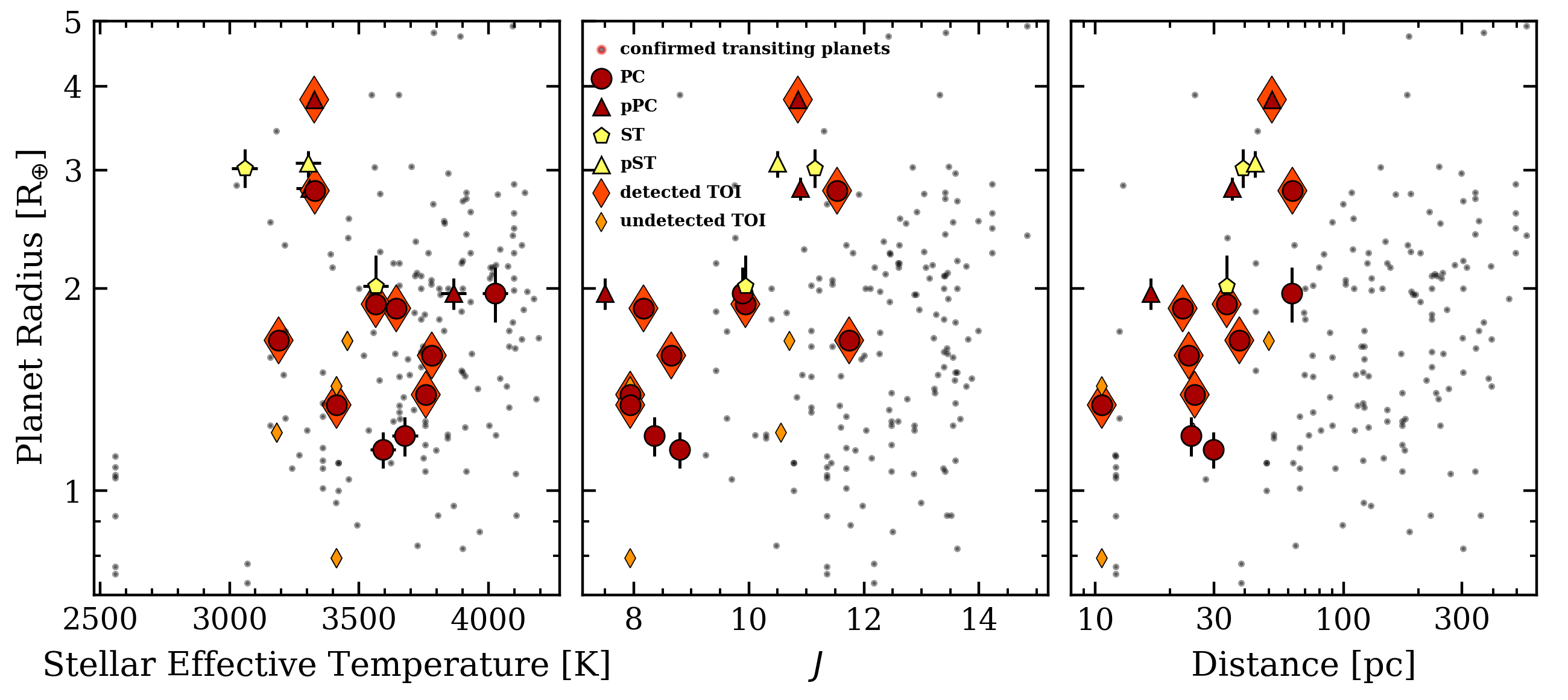}
  \caption{The planetary radii of our 16 \pipeline{} candidates as functions of their host stellar
    effective temperatures, $J$-band magnitudes, and \gaia{} distances. Our candidates
    are compared to the population of confirmed transiting planets around cool stars (\teff{} $<4200$
    K) from the NASA Exoplanet Archive which are depicted with small black circles.
    The legend labels are planet candidates (PC), 
    putative planet candidates (pPC), 
    single transit events (ST), and  putative single transits (pST).
    TOIs which are also detected by \pipeline{} are highlighted with
    orange diamonds surrounding the associated PC marker. TOIs remaining undetected by \pipeline{}
    are depicted as small orange diamonds.}
  \label{fig:planetstar}
\end{figure*}

The distribution of our candidates versus stellar parameters of interest are shown in
Fig.~\ref{fig:planetstar}. All candidates are detected around stars hotter than 3000 K which
approximately corresponds to stars earlier than M5.5V \citep{pecaut13}. Most candidates
are detected around stars with \teff{} $\in [3200,3900]$ K which is largely consistent with the
population of confirmed transiting planets recovered from the NASA Exoplanet Archive on
December 13, 2018 \citep{akeson13}, modulo the TRAPPIST-1 planets \citep{gillon17}, GJ 1214b
\citep{charbonneau09}, and the Kepler-42 planets \citep{muirhead12b}. The median effective
temperature of the candidate-hosting TICs in our sample is 3560 K. A notable dearth of PCs
with $r_p \gtrsim 2$ R$_{\oplus}$ exists around stars hotter than $\sim 3500$ K. The cause of
this is unlikely to be due to sensitivity losses around these relatively hot (and correspondingly
bright) stars and is instead likely attributable to the sharp decrease in the number of stars
within our sample at \teff{} $\sim 3550$ K (see Fig.~\ref{fig:stars}).

The distributions of previously confirmed planets
and our candidates versus $J$-band magnitude reveals that many ($\sim 10/16$) of our candidates
orbit systematically brighter stars than the majority of confirmed transiting planets with
$J<10$. This directly demonstrates the power of a survey mission like \tess{} at discovering
transiting planets orbiting nearby bright stars which are amenable to forthcoming detailed
characterization efforts. The remaining six candidates still orbit moderately
bright stars with $J \in [10,12]$. The median $J$ of our candidate-hosting TICs is 9.9.
From the number of \pipeline{} candidates as a function of $J$,
it is clear that the \pipeline{} sensitivity to planets orbiting low mass TICs starts to drop
off around $J \gtrsim 12$.
The large photometric uncertainties in this regime are largely dominated by photon-noise
from the target star, zodiacal light, and unresolved background stars \citep{ricker15}.

The distance distribution of our candidates is also included in Fig.~\ref{fig:planetstar}. All of
our candidates are found between 10-65 pc with a median distance of 34 pc. Our sample includes
five candidates within 25 pc as well as the closest transiting PC around a low mass star to date:
TIC 307210830.01 (TOI-175.01) at 10.6 pc which is also hypothesized to contain two additional TOIs
not detected by \pipeline{} (see Sect.~\ref{sect:undet}). Among our eight new candidates not
released as TOIs, we detect two candidates within 25 pc. These include the pPC 100103200.01,
and the PC 206660104.01. Barring the rejection of the candidates in
these systems and the TOIs around TIC 307210830, these three systems of five planet candidates
represent 3/8 of the
closest transiting planetary systems around low mass stars along with GJ 1132 \citep{berta15,bonfils18},
TRAPPIST-1 \citep{gillon17,luger17}, LHS 1140 \citep{dittmann17a,ment18}, GJ 1214 \citep{charbonneau09}, and
GJ 3470 \citep{bonfils12}.

\section{Discussion} \label{sect:disc}
\subsection{Discussions of individual systems with planet candidates} \label{sect:indiv}
\emph{OI 12421862.01}. This PC has already been reported as TOI-198.01. The \pipeline{} planet parameters
are consistent with those of TOI-198.01 with a 20.4 day orbital period and measured
radius of 1.6 R$_{\oplus}$, making it an interesting target for probing the photoevaporation valley
around M dwarfs.

\emph{OI 47484268.01}. This pPC based on its moderate FPP, has already been reported as TOI-226.01.
The \pipeline{} orbital period is consistent with that of TOI-226.01 but we derive a 10\% smaller
planet radius of 3.8 R$_{\oplus}$ despite the star's refined radius being 15\% larger. At 1.4 S$_{\oplus}$,
this pPC orbits within the `recent-Venus' habitable zone but remains an attractive target for rapid RV
characterization owing to its expected large mass compared to most of the smaller candidate planets (see
Sect.~\ref{sect:rv}).

\emph{OI 49678165.01}. A new candidate ST event with an estimated period between 21-85 days and a radius of
3 R$_{\oplus}$. This candidate is likely the coldest object in our sample with a MAP $S=0.3$ S$_{\oplus}$
and a corresponding equilibrium temperature of $T_{\text{eq}}= 212$ K assuming zero albedo and perfect heat redistribution.

\emph{OI 92444219.01}. A pST based on its moderate FPP. This cool 3 R$_{\oplus}$ candidate is
has an estimated period between 22-95 days placing it at $S=0.5$ S$_{\oplus}$ with
an equilibrium temperature $T_{\text{eq}}= 238$ K.

\emph{OI 100103200.01}. The pPC with ultra low FPP but is seen to likely be contaminated by the nearby
TIC 100103201 based on the \gaia{} astrometry. This pPC orbits a $J=7.5$ star with an 18.4 day period and
whose size is 2 R$_{\oplus}$ making it a attractive target for both RV characterization and possibly
transmission spectroscopy observations with \emph{JWST} if impending difficulties of observing
bright targets can be strategically mitigated (see Sect.~\ref{sect:atmospheres}).

\emph{OI 206660104.01}. A new terrestrial-sized (1.2 R$_{\oplus}$) PC at 13.4 days. This PC is the
second smallest candidate recovered by \pipeline{} and is smaller than most (3/4) of the TOIs missed by
\pipeline{} (see Sect.~\ref{sect:undet}).

\emph{OI 231702397.01}. This PC has already been reported as TOI-122.01. The \pipeline{} orbital
period is consistent with that of TOI-122.01 although we derive a 15\% larger radius (2.8 R$_{\oplus}$)
corresponding to the 13\% larger stellar radius in our sample. Given its large size, the expected
mass from the \cite{chen17} mass-radius relation is 8.26 M$_{\oplus}$ making it an attractive target
for rapid RV follow-up despite being relatively dim with $J=11.5$.

\emph{OI 234994474.01}: This PC has been reported as TOI-134.01 and is being validated with RVs
from HARPS and PFS \citep{astudillodefru19}. The \pipeline{} parameters are largely consistent with the TOI-134.01
parameters albeit with a slightly smaller radius of 1.38 R$_{\oplus}$ (12\% reduction) for
an unchanged stellar radius. Being by-far the hottest target in candidate sample ($T_{\text{eq}}=965$ K)
and orbiting an early M dwarf with $J=7.9$, this PC is one of the best targets for any further RV
characterization to search for longer period companions, transmission spectroscopy, and even thermal
emission spectroscopy. Regarding the latter, TOI-134.01
is even more favorable than the previously most attractive
terrestrial-sized planet for such observations: GJ 1132b \citep{morley17}.

\emph{OI 260004324.01}. A new terrestrial-sized PC which is the smallest in our sample
at 1.15 R$_{\oplus}$ with a 3.8 day period. Its small size and small expected mass of 1.6 M$_{\oplus}$
around a 0.56 M$_{\oplus}$ early M dwarf ($J=8.8$) will make this a slightly more challenging target
for RV follow-up but may still be of interest for probing the low mass end of the 50 planets
smaller than 4 R$_{\oplus}$ targeted for completion of the \tess{} level one science requirement.

\emph{OI 262530407.01}. This PC has already been reported as TOI-177.01. The \pipeline{} orbital
period is consistent with that of TOI-177.01 although we find a 12\% smaller radius of 1.87
$R_{\oplus}$ for an unchanged stellar radius. The short period and host star brightness ($J=8.17$)
make this PC an exceptional candidate for transmission spectroscopy observations and is one
that is less affected by the \emph{NIRISS} bright limit of its SOSS mode compared to other
close-in super-Earth-sized planets in this catalog.

\emph{OI 278661431.01}. A new pPC based on its moderate FFP. If validated, the candidate
would have an orbital period of 17 days and a radius of 2.8 $R_{\oplus}$. This pPC is temperate
at $S=1.1$ S$_{\oplus}$ and near to the `water-loss' HZ inner edge. This pPC orbits one of the
cooler TICs in our candidate-hosting sample (\teff{} $=3300$ K) with a correspondingly small
radius of 0.28 R$_{\odot}$, thus making it an attractive target for transmission spectroscopy
observations.

\emph{OI 305048087.01}. This PC has already been reported as TOI-237.01. The \pipeline{} orbital
period is consistent with the 5.4 day period and 1.7 R$_{\oplus}$ radius of TOI-177.01.

\emph{OI 307210830.01}. This PC is one of the three reported TOIs around TIC 307210830 (i.e. 175-01,
02, 03). Only TOI-175.01 is detected by \pipeline{} for the reasons discussed in Sect.~\ref{sect:undet}.
The \pipeline{} planet parameters for this, the middle planet in this candidate three planet
system, are all consistent with those for TOI-175.01. This PC is in the closest planetary system
in our sample and is correspondingly an attractive target for RV characterization of individual masses
and the RV analysis of the  possible resonant pair 175.01/175.02 for comparisons to TTV analyzes
from follow-up photometry. This PC is also a viable target for the atmospheric characterization of
a terrestrial-sized planet (1.3 R$_{\oplus}$).

\emph{OI 415969908.01}. This PC has already been reported as TOI-233.01. The \pipeline{} orbital
period is consistent with that of TOI-233.01 but finds a 26\% smaller radius (1.9 R$_{\oplus}$)
in part because of the refined stellar radius being reduced by 21\% in our sample.

\emph{OI 415969908.02}. The ST event detected in the light of TIC 415969908 which already
hosts the aforementioned candidate TOI-233.01 at 11.7 days. The estimated period of OI 415969908.02 is 32-108 days
which effectively spans the conservative HZ limits from \cite{kopparapu13} and whose MAP value is
53 days or $S=0.5$ S$_{\oplus}$. The measured radius is 2 R$_{\oplus}$ ($r_p/Rs=0.05$) making this
ST event the most difficult to follow-up from the ground among the three ST events in our candidate
catalog.

\emph{OI 441056702.01}. A new PC with an orbital period of 6.3 days and radius of 2 R$_{\oplus}$
thus potentially being located within the photoevaporation valley around M dwarfs. Orbiting a moderately bright
early-M to late-K dwarf (\teff{} = 4030 K) and with an expected mass of 4.5 M$_{\oplus}$, this PC
represents another attractive target for RV characterization aimed at addressing the \tess{} level
one science requirement.

\subsection{TOIs undetected by \pipeline{}} \label{sect:undet}
Our stellar sample using refined stellar parameters based on \gaia{} distances,
contains ten TICs with TOIs listed as part of the TESS alerts\footnote{As of December 19, 2018.}.
These ten systems host a total of twelve TOIs in nine single PC systems
and TIC 307210830 which hosts three TOIs. The \pipeline{} results presented in this paper
include the independent detection of 8/12 TOIs with an additional ST event around TIC 415969908 which
already hosts TOI-233.01. The discrepancies between the TESS alerts and our \pipeline{} results
are described in what follows.

\emph{TIC 259962054}. This TIC was observed in the consecutive \tess{} sectors 1 and 2.
The TOI-203.01 has an orbital period of 52 days, longer than any repeating
candidate in our catalog. A signal at $\sim 52.1$ days is found in the \pipeline{} linear and
periodic search stages with an orbital phase that is consistent with that reported for TOI-203.01.
This suggests that the TOI-203.01 transit-like signal at 52 days does exist in the light curve.
The phase-folded light curve satisfies all but the second criterion from the automated vetting
stage (see Sect.~\ref{sect:autovetting}) with S/N$_{\text{transit}}=5.6<c_1=8.4$. This is principally
because the fitted transit depth within \pipeline{} is $Z=1839$ ppm  which is just
$\sim 73$\% of the TOI's reported depth thus making it difficult to confirm the nature of the
repeating signal as due to a transiting planet with just two transits observed.

\emph{TIC 307210830}. This system contains three TOIs (175.01, 175.02, 175.03) at 3.69, 7.45,
and 2.25 days respectively. The innermost planet candidate is not found during the linear and
periodic search stages. This is likely caused by the candidate's depth of 571 ppm
(as reported in the TESS alert portal) being less than the median photometric uncertainty of
its light curve (i.e. median($\sigma_{\text{f}}$) $\sim 770$ ppm). The two remaining planet candidates were each seen
to be detected in the \pipeline{} linear search stage owing to their $\sim 3$ times larger
transit depths. However, this candidate pair has an orbital period ratio that is within 1\%
of a 2:1 resonant configuration. Recall that pairs of periods of interest which are that close to
an integer period ratio will have one of those periods automatically discarded in the periodic search
stage due to the detected periodic signals likely being aliases of the each other rather than being
due to two separate planetary candidates.

\emph{TIC 316937670}. TOI-221.01 has an orbital period of 0.624 days and a low transit depth
of 954 ppm. By adopting the reported TOI-221.01 transit depth and duration, we estimate
CDPP$_{\text{transit}}$ and find that $Z/\text{CDPP}_{\text{transit}} \sim 1.1$. Because of this,
the results of the \pipeline{} linear search only detect
a single transit-event above the S/N threshold such that
no periodic events can be found. If the linear search S/N threshold is lowered
such that multiple transits from TOI-221.01 are detected in the linear and periodic search stages, then
the expected S/N$_{\text{transit}}$ is $\sim 7.2$ which would still be insufficient to detect the PC in
a single \tess{} sector. We are further discouraged by the prospect of lowering the linear search S/N
threshold as this would drastically increase the number of FPs purely from the noise.

\subsection{Prospects for follow-up observations}
\subsubsection{Mass characterization via precision radial velocities} \label{sect:rv}
The \tess{} level one science requirement is to deliver at least 50 planets smaller than 4
R$_{\oplus}$ with measured masses via precision radial velocity (RV) follow-up. All sixteen of our
candidates have a measured radius consistent with being $<4$ R$_{\oplus}$ (see Table~\ref{table:planets}).
Using the empirical mass-radius relation for small planets $<14.26$ R$_{\oplus}$ from \cite{chen17},
we compute the maximum likelihood masses $m_p$ of our planet candidates and single transit events
to then infer their expected RV semi-amplitudes using 

\begin{multline}
  K_{\text{RV}} = 2.4 \text{ m/s} \left( \frac{m_p}{5 \text{ M}_{\oplus}} \right)
  \left( \frac{M_s}{0.5 \text{ M}_{\odot}} \right)^{-2/3} \\
  \left( \frac{P}{10\text{ days}} \right)^{-1/3}.
\end{multline}

\begin{figure}
  \centering
  \includegraphics{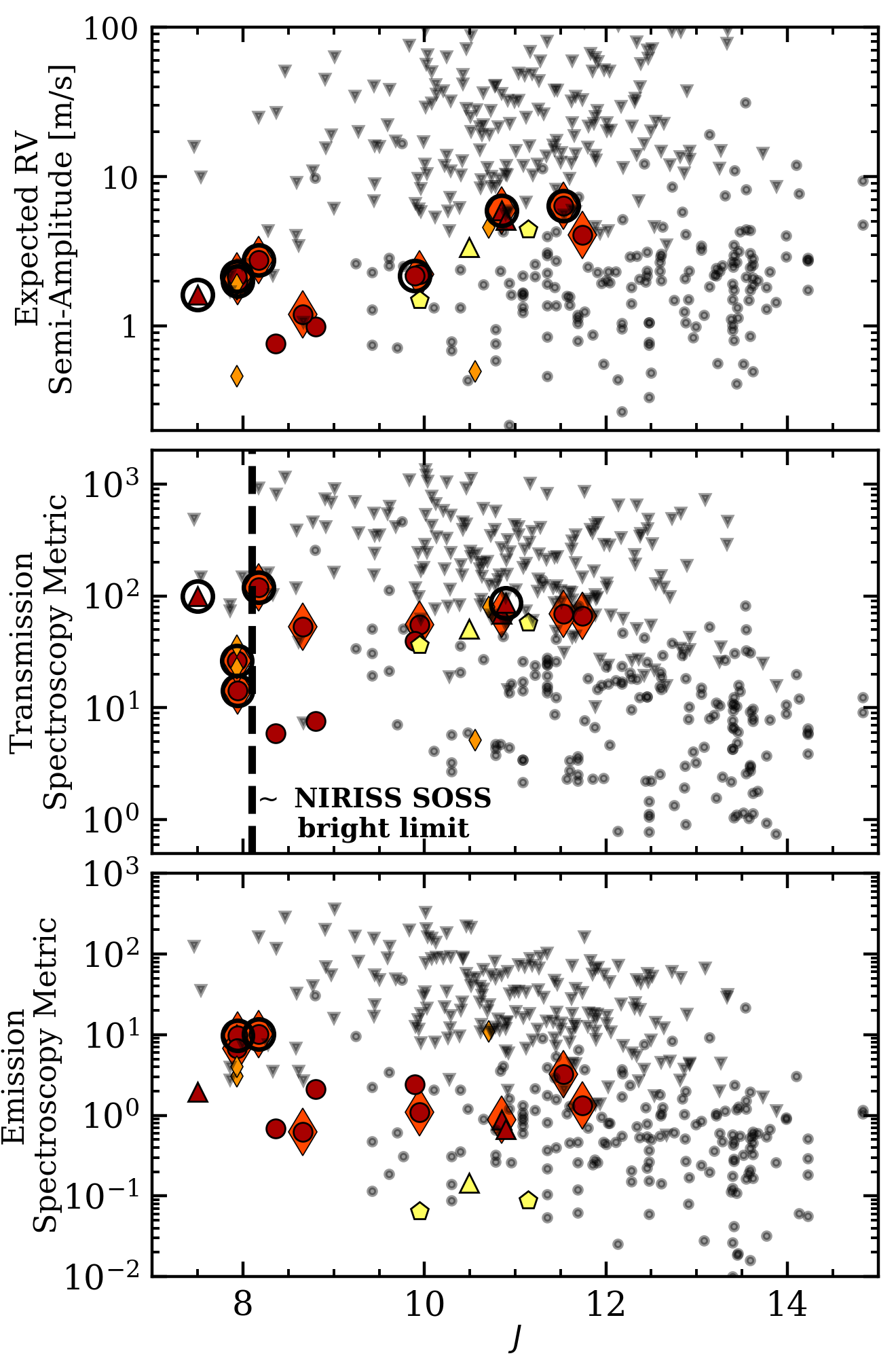}
  \caption{Expected values of the RV semi-amplitude (upper panel), transmission spectroscopy metric (middle panel),
    and emission spectroscopy metric (lower panel) for our set of 16 candidates as a function of $J$-band
    magnitude. The marker legend for our candidates is identical to those in Figs.~\ref{fig:planets}
    and~\ref{fig:planetstar} and are not included here for visual clarity. Candidate planets
    exceeding parameter cutoffs (Eqs.~\ref{eq:Omega1} and~\ref{eq:Omega2} for $\Omega$ and see \citealt{kempton18} for
    the TSM and ESM cutoffs) are highly favorable targets for follow-up observations and are highlighted by black rings.
    Also plotted are the expected values for simulated \tess{} planets around cool stars (\teff{} $< 4200$ K) from
    \cite{barclay18} (black inverted triangles) and for confirmed transiting planets around cool stars from the NASA
    Exoplanet Archive (black circles). To first order, our sample of candidate planets straddles the boundary
    between the expected \tess{} population and the population of known transiting planets.}
  \label{fig:followup}
\end{figure}

\noindent These values are shown in the first panel of Fig~\ref{fig:followup} versus $J$-band magnitude
and are accompanied by the simulated \tess{} yield from \cite{barclay18} and the set of
confirmed transiting planets from the NASA Exoplanet Archive.
Both of the latter samples are restricted to cool host stars (i.e. \teff{}
$<4200$ K). Many existing high performance RV spectrographs are stable at the level of a few cm/s but
RV observations are often limited by photon noise and intrinsic stellar activity at the level of 1 to a few
m/s, even for relatively inactive stars \citep{fischer16}. 
The majority of our objects of interest have expected $K_{\text{RV}}$ values in excess of this typical RV
sensitivity limit and are reported in Table~\ref{table:followup}. 

\cite{cloutier18b} calculated the observing time required to complete the \tess{} level on science requirement
based on the expected \tess{} yield from \cite{sullivan15}. The 50 TOIs requiring the shortest time
commitment to characterizing their planet masses at $5\sigma$ with RVs satisfy the following empirically-derived
conditions: 

\begin{align}
  J &< 11.7 \text{ and} \label{eq:Omega1} \\
  \Omega &> 0.14J - 0.35, \label{eq:Omega2}
\end{align}

\noindent where

\begin{equation}
\Omega = \left( \frac{r_p}{\text{ R}_{\oplus}} \right) \left( \frac{P}{\text{day}}  \right)^{-1/3}
\end{equation}

\noindent is a proxy for $K_{\text{RV}}$ that can be computed from transit-derived parameters.
Seven out of sixteen of our candidates satisfy Eqs.~\ref{eq:Omega1} and~\ref{eq:Omega2} and are highlighted
in Table~\ref{table:followup} and in Fig.~\ref{fig:followup}.
If astrophysical false positive scenarios can be ruled out for these OIs, then
they represent highly favorable targets for RV follow-up observations and the rapid completion of the
\tess{} level one science requirement. That is, assuming that the stars are not active which is an
important charactistic to consider for RV follow-up \citep{moutou17} and is one that is not taken into
account in Eqs.~\ref{eq:Omega1} and~\ref{eq:Omega2}.

\subsubsection{Atmospheric characterization} \label{sect:atmospheres}
\tess{} will provide many of the best transiting exoplanets for atmospheric characterization in the
near future. \cite{kempton18} presented a framework to prioritize transiting planets for
either transmission spectroscopy or emission spectroscopy observations with dedicated missions
like \emph{JWST} and \emph{ARIEL}. This framework consists of analytical metrics which quantify
the expected S/N ratio of transmission and emission signals from planetary atmospheres.

The transmission spectroscopy metric from \cite{kempton18} is

\begin{equation}
  \text{TSM } = f \cdot 10^{-0.2J} \cdot \left( \frac{r_p^3 T_{\text{eq}}}{m_p R_s^2} \right)
  \label{eq:tsm}
\end{equation}

\noindent The TSM represents the expected S/N of 10 hour observing programs with \emph{JWST}/\emph{NIRISS}
assuming fixed atmospheric compositions for different planet types, cloud-free atmospheres, and a
deterministic planet mass-radius relation. The planetary equilibrium temperature in Eq.~\ref{eq:tsm}
is $T_{\text{eq}} = T_{\text{eff}} \sqrt{R_s/2a}$ where $a$ is the planet's semimajor axis and is calculated
assuming zero albedo and full heat redistribution over the planetary surface. The scale factor $f$ is
used to make the TSM non-dimensional and is used to correct discrepancies between the analytical TSM
and the detailed simulations from \cite{louie18} using the \emph{NIRISS} simulator for Single Object
Slitless Spectroscopy (SOSS) observations. Values of $f$ are reported in \cite{kempton18} each of four
planet types separately: terrestrials ($r_p < 1.5$ R$_{\oplus}$), super-Earths ($1.5 < r_p/\text{R}_{\oplus} < 2.75$) 
sub-Neptunes ($2.75 < r_p/\text{R}_{\oplus} < 4$), and giants ($4 < r_p/\text{R}_{\oplus} < 10$). 
We calculate the TSM for our 16 candidates and report those values in Table~\ref{table:followup} and in
Fig.~\ref{fig:followup}. \cite{kempton18} highly recommends planets for atmospheric characterization
(and a-priori RV characterization) based on their TSM values relative to their derived cutoffs (see
their Table 1). Five out of sixteen candidates exceed the threshold TSM cutoff as highlighted in Table~\ref{table:followup}
and should be considered for confirmation as they represent highly attractive targets for transmission spectroscopy
observations. We note however that host stars with $J\lesssim 8.1$ start to approach the \emph{NIRISS}
bright limit (depending on its spectral type) and will require specialized fast readout modes and/or the use
of detector subarrays to be observed with \emph{NIRISS} in its SOSS mode \citep{beichman14}.

Similarly to the TSM, \cite{kempton18} defined the thermal emission spectroscopy metric as

\begin{equation}
  \text{ESM } = c \cdot 10^{-0.2K_{\text{S}}} \cdot \frac{B_{7.5}(T_{\text{day}})}{B_{7.5}(T_{\text{eff}})} \cdot
  \left( \frac{r_p}{R_s} \right)^2
\end{equation}

\noindent where $B_{7.5}(T)$ is the Planck function of spectral irradiance evaluated for a given temperature
$T$ at 7.5 $\mu$m and $T_{\text{day}}$ is the planet's day-side temperature assumed to be $1.1T_{\text{eq}}$.
The constant $c=4.29\times 10^6$ is used to scale the ESM to yield the S/N of the reference planet GJ 1132b
\citep{berta15,dittmann17b} in the center of the \emph{MIRI} low resolution spectroscopy bandpass at 7.5 $\mu$m. 
We calculate the ESM for our 16 candidates and report those values in Table~\ref{table:followup} and in
Fig.~\ref{fig:followup}. \citep{kempton18} advocate that planets with ESM $>$ ESM$_{\text{GJ 1132}}=7.5$ should
be considered favorable targets for thermal emission spectroscopy observations with \emph{MIRI}. Within our
candidate sample, 2/16 candidates exceed this threshold ESM and represent some of the best targets to-date for
the characterization of a terrestrial or super-Earth's atmosphere with emission spectroscopy for the first time.

\subsection{Comparison to yield simulations}
\cite{ballard18} performed a set of yield simulations focusing on M dwarfs not unlike the
stellar population considered in this study. \cite{ballard18} derive an ensemble completeness
function for M dwarfs observed by \tess{} based on the simulated \tess{} yield
from \cite{sullivan15} which includes details of the \tess{} footprint, systematics, the
photometric error budget, and FP likelihoods. The expected \tess{} yield around M dwarfs is
then derived by applying the completeness as a function of $P$ and $r_p$ to M dwarf planet
occurrence rates. Said occurrence rate are
derived from \cite{dressing15a} and corrected for the eccentricity
distribution \citep{limbach15}, dynamical stability \citep{fabrycky12}, and multiplicity
effects according to the `Kepler dichotomy' \citep{ballard16} of M dwarf planet populations:
either high multiplicity systems ($N>5$) with low mutual inclinations or systems with lower
multiplicity ($N\sim 1-2$) and high mutual inclinations. The resulting \tess{} yield
around M1-M4 dwarfs is predicted to be $990 \pm 350$ planets. 

The following back-of-the-envelope calculation reveals how the expected
M dwarf planet population to be discovered with \tess{} compares to our \pipeline{} results
from the first two \tess{} sectors. First we note
that the \cite{sullivan15} stellar population is a synthetic one derived from \texttt{TRILEGAL}
galaxy model \citep{girardi05}. It contains 200,000 stars targeted by \tess{} which is
effectively the size of the TIC \citep{stassun17}. Using the stellar parameters from the TIC-7
we find 53204 TIC M1-M4 dwarfs (\teff{} $\in [3200-3700]$ K; \citealt{pecaut13}).
Of these, 1624 and 1869 are targeted in sectors 1 and 2 respectively with 2849
being unique TICs. However unlike in the TIC-7,
our stellar sample is derived using parallaxes from the \gaia{} DR2 which results in
a distinct population of just 1149 M1-M4 dwarfs observed in either or both of \tess{} sectors 1
and 2. We use a simple correction factor to the expected \tess{} yield to account for the
fractionally fewer M1-M4 stars that we target for transit searches compared to the TIC-7.
This factor is $f=1149/2849 = 0.40 \pm 0.01$ where the $f$ uncertainty
is propagated from Poisson statistics. 

The predicted number of M dwarf \tess{} planets discovered in sectors 1 and 2 is
$990w \sim 53 \pm 19$ where $w=2849/53204$ is the fraction of all M1-M4 dwarfs targeted in
those sectors. Correcting this expected number of planets from the first two \tess{} sectors
by the $f$ times fewer M1-M4 dwarfs in our stellar sample compared to in the TIC-7,
we find that we are expected to detect
$\sim 21 \pm 8$ planets. Note that this calculation inherently assumes that the detection
completeness of \pipeline{} is equivalent to the ensemble completeness derived in \cite{ballard18}. 

If we include putative planet candidates, then recall that the
total number of PCs detected around M1-M4 dwarfs in this study is sixteen (see
Fig.~\ref{fig:planetstar}). If each of these candidates become validated planets that this
many planets would consistent with the expected number of M dwarf planet detections from
\cite{ballard18}. Although, our yield is somewhat on the lower end of what is expected.
If the TESS alert TOIs around M1-M4 dwarfs that remain undetected by \pipeline{} in this study are
included, then \tess{} has discovered 19 M1-M4 PCs in its first two sectors.
The consistency between the \pipeline{} yield and the expected \tess{} yield speaks highly to \tess{'}s
overall performance compared to its expected completeness from \cite{sullivan15} and
\cite{ballard18}, as well as to the outstanding performance that the \tess{} mission has
already achieved so early-on in its lifetime.

\section*{Acknowledgements}
I would like to thank each of
Jo Bovy,
Ren\'e Doyon,
Xu (Chelsea) Huang,
David Latham,
Kristen Menou,
Norm Murray,
Adiv Paradise, and
Marten van Kerkwijk
for both motivating discussions as well as informative ones that improved the content of this
paper as well as my understanding of the data from \gaia{} and \tess{} used throughout.
I also thank the Canadian Institute for Theoretical Astrophysics for use of the Sunnyvale
computing cluster throughout this work. 
I would also like to acknowledge that this
work was performed on land traditionally inhabited by the
Wendat, the Anishnaabeg, Haudenosaunee, Metis, and the
Mississaugas of the New Credit First Nation.

To my wife, for preparing the majority of our meals since the \tess{} data release and
for turning the volume down on the TV during evening Leaf games so that I could hear myself
think. 

This work is supported in-part by the Natural Sciences
and Engineering Council of Canada (NSERC).

Special thanks go to NASA and to the entire TESS team for their exhaustive efforts over the
years in developing such an amazing observatory that will serve the exoplanet community
and dreamers alike for years to come.

This research has made use of the NASA Exoplanet Archive, which is operated by the California
Institute of Technology, under contract with the National Aeronautics and Space Administration
under the Exoplanet Exploration Program.

This work presents results from the European Space Agency (ESA) space mission \gaia{.} \gaia{}
data are being processed by the Gaia Data Processing and Analysis Consortium (DPAC). Funding for
the DPAC is provided by national institutions, in particular the institutions participating in
the Gaia MultiLateral Agreement (MLA). The Gaia mission website is
\url{https://www.cosmos.esa.int/gaia}.
The Gaia archive website is \url{https://archives.esac.esa.int/gaia}.

This paper includes data collected by the TESS mission, which are publicly available
from the Mikulski Archive for Space Telescopes (MAST).
Some of the data presented in this paper were obtained from the Mikulski Archive for Space
Telescopes (MAST). STScI is operated by the Association of Universities for Research in Astronomy,
Inc., under NASA contract NAS5-26555. 

\bibliographystyle{mnras}
\bibliography{refs}

\begin{thebibliography}{}
\makeatletter
\relax
\def\mn@urlcharsother{\let\do\@makeother \do\$\do\&\do\#\do\^\do\_\do\%\do\~}
\def\mn@doi{\begingroup\mn@urlcharsother \@ifnextchar [ {\mn@doi@}
  {\mn@doi@[]}}
\def\mn@doi@[#1]#2{\def\@tempa{#1}\ifx\@tempa\@empty \href
  {http://dx.doi.org/#2} {doi:#2}\else \href {http://dx.doi.org/#2} {#1}\fi
  \endgroup}
\def\mn@eprint#1#2{\mn@eprint@#1:#2::\@nil}
\def\mn@eprint@arXiv#1{\href {http://arxiv.org/abs/#1} {{\tt arXiv:#1}}}
\def\mn@eprint@dblp#1{\href {http://dblp.uni-trier.de/rec/bibtex/#1.xml}
  {dblp:#1}}
\def\mn@eprint@#1:#2:#3:#4\@nil{\def\@tempa {#1}\def\@tempb {#2}\def\@tempc
  {#3}\ifx \@tempc \@empty \let \@tempc \@tempb \let \@tempb \@tempa \fi \ifx
  \@tempb \@empty \def\@tempb {arXiv}\fi \@ifundefined
  {mn@eprint@\@tempb}{\@tempb:\@tempc}{\expandafter \expandafter \csname
  mn@eprint@\@tempb\endcsname \expandafter{\@tempc}}}

\bibitem[\protect\citeauthoryear{{Adams}, {Jackson}  \& {Endl}}{{Adams}
  et~al.}{2016}]{adams16}
{Adams} E.~R.,  {Jackson} B.,   {Endl} M.,  2016, \mn@doi [\aj]
  {10.3847/0004-6256/152/2/47}, \href
  {http://adsabs.harvard.edu/abs/2016AJ....152...47A} {152, 47}

\bibitem[\protect\citeauthoryear{{Aigrain}, {Hodgkin}, {Irwin}, {Lewis}  \&
  {Roberts}}{{Aigrain} et~al.}{2015}]{aigrain15}
{Aigrain} S.,  {Hodgkin} S.~T.,  {Irwin} M.~J.,  {Lewis} J.~R.,   {Roberts}
  S.~J.,  2015, \mn@doi [\mnras] {10.1093/mnras/stu2638}, \href
  {http://adsabs.harvard.edu/abs/2015MNRAS.447.2880A} {447, 2880}

\bibitem[\protect\citeauthoryear{{Aigrain}, {Parviainen}  \& {Pope}}{{Aigrain}
  et~al.}{2016}]{aigrain16}
{Aigrain} S.,  {Parviainen} H.,   {Pope} B.~J.~S.,  2016, \mn@doi [\mnras]
  {10.1093/mnras/stw706}, \href
  {https://ui.adsabs.harvard.edu/#abs/2016MNRAS.459.2408A} {459, 2408}

\bibitem[\protect\citeauthoryear{{Akeson} et~al.,}{{Akeson}
  et~al.}{2013}]{akeson13}
{Akeson} R.~L.,  et~al., 2013, \mn@doi [\pasp] {10.1086/672273}, \href
  {http://adsabs.harvard.edu/abs/2013PASP..125..989A} {125, 989}

\bibitem[\protect\citeauthoryear{{Angus}, {Morton}, {Aigrain}, {Foreman-Mackey}
   \& {Rajpaul}}{{Angus} et~al.}{2018}]{angus18}
{Angus} R.,  {Morton} T.,  {Aigrain} S.,  {Foreman-Mackey} D.,   {Rajpaul} V.,
  2018, \mn@doi [\mnras] {10.1093/mnras/stx2109}, \href
  {http://adsabs.harvard.edu/abs/2018MNRAS.474.2094A} {474, 2094}

\bibitem[\protect\citeauthoryear{{Astraatmadja} \&
  {Bailer-Jones}}{{Astraatmadja} \& {Bailer-Jones}}{2016}]{astraatmadja16}
{Astraatmadja} T.~L.,  {Bailer-Jones} C.~A.~L.,  2016, \mn@doi [\apj]
  {10.3847/0004-637X/832/2/137}, \href
  {http://adsabs.harvard.edu/abs/2016ApJ...832..137A} {832, 137}

\bibitem[\protect\citeauthoryear{{Astudillo-Defru} et~al.,}{{Astudillo-Defru}
  et~al.}{2019}]{astudillodefru19}
{Astudillo-Defru} N.,  et~al., 2019, preprint (\mn@eprint {arXiv} {in prep.})

\bibitem[\protect\citeauthoryear{{Bailer-Jones}}{{Bailer-Jones}}{2015}]{bailorjones15}
{Bailer-Jones} C.~A.~L.,  2015, \mn@doi [\pasp] {10.1086/683116}, \href
  {http://adsabs.harvard.edu/abs/2015PASP..127..994B} {127, 994}

\bibitem[\protect\citeauthoryear{{Bailer-Jones}, {Rybizki}, {Fouesneau},
  {Mantelet}  \& {Andrae}}{{Bailer-Jones} et~al.}{2018}]{bailorjones18}
{Bailer-Jones} C.~A.~L.,  {Rybizki} J.,  {Fouesneau} M.,  {Mantelet} G.,
  {Andrae} R.,  2018, \mn@doi [\aj] {10.3847/1538-3881/aacb21}, \href
  {http://adsabs.harvard.edu/abs/2018AJ....156...58B} {156, 58}

\bibitem[\protect\citeauthoryear{{Ballard}}{{Ballard}}{2018}]{ballard18}
{Ballard} S.,  2018, preprint, \href
  {http://adsabs.harvard.edu/abs/2018arXiv180104949B} {} (\mn@eprint {arXiv}
  {1801.04949})

\bibitem[\protect\citeauthoryear{{Ballard} \& {Johnson}}{{Ballard} \&
  {Johnson}}{2016}]{ballard16}
{Ballard} S.,  {Johnson} J.~A.,  2016, \mn@doi [\apj]
  {10.3847/0004-637X/816/2/66}, \href
  {https://ui.adsabs.harvard.edu/\#abs/2016ApJ...816...66B} {816, 66}

\bibitem[\protect\citeauthoryear{{Barclay}, {Pepper}  \& {Quintana}}{{Barclay}
  et~al.}{2018}]{barclay18}
{Barclay} T.,  {Pepper} J.,   {Quintana} E.~V.,  2018, \mn@doi [The
  Astrophysical Journal Supplement Series] {10.3847/1538-4365/aae3e9}, \href
  {https://ui.adsabs.harvard.edu/\#abs/2018ApJS..239....2B} {239, 2}

\bibitem[\protect\citeauthoryear{{Bastien}, {Stassun}, {Basri}  \&
  {Pepper}}{{Bastien} et~al.}{2013}]{bastien13}
{Bastien} F.~A.,  {Stassun} K.~G.,  {Basri} G.,   {Pepper} J.,  2013, \mn@doi
  [\nat] {10.1038/nature12419}, \href
  {http://adsabs.harvard.edu/abs/2013Natur.500..427B} {500, 427}

\bibitem[\protect\citeauthoryear{{Batalha} et~al.,}{{Batalha}
  et~al.}{2010}]{batalha10}
{Batalha} N.~M.,  et~al., 2010, \mn@doi [\apjl] {10.1088/2041-8205/713/2/L103},
  \href {http://adsabs.harvard.edu/abs/2010ApJ...713L.103B} {713, L103}

\bibitem[\protect\citeauthoryear{{Beichman} et~al.,}{{Beichman}
  et~al.}{2014}]{beichman14}
{Beichman} C.,  et~al., 2014, \mn@doi [\pasp] {10.1086/679566}, \href
  {http://adsabs.harvard.edu/abs/2014PASP..126.1134B} {126, 1134}

\bibitem[\protect\citeauthoryear{{Benedict} et~al.,}{{Benedict}
  et~al.}{2016}]{benedict16}
{Benedict} G.~F.,  et~al., 2016, \mn@doi [\aj] {10.3847/0004-6256/152/5/141},
  \href {http://adsabs.harvard.edu/abs/2016AJ....152..141B} {152, 141}

\bibitem[\protect\citeauthoryear{{Berger}, {Huber}, {Gaidos}  \& {van
  Saders}}{{Berger} et~al.}{2018}]{berger18}
{Berger} T.~A.,  {Huber} D.,  {Gaidos} E.,   {van Saders} J.~L.,  2018, \mn@doi
  [\apj] {10.3847/1538-4357/aada83}, \href
  {https://ui.adsabs.harvard.edu/#abs/2018ApJ...866...99B} {866, 99}

\bibitem[\protect\citeauthoryear{{Berta-Thompson} et~al.,}{{Berta-Thompson}
  et~al.}{2015}]{berta15}
{Berta-Thompson} Z.~K.,  et~al., 2015, \mn@doi [\nat] {10.1038/nature15762},
  \href {http://adsabs.harvard.edu/abs/2015Natur.527..204B} {527, 204}

\bibitem[\protect\citeauthoryear{{Bonfils} et~al.,}{{Bonfils}
  et~al.}{2012}]{bonfils12}
{Bonfils} X.,  et~al., 2012, \mn@doi [\aap] {10.1051/0004-6361/201219623},
  \href {http://adsabs.harvard.edu/abs/2012A%26A...546A..27B} {546, A27}

\bibitem[\protect\citeauthoryear{{Bonfils} et~al.,}{{Bonfils}
  et~al.}{2018}]{bonfils18}
{Bonfils} X.,  et~al., 2018, preprint, \href
  {http://adsabs.harvard.edu/abs/2018arXiv180603870B} {} (\mn@eprint {arXiv}
  {1806.03870})

\bibitem[\protect\citeauthoryear{{Bouma}, {Winn}, {Kosiarek}  \&
  {McCullough}}{{Bouma} et~al.}{2017}]{bouma17}
{Bouma} L.~G.,  {Winn} J.~N.,  {Kosiarek} J.,   {McCullough} P.~R.,  2017,
  preprint, \href {http://adsabs.harvard.edu/abs/2017arXiv170508891B} {}
  (\mn@eprint {arXiv} {1705.08891})

\bibitem[\protect\citeauthoryear{{Bovy}, {Rix}, {Green}, {Schlafly}  \&
  {Finkbeiner}}{{Bovy} et~al.}{2016}]{bovy16}
{Bovy} J.,  {Rix} H.-W.,  {Green} G.~M.,  {Schlafly} E.~F.,   {Finkbeiner}
  D.~P.,  2016, \mn@doi [\apj] {10.3847/0004-637X/818/2/130}, \href
  {http://adsabs.harvard.edu/abs/2016ApJ...818..130B} {818, 130}

\bibitem[\protect\citeauthoryear{{Broeg} et~al.,}{{Broeg}
  et~al.}{2013}]{broeg13}
{Broeg} C.,  et~al., 2013, in European Physical Journal Web of Conferences. p.
  03005 (\mn@eprint {arXiv} {1305.2270}), \mn@doi{10.1051/epjconf/20134703005}

\bibitem[\protect\citeauthoryear{{Bryson} et~al.,}{{Bryson}
  et~al.}{2013}]{bryson13}
{Bryson} S.~T.,  et~al., 2013, \mn@doi [\pasp] {10.1086/671767}, \href
  {http://adsabs.harvard.edu/abs/2013PASP..125..889B} {125, 889}

\bibitem[\protect\citeauthoryear{{Buchner} et~al.,}{{Buchner}
  et~al.}{2014}]{buchner14}
{Buchner} J.,  et~al., 2014, \mn@doi [\aap] {10.1051/0004-6361/201322971},
  \href {http://adsabs.harvard.edu/abs/2014A%26A...564A.125B} {564, A125}

\bibitem[\protect\citeauthoryear{{Charbonneau} et~al.,}{{Charbonneau}
  et~al.}{2009}]{charbonneau09}
{Charbonneau} D.,  et~al., 2009, \mn@doi [\nat] {10.1038/nature08679}, \href
  {http://adsabs.harvard.edu/abs/2009Natur.462..891C} {462, 891}

\bibitem[\protect\citeauthoryear{{Chen} \& {Kipping}}{{Chen} \&
  {Kipping}}{2017}]{chen17}
{Chen} J.,  {Kipping} D.,  2017, \mn@doi [\apj] {10.3847/1538-4357/834/1/17},
  \href {https://ui.adsabs.harvard.edu/\#abs/2017ApJ...834...17C} {834, 17}

\bibitem[\protect\citeauthoryear{{Christiansen} et~al.,}{{Christiansen}
  et~al.}{2012}]{christiansen12}
{Christiansen} J.~L.,  et~al., 2012, \mn@doi [\pasp] {10.1086/668847}, \href
  {http://adsabs.harvard.edu/abs/2012PASP..124.1279C} {124, 1279}

\bibitem[\protect\citeauthoryear{{Christiansen} et~al.,}{{Christiansen}
  et~al.}{2013}]{christiansen13}
{Christiansen} J.~L.,  et~al., 2013, \mn@doi [\apjs]
  {10.1088/0067-0049/207/2/35}, \href
  {http://adsabs.harvard.edu/abs/2013ApJS..207...35C} {207, 35}

\bibitem[\protect\citeauthoryear{{Christiansen} et~al.,}{{Christiansen}
  et~al.}{2015}]{christiansen15}
{Christiansen} J.~L.,  et~al., 2015, \mn@doi [\apj]
  {10.1088/0004-637X/810/2/95}, \href
  {http://adsabs.harvard.edu/abs/2015ApJ...810...95C} {810, 95}

\bibitem[\protect\citeauthoryear{{Christiansen} et~al.,}{{Christiansen}
  et~al.}{2016}]{christiansen16}
{Christiansen} J.~L.,  et~al., 2016, \mn@doi [\apj]
  {10.3847/0004-637X/828/2/99}, \href
  {http://adsabs.harvard.edu/abs/2016ApJ...828...99C} {828, 99}

\bibitem[\protect\citeauthoryear{{Claret}}{{Claret}}{2017}]{claret17}
{Claret} A.,  2017, \mn@doi [\aap] {10.1051/0004-6361/201629705}, \href
  {https://ui.adsabs.harvard.edu/#abs/2017A&A...600A..30C} {600, A30}

\bibitem[\protect\citeauthoryear{{Cloutier}, {Doyon}, {Bouchy}  \&
  {H{\'e}brard}}{{Cloutier} et~al.}{2018}]{cloutier18b}
{Cloutier} R.,  {Doyon} R.,  {Bouchy} F.,   {H{\'e}brard} G.,  2018, \mn@doi
  [\aj] {10.3847/1538-3881/aacea9}, \href
  {http://adsabs.harvard.edu/abs/2018AJ....156...82C} {156, 82}

\bibitem[\protect\citeauthoryear{{Collier Cameron} et~al.,}{{Collier Cameron}
  et~al.}{2007}]{collier07}
{Collier Cameron} A.,  et~al., 2007, \mn@doi [\mnras]
  {10.1111/j.1365-2966.2007.12195.x}, \href
  {http://adsabs.harvard.edu/abs/2007MNRAS.380.1230C} {380, 1230}

\bibitem[\protect\citeauthoryear{{Cooke}, {Pollacco}, {West}, {McCormac}  \&
  {Wheatley}}{{Cooke} et~al.}{2018}]{cooke18}
{Cooke} B.~F.,  {Pollacco} D.,  {West} R.,  {McCormac} J.,   {Wheatley} P.~J.,
  2018, \mn@doi [\aap] {10.1051/0004-6361/201834014}, \href
  {https://ui.adsabs.harvard.edu/#abs/2018A&A...619A.175C} {619, A175}

\bibitem[\protect\citeauthoryear{{Crossfield} et~al.,}{{Crossfield}
  et~al.}{2015}]{crossfield15}
{Crossfield} I.~J.~M.,  et~al., 2015, \mn@doi [\apj]
  {10.1088/0004-637X/804/1/10}, \href
  {http://adsabs.harvard.edu/abs/2015ApJ...804...10C} {804, 10}

\bibitem[\protect\citeauthoryear{{Crossfield} et~al.,}{{Crossfield}
  et~al.}{2018}]{crossfield18}
{Crossfield} I.~J.~M.,  et~al., 2018, \mn@doi [\apjs]
  {10.3847/1538-4365/aae155}, \href
  {http://adsabs.harvard.edu/abs/2018ApJS..239....5C} {239, 5}

\bibitem[\protect\citeauthoryear{{Cutri} et~al.,}{{Cutri}
  et~al.}{2003}]{cutri03}
{Cutri} R.~M.,  et~al., 2003, {2MASS All Sky Catalog of point sources.}

\bibitem[\protect\citeauthoryear{{Delfosse}, {Forveille}, {S{\'e}gransan},
  {Beuzit}, {Udry}, {Perrier}  \& {Mayor}}{{Delfosse}
  et~al.}{2000}]{delfosse00}
{Delfosse} X.,  {Forveille} T.,  {S{\'e}gransan} D.,  {Beuzit} J.-L.,  {Udry}
  S.,  {Perrier} C.,   {Mayor} M.,  2000, \aap, \href
  {http://adsabs.harvard.edu/abs/2000A%26A...364..217D} {364, 217}

\bibitem[\protect\citeauthoryear{{Dittmann}, {Irwin}, {Charbonneau},
  {Berta-Thompson}  \& {Newton}}{{Dittmann} et~al.}{2017a}]{dittmann17b}
{Dittmann} J.~A.,  {Irwin} J.~M.,  {Charbonneau} D.,  {Berta-Thompson} Z.~K.,
  {Newton} E.~R.,  2017a, \mn@doi [\aj] {10.3847/1538-3881/aa855b}, \href
  {https://ui.adsabs.harvard.edu/\#abs/2017AJ....154..142D} {154, 142}

\bibitem[\protect\citeauthoryear{{Dittmann} et~al.,}{{Dittmann}
  et~al.}{2017b}]{dittmann17a}
{Dittmann} J.~A.,  et~al., 2017b, \mn@doi [\nat] {10.1038/nature22055}, \href
  {http://adsabs.harvard.edu/abs/2017Natur.544..333D} {544, 333}

\bibitem[\protect\citeauthoryear{{Dressing} \& {Charbonneau}}{{Dressing} \&
  {Charbonneau}}{2013}]{dressing13}
{Dressing} C.~D.,  {Charbonneau} D.,  2013, \mn@doi [\apj]
  {10.1088/0004-637X/767/1/95}, \href
  {http://adsabs.harvard.edu/abs/2013ApJ...767...95D} {767, 95}

\bibitem[\protect\citeauthoryear{{Dressing} \& {Charbonneau}}{{Dressing} \&
  {Charbonneau}}{2015}]{dressing15a}
{Dressing} C.~D.,  {Charbonneau} D.,  2015, \mn@doi [\apj]
  {10.1088/0004-637X/807/1/45}, \href
  {http://adsabs.harvard.edu/abs/2015ApJ...807...45D} {807, 45}

\bibitem[\protect\citeauthoryear{{Drimmel}, {Cabrera-Lavers}  \&
  {L{\'o}pez-Corredoira}}{{Drimmel} et~al.}{2003}]{drimmel03}
{Drimmel} R.,  {Cabrera-Lavers} A.,   {L{\'o}pez-Corredoira} M.,  2003, \mn@doi
  [\aap] {10.1051/0004-6361:20031070}, \href
  {http://adsabs.harvard.edu/abs/2003A%26A...409..205D} {409, 205}

\bibitem[\protect\citeauthoryear{{Esposito} et~al.,}{{Esposito}
  et~al.}{2018}]{esposito18}
{Esposito} M.,  et~al., 2018, arXiv e-prints, \href
  {https://ui.adsabs.harvard.edu/\#abs/2018arXiv181205881E} {p.
  arXiv:1812.05881}

\bibitem[\protect\citeauthoryear{{Evans} et~al.,}{{Evans}
  et~al.}{2018}]{evans18}
{Evans} D.~W.,  et~al., 2018, \mn@doi [\aap] {10.1051/0004-6361/201832756},
  \href {https://ui.adsabs.harvard.edu/#abs/2018A&A...616A...4E} {616, A4}

\bibitem[\protect\citeauthoryear{{Fabrycky} et~al.,}{{Fabrycky}
  et~al.}{2012}]{fabrycky12}
{Fabrycky} D.~C.,  et~al., 2012, \mn@doi [\apj] {10.1088/0004-637X/750/2/114},
  \href {http://adsabs.harvard.edu/abs/2012ApJ...750..114F} {750, 114}

\bibitem[\protect\citeauthoryear{{Fischer} et~al.,}{{Fischer}
  et~al.}{2016}]{fischer16}
{Fischer} D.~A.,  et~al., 2016, \mn@doi [\pasp]
  {10.1088/1538-3873/128/964/066001}, \href
  {http://adsabs.harvard.edu/abs/2016PASP..128f6001F} {128, 066001}

\bibitem[\protect\citeauthoryear{{Foreman-Mackey}, {Hogg}, {Lang}  \&
  {Goodman}}{{Foreman-Mackey} et~al.}{2013}]{foremanmackey13}
{Foreman-Mackey} D.,  {Hogg} D.~W.,  {Lang} D.,   {Goodman} J.,  2013, \mn@doi
  [\pasp] {10.1086/670067}, \href
  {http://adsabs.harvard.edu/abs/2013PASP..125..306F} {125, 306}

\bibitem[\protect\citeauthoryear{{Foreman-Mackey}, {Montet}, {Hogg}, {Morton},
  {Wang}  \& {Sch{\"o}lkopf}}{{Foreman-Mackey} et~al.}{2015}]{foremanmackey15a}
{Foreman-Mackey} D.,  {Montet} B.~T.,  {Hogg} D.~W.,  {Morton} T.~D.,  {Wang}
  D.,   {Sch{\"o}lkopf} B.,  2015, \mn@doi [\apj]
  {10.1088/0004-637X/806/2/215}, \href
  {http://adsabs.harvard.edu/abs/2015ApJ...806..215F} {806, 215}

\bibitem[\protect\citeauthoryear{{Fulton} \& {Petigura}}{{Fulton} \&
  {Petigura}}{2018}]{fulton18}
{Fulton} B.~J.,  {Petigura} E.~A.,  2018, \mn@doi [\aj]
  {10.3847/1538-3881/aae828}, \href
  {http://adsabs.harvard.edu/abs/2018AJ....156..264F} {156, 264}

\bibitem[\protect\citeauthoryear{{Gaia Collaboration} et~al.,}{{Gaia
  Collaboration} et~al.}{2016a}]{gaia16}
{Gaia Collaboration} et~al., 2016a, \mn@doi [\aap]
  {10.1051/0004-6361/201629272}, \href
  {http://adsabs.harvard.edu/abs/2016A%26A...595A...1G} {595, A1}

\bibitem[\protect\citeauthoryear{{Gaia Collaboration} et~al.,}{{Gaia
  Collaboration} et~al.}{2016b}]{brown16}
{Gaia Collaboration} et~al., 2016b, \mn@doi [\aap]
  {10.1051/0004-6361/201629512}, \href
  {https://ui.adsabs.harvard.edu/#abs/2016A&A...595A...2G} {595, A2}

\bibitem[\protect\citeauthoryear{{Gaia Collaboration} et~al.,}{{Gaia
  Collaboration} et~al.}{2018}]{gaia18}
{Gaia Collaboration} et~al., 2018, \mn@doi [\aap]
  {10.1051/0004-6361/201833051}, \href
  {http://adsabs.harvard.edu/abs/2018A%26A...616A...1G} {616, A1}

\bibitem[\protect\citeauthoryear{{Gaidos}, {Kitzmann}  \& {Heng}}{{Gaidos}
  et~al.}{2017}]{gaidos17}
{Gaidos} E.,  {Kitzmann} D.,   {Heng} K.,  2017, \mn@doi [\mnras]
  {10.1093/mnras/stx615}, \href
  {https://ui.adsabs.harvard.edu/#abs/2017MNRAS.468.3418G} {468, 3418}

\bibitem[\protect\citeauthoryear{{Gandolfi} et~al.,}{{Gandolfi}
  et~al.}{2018}]{gandolfi18}
{Gandolfi} D.,  et~al., 2018, \mn@doi [\aap] {10.1051/0004-6361/201834289},
  \href {http://adsabs.harvard.edu/abs/2018A%26A...619L..10G} {619, L10}

\bibitem[\protect\citeauthoryear{{Gillon} et~al.,}{{Gillon}
  et~al.}{2017}]{gillon17}
{Gillon} M.,  et~al., 2017, \mn@doi [\nat] {10.1038/nature21360}, \href
  {http://adsabs.harvard.edu/abs/2017Natur.542..456G} {542, 456}

\bibitem[\protect\citeauthoryear{{Ginsburg} et~al.,}{{Ginsburg}
  et~al.}{2017}]{ginsburg17}
{Ginsburg} A.,  et~al., 2017, {Astroquery: Access to online data resources},
  Astrophysics Source Code Library (\mn@eprint {ascl} {1708.004})

\bibitem[\protect\citeauthoryear{{Girardi}, {Groenewegen}, {Hatziminaoglou}  \&
  {da Costa}}{{Girardi} et~al.}{2005}]{girardi05}
{Girardi} L.,  {Groenewegen} M.~A.~T.,  {Hatziminaoglou} E.,   {da Costa} L.,
  2005, \mn@doi [\aap] {10.1051/0004-6361:20042352}, \href
  {http://adsabs.harvard.edu/abs/2005A%26A...436..895G} {436, 895}

\bibitem[\protect\citeauthoryear{{Green} et~al.,}{{Green}
  et~al.}{2015}]{green15}
{Green} G.~M.,  et~al., 2015, \mn@doi [\apj] {10.1088/0004-637X/810/1/25},
  \href {http://adsabs.harvard.edu/abs/2015ApJ...810...25G} {810, 25}

\bibitem[\protect\citeauthoryear{{Green} et~al.,}{{Green}
  et~al.}{2018}]{green18}
{Green} G.~M.,  et~al., 2018, \mn@doi [\mnras] {10.1093/mnras/sty1008}, \href
  {http://adsabs.harvard.edu/abs/2018MNRAS.478..651G} {478, 651}

\bibitem[\protect\citeauthoryear{{G{\"u}nther}, {Queloz}, {Demory}  \&
  {Bouchy}}{{G{\"u}nther} et~al.}{2017}]{gunther17}
{G{\"u}nther} M.~N.,  {Queloz} D.,  {Demory} B.-O.,   {Bouchy} F.,  2017,
  \mn@doi [\mnras] {10.1093/mnras/stw2908}, \href
  {http://adsabs.harvard.edu/abs/2017MNRAS.465.3379G} {465, 3379}

\bibitem[\protect\citeauthoryear{{Hawley}, {Davenport}, {Kowalski},
  {Wisniewski}, {Hebb}, {Deitrick}  \& {Hilton}}{{Hawley}
  et~al.}{2014}]{hawley14}
{Hawley} S.~L.,  {Davenport} J.~R.~A.,  {Kowalski} A.~F.,  {Wisniewski} J.~P.,
  {Hebb} L.,  {Deitrick} R.,   {Hilton} E.~J.,  2014, \mn@doi [\apj]
  {10.1088/0004-637X/797/2/121}, \href
  {http://adsabs.harvard.edu/abs/2014ApJ...797..121H} {797, 121}

\bibitem[\protect\citeauthoryear{{Huang} et~al.,}{{Huang}
  et~al.}{2018a}]{huang18b}
{Huang} C.~X.,  et~al., 2018a, arXiv e-prints, \href
  {https://ui.adsabs.harvard.edu/#abs/2018arXiv180711129H} {p.
  arXiv:1807.11129}

\bibitem[\protect\citeauthoryear{{Huang} et~al.,}{{Huang}
  et~al.}{2018b}]{huang18a}
{Huang} C.~X.,  et~al., 2018b, arXiv e-prints, \href
  {https://ui.adsabs.harvard.edu/#abs/2018arXiv180905967H} {p.
  arXiv:1809.05967}

\bibitem[\protect\citeauthoryear{{Jenkins} et~al.,}{{Jenkins}
  et~al.}{2010}]{jenkins10}
{Jenkins} J.~M.,  et~al., 2010, in Software and Cyberinfrastructure for
  Astronomy. p. 77400D, \mn@doi{10.1117/12.856764}

\bibitem[\protect\citeauthoryear{{Jenkins} et~al.,}{{Jenkins}
  et~al.}{2016}]{jenkins16}
{Jenkins} J.~M.,  et~al., 2016, in Software and Cyberinfrastructure for
  Astronomy IV. p. 99133E, \mn@doi{10.1117/12.2233418}

\bibitem[\protect\citeauthoryear{{Johnson} et~al.,}{{Johnson}
  et~al.}{2012}]{johnson12}
{Johnson} J.~A.,  et~al., 2012, \mn@doi [\aj] {10.1088/0004-6256/143/5/111},
  \href {http://adsabs.harvard.edu/abs/2012AJ....143..111J} {143, 111}

\bibitem[\protect\citeauthoryear{{Kasting}, {Whitmire}  \&
  {Reynolds}}{{Kasting} et~al.}{1993}]{kasting93}
{Kasting} J.~F.,  {Whitmire} D.~P.,   {Reynolds} R.~T.,  1993, \mn@doi
  [\icarus] {10.1006/icar.1993.1010}, \href
  {http://adsabs.harvard.edu/abs/1993Icar..101..108K} {101, 108}

\bibitem[\protect\citeauthoryear{{Kempton} et~al.,}{{Kempton}
  et~al.}{2018}]{kempton18}
{Kempton} E.~M.-R.,  et~al., 2018, preprint, \href
  {http://adsabs.harvard.edu/abs/2018arXiv180503671K} {} (\mn@eprint {arXiv}
  {1805.03671})

\bibitem[\protect\citeauthoryear{{Kopparapu} et~al.,}{{Kopparapu}
  et~al.}{2013}]{kopparapu13}
{Kopparapu} R.~K.,  et~al., 2013, \mn@doi [\apj] {10.1088/0004-637X/765/2/131},
  \href {http://adsabs.harvard.edu/abs/2013ApJ...765..131K} {765, 131}

\bibitem[\protect\citeauthoryear{{Kov{\'a}cs}, {Zucker}  \&
  {Mazeh}}{{Kov{\'a}cs} et~al.}{2002}]{kovacs02}
{Kov{\'a}cs} G.,  {Zucker} S.,   {Mazeh} T.,  2002, \mn@doi [\aap]
  {10.1051/0004-6361:20020802}, \href
  {http://adsabs.harvard.edu/abs/2002A%26A...391..369K} {391, 369}

\bibitem[\protect\citeauthoryear{{Kreidberg}}{{Kreidberg}}{2015}]{kreidberg15}
{Kreidberg} L.,  2015, \mn@doi [Publications of the Astronomical Society of the
  Pacific] {10.1086/683602}, \href
  {https://ui.adsabs.harvard.edu/#abs/2015PASP..127.1161K} {127, 1161}

\bibitem[\protect\citeauthoryear{{Leggett}}{{Leggett}}{1992}]{leggett92}
{Leggett} S.~K.,  1992, \mn@doi [\apjs] {10.1086/191720}, \href
  {http://adsabs.harvard.edu/abs/1992ApJS...82..351L} {82, 351}

\bibitem[\protect\citeauthoryear{{Li}, {Tenenbaum}, {Twicken}, {Burke},
  {Jenkins}, {Quintana}, {Rowe}  \& {Seader}}{{Li} et~al.}{2018}]{li18}
{Li} J.,  {Tenenbaum} P.,  {Twicken} J.~D.,  {Burke} C.~J.,  {Jenkins} J.~M.,
  {Quintana} E.~V.,  {Rowe} J.~F.,   {Seader} S.~E.,  2018, arXiv e-prints,
  \href {https://ui.adsabs.harvard.edu/#abs/2018arXiv181200103L} {p.
  arXiv:1812.00103}

\bibitem[\protect\citeauthoryear{{Limbach} \& {Turner}}{{Limbach} \&
  {Turner}}{2015}]{limbach15}
{Limbach} M.~A.,  {Turner} E.~L.,  2015, \mn@doi [Proceedings of the National
  Academy of Science] {10.1073/pnas.1406545111}, \href
  {http://adsabs.harvard.edu/abs/2015PNAS..112...20L} {112, 20}

\bibitem[\protect\citeauthoryear{{Lindegren} et~al.,}{{Lindegren}
  et~al.}{2018}]{lindegren18}
{Lindegren} L.,  et~al., 2018, \mn@doi [\aap] {10.1051/0004-6361/201832727},
  \href {http://adsabs.harvard.edu/abs/2018A%26A...616A...2L} {616, A2}

\bibitem[\protect\citeauthoryear{{Livingston} et~al.,}{{Livingston}
  et~al.}{2018}]{livingston18}
{Livingston} J.~H.,  et~al., 2018, \mn@doi [\aj] {10.3847/1538-3881/aaccde},
  \href {https://ui.adsabs.harvard.edu/\#abs/2018AJ....156...78L} {156, 78}

\bibitem[\protect\citeauthoryear{{Louie}, {Deming}, {Albert}, {Bouma}, {Bean}
  \& {Lopez-Morales}}{{Louie} et~al.}{2018}]{louie18}
{Louie} D.~R.,  {Deming} D.,  {Albert} L.,  {Bouma} L.~G.,  {Bean} J.,
  {Lopez-Morales} M.,  2018, \mn@doi [\pasp] {10.1088/1538-3873/aaa87b}, \href
  {http://adsabs.harvard.edu/abs/2018PASP..130d4401L} {130, 044401}

\bibitem[\protect\citeauthoryear{{Luger} et~al.,}{{Luger}
  et~al.}{2017}]{luger17}
{Luger} R.,  et~al., 2017, \mn@doi [Nature Astronomy]
  {10.1038/s41550-017-0129}, \href
  {http://adsabs.harvard.edu/abs/2017NatAs...1E.129L} {1, 0129}

\bibitem[\protect\citeauthoryear{{Lundkvist} et~al.,}{{Lundkvist}
  et~al.}{2016}]{lundkvist16}
{Lundkvist} M.~S.,  et~al., 2016, \mn@doi [Nature Communications]
  {10.1038/ncomms11201}, \href
  {https://ui.adsabs.harvard.edu/\#abs/2016NatCo...711201L} {7, 11201}

\bibitem[\protect\citeauthoryear{{Luri} et~al.,}{{Luri} et~al.}{2018}]{luri18}
{Luri} X.,  et~al., 2018, \mn@doi [\aap] {10.1051/0004-6361/201832964}, \href
  {http://adsabs.harvard.edu/abs/2018A%26A...616A...9L} {616, A9}

\bibitem[\protect\citeauthoryear{{Mandel} \& {Agol}}{{Mandel} \&
  {Agol}}{2002}]{mandel02}
{Mandel} K.,  {Agol} E.,  2002, \mn@doi [\apjl] {10.1086/345520}, \href
  {http://adsabs.harvard.edu/abs/2002ApJ...580L.171M} {580, L171}

\bibitem[\protect\citeauthoryear{{Mann}, {Brewer}, {Gaidos}, {L{\'e}pine}  \&
  {Hilton}}{{Mann} et~al.}{2013}]{mann13}
{Mann} A.~W.,  {Brewer} J.~M.,  {Gaidos} E.,  {L{\'e}pine} S.,   {Hilton}
  E.~J.,  2013, \mn@doi [\aj] {10.1088/0004-6256/145/2/52}, \href
  {https://ui.adsabs.harvard.edu/#abs/2013AJ....145...52M} {145, 52}

\bibitem[\protect\citeauthoryear{{Mann}, {Feiden}, {Gaidos}, {Boyajian}  \&
  {von Braun}}{{Mann} et~al.}{2015}]{mann15}
{Mann} A.~W.,  {Feiden} G.~A.,  {Gaidos} E.,  {Boyajian} T.,   {von Braun} K.,
  2015, \mn@doi [\apj] {10.1088/0004-637X/804/1/64}, \href
  {http://adsabs.harvard.edu/abs/2015ApJ...804...64M} {804, 64}

\bibitem[\protect\citeauthoryear{{Marshall}, {Robin}, {Reyl{\'e}}, {Schultheis}
   \& {Picaud}}{{Marshall} et~al.}{2006}]{marshall06}
{Marshall} D.~J.,  {Robin} A.~C.,  {Reyl{\'e}} C.,  {Schultheis} M.,   {Picaud}
  S.,  2006, \mn@doi [\aap] {10.1051/0004-6361:20053842}, \href
  {http://adsabs.harvard.edu/abs/2006A%26A...453..635M} {453, 635}

\bibitem[\protect\citeauthoryear{{Ment} et~al.,}{{Ment} et~al.}{2018}]{ment18}
{Ment} K.,  et~al., 2018, arXiv e-prints, \href
  {https://ui.adsabs.harvard.edu/\#abs/2018arXiv180800485M} {p.
  arXiv:1808.00485}

\bibitem[\protect\citeauthoryear{{Miller-Ricci}, {Seager}  \&
  {Sasselov}}{{Miller-Ricci} et~al.}{2009}]{millerricci09}
{Miller-Ricci} E.,  {Seager} S.,   {Sasselov} D.,  2009, \mn@doi [\apj]
  {10.1088/0004-637X/690/2/1056}, \href
  {https://ui.adsabs.harvard.edu/#abs/2009ApJ...690.1056M} {690, 1056}

\bibitem[\protect\citeauthoryear{{Moffett}}{{Moffett}}{1974}]{moffett74}
{Moffett} T.~J.,  1974, \mn@doi [\apjs] {10.1086/190330}, \href
  {http://adsabs.harvard.edu/abs/1974ApJS...29....1M} {29, 1}

\bibitem[\protect\citeauthoryear{{Montet} et~al.,}{{Montet}
  et~al.}{2015}]{montet15}
{Montet} B.~T.,  et~al., 2015, \mn@doi [\apj] {10.1088/0004-637X/809/1/25},
  \href {http://adsabs.harvard.edu/abs/2015ApJ...809...25M} {809, 25}

\bibitem[\protect\citeauthoryear{{Morley}, {Kreidberg}, {Rustamkulov},
  {Robinson}  \& {Fortney}}{{Morley} et~al.}{2017}]{morley17}
{Morley} C.~V.,  {Kreidberg} L.,  {Rustamkulov} Z.,  {Robinson} T.,   {Fortney}
  J.~J.,  2017, \mn@doi [\apj] {10.3847/1538-4357/aa927b}, \href
  {https://ui.adsabs.harvard.edu/#abs/2017ApJ...850..121M} {850, 121}

\bibitem[\protect\citeauthoryear{{Morton}}{{Morton}}{2012}]{morton12}
{Morton} T.~D.,  2012, \mn@doi [\apj] {10.1088/0004-637X/761/1/6}, \href
  {http://adsabs.harvard.edu/abs/2012ApJ...761....6M} {761, 6}

\bibitem[\protect\citeauthoryear{{Morton}}{{Morton}}{2015}]{morton15}
{Morton} T.~D.,  2015, {VESPA: False positive probabilities calculator},
  Astrophysics Source Code Library (\mn@eprint {ascl} {1503.011})

\bibitem[\protect\citeauthoryear{{Morton} \& {Swift}}{{Morton} \&
  {Swift}}{2014}]{morton14}
{Morton} T.~D.,  {Swift} J.,  2014, \mn@doi [\apj]
  {10.1088/0004-637X/791/1/10}, \href
  {http://adsabs.harvard.edu/abs/2014ApJ...791...10M} {791, 10}

\bibitem[\protect\citeauthoryear{{Moutou} et~al.,}{{Moutou}
  et~al.}{2017}]{moutou17}
{Moutou} C.,  et~al., 2017, \mn@doi [\mnras] {10.1093/mnras/stx2306}, \href
  {http://adsabs.harvard.edu/abs/2017MNRAS.472.4563M} {472, 4563}

\bibitem[\protect\citeauthoryear{{Muirhead} et~al.,}{{Muirhead}
  et~al.}{2012}]{muirhead12b}
{Muirhead} P.~S.,  et~al., 2012, \mn@doi [\apj] {10.1088/0004-637X/747/2/144},
  \href {http://adsabs.harvard.edu/abs/2012ApJ...747..144M} {747, 144}

\bibitem[\protect\citeauthoryear{{Muirhead}, {Dressing}, {Mann}, {Rojas-Ayala},
  {L{\'e}pine}, {Paegert}, {De Lee}  \& {Oelkers}}{{Muirhead}
  et~al.}{2018}]{muirhead18}
{Muirhead} P.~S.,  {Dressing} C.~D.,  {Mann} A.~W.,  {Rojas-Ayala} B.,
  {L{\'e}pine} S.,  {Paegert} M.,  {De Lee} N.,   {Oelkers} R.,  2018, \mn@doi
  [\aj] {10.3847/1538-3881/aab710}, \href
  {https://ui.adsabs.harvard.edu/#abs/2018AJ....155..180M} {155, 180}

\bibitem[\protect\citeauthoryear{{Newton}, {Charbonneau}, {Irwin},
  {Berta-Thompson}, {Rojas-Ayala}, {Covey}  \& {Lloyd}}{{Newton}
  et~al.}{2014}]{newton14}
{Newton} E.~R.,  {Charbonneau} D.,  {Irwin} J.,  {Berta-Thompson} Z.~K.,
  {Rojas-Ayala} B.,  {Covey} K.,   {Lloyd} J.~P.,  2014, \mn@doi [\aj]
  {10.1088/0004-6256/147/1/20}, \href
  {http://adsabs.harvard.edu/abs/2014AJ....147...20N} {147, 20}

\bibitem[\protect\citeauthoryear{{Pecaut} \& {Mamajek}}{{Pecaut} \&
  {Mamajek}}{2013}]{pecaut13}
{Pecaut} M.~J.,  {Mamajek} E.~E.,  2013, \mn@doi [\apjs]
  {10.1088/0067-0049/208/1/9}, \href
  {http://adsabs.harvard.edu/abs/2013ApJS..208....9P} {208, 9}

\bibitem[\protect\citeauthoryear{{Petigura}, {Marcy}  \& {Howard}}{{Petigura}
  et~al.}{2013}]{petigura13a}
{Petigura} E.~A.,  {Marcy} G.~W.,   {Howard} A.~W.,  2013, \mn@doi [\apj]
  {10.1088/0004-637X/770/1/69}, \href
  {http://adsabs.harvard.edu/abs/2013ApJ...770...69P} {770, 69}

\bibitem[\protect\citeauthoryear{{Ricker} et~al.,}{{Ricker}
  et~al.}{2015}]{ricker15}
{Ricker} G.~R.,  et~al., 2015, \mn@doi [Journal of Astronomical Telescopes,
  Instruments, and Systems] {10.1117/1.JATIS.1.1.014003}, \href
  {http://adsabs.harvard.edu/abs/2015JATIS...1a4003R} {1, 014003}

\bibitem[\protect\citeauthoryear{{Rogers}}{{Rogers}}{2015}]{rogers15}
{Rogers} L.~A.,  2015, \mn@doi [\apj] {10.1088/0004-637X/801/1/41}, \href
  {http://adsabs.harvard.edu/abs/2015ApJ...801...41R} {801, 41}

\bibitem[\protect\citeauthoryear{{Sanchis-Ojeda}, {Rappaport}, {Winn},
  {Kotson}, {Levine}  \& {El Mellah}}{{Sanchis-Ojeda}
  et~al.}{2014}]{sanchisojeda14}
{Sanchis-Ojeda} R.,  {Rappaport} S.,  {Winn} J.~N.,  {Kotson} M.~C.,  {Levine}
  A.,   {El Mellah} I.,  2014, \mn@doi [\apj] {10.1088/0004-637X/787/1/47},
  \href {http://adsabs.harvard.edu/abs/2014ApJ...787...47S} {787, 47}

\bibitem[\protect\citeauthoryear{{Scargle}}{{Scargle}}{1982}]{scargle82}
{Scargle} J.~D.,  1982, \mn@doi [\apj] {10.1086/160554}, \href
  {http://adsabs.harvard.edu/abs/1982ApJ...263..835S} {263, 835}

\bibitem[\protect\citeauthoryear{{Schlafly} \& {Finkbeiner}}{{Schlafly} \&
  {Finkbeiner}}{2011}]{schlafly11}
{Schlafly} E.~F.,  {Finkbeiner} D.~P.,  2011, \mn@doi [\apj]
  {10.1088/0004-637X/737/2/103}, \href
  {http://adsabs.harvard.edu/abs/2011ApJ...737..103S} {737, 103}

\bibitem[\protect\citeauthoryear{{Seager} \& {Mall{\'e}n-Ornelas}}{{Seager} \&
  {Mall{\'e}n-Ornelas}}{2003}]{seager03}
{Seager} S.,  {Mall{\'e}n-Ornelas} G.,  2003, \mn@doi [\apj] {10.1086/346105},
  \href {https://ui.adsabs.harvard.edu/\#abs/2003ApJ...585.1038S} {585, 1038}

\bibitem[\protect\citeauthoryear{{Shan}, {Johnson}  \& {Morton}}{{Shan}
  et~al.}{2015}]{shan15}
{Shan} Y.,  {Johnson} J.~A.,   {Morton} T.~D.,  2015, \mn@doi [\apj]
  {10.1088/0004-637X/813/1/75}, \href
  {https://ui.adsabs.harvard.edu/#abs/2015ApJ...813...75S} {813, 75}

\bibitem[\protect\citeauthoryear{{Smith} et~al.,}{{Smith}
  et~al.}{2012}]{smith12}
{Smith} J.~C.,  et~al., 2012, \mn@doi [\pasp] {10.1086/667697}, \href
  {http://adsabs.harvard.edu/abs/2012PASP..124.1000S} {124, 1000}

\bibitem[\protect\citeauthoryear{{Stassun} et~al.,}{{Stassun}
  et~al.}{2018}]{stassun17}
{Stassun} K.~G.,  et~al., 2018, \mn@doi [\aj] {10.3847/1538-3881/aad050}, \href
  {https://ui.adsabs.harvard.edu/#abs/2018AJ....156..102S} {156, 102}

\bibitem[\protect\citeauthoryear{{Stefansson} et~al.,}{{Stefansson}
  et~al.}{2017}]{stefansson17}
{Stefansson} G.,  et~al., 2017, \mn@doi [\apj] {10.3847/1538-4357/aa88aa},
  \href {https://ui.adsabs.harvard.edu/#abs/2017ApJ...848....9S} {848, 9}

\bibitem[\protect\citeauthoryear{{Sullivan} et~al.,}{{Sullivan}
  et~al.}{2015}]{sullivan15}
{Sullivan} P.~W.,  et~al., 2015, \mn@doi [\apj] {10.1088/0004-637X/809/1/77},
  \href {http://adsabs.harvard.edu/abs/2015ApJ...809...77S} {809, 77}

\bibitem[\protect\citeauthoryear{{Torres}, {Andersen}  \&
  {Gim{\'e}nez}}{{Torres} et~al.}{2010}]{torres10}
{Torres} G.,  {Andersen} J.,   {Gim{\'e}nez} A.,  2010, \mn@doi [\aapr]
  {10.1007/s00159-009-0025-1}, \href
  {http://adsabs.harvard.edu/abs/2010A%26ARv..18...67T} {18, 67}

\bibitem[\protect\citeauthoryear{{Trifonov}, {Rybizki}  \&
  {K{\"u}rster}}{{Trifonov} et~al.}{2018}]{trifonov18}
{Trifonov} T.,  {Rybizki} J.,   {K{\"u}rster} M.,  2018, arXiv e-prints, \href
  {https://ui.adsabs.harvard.edu/\#abs/2018arXiv181204501T} {p.
  arXiv:1812.04501}

\bibitem[\protect\citeauthoryear{{Twicken} et~al.,}{{Twicken}
  et~al.}{2018}]{twicken18}
{Twicken} J.~D.,  et~al., 2018, \mn@doi [Publications of the Astronomical
  Society of the Pacific] {10.1088/1538-3873/aab694}, \href
  {https://ui.adsabs.harvard.edu/#abs/2018PASP..130f4502T} {130, 064502}

\bibitem[\protect\citeauthoryear{{Vanderburg} \& {Johnson}}{{Vanderburg} \&
  {Johnson}}{2014}]{vanderburg14}
{Vanderburg} A.,  {Johnson} J.~A.,  2014, \mn@doi [\pasp] {10.1086/678764},
  \href {http://adsabs.harvard.edu/abs/2014PASP..126..948V} {126, 948}

\bibitem[\protect\citeauthoryear{{Vanderspek} et~al.,}{{Vanderspek}
  et~al.}{2018}]{vanderspek18}
{Vanderspek} R.,  et~al., 2018, arXiv e-prints, \href
  {https://ui.adsabs.harvard.edu/#abs/2018arXiv180907242V} {p.
  arXiv:1809.07242}

\bibitem[\protect\citeauthoryear{{Walkowicz} et~al.,}{{Walkowicz}
  et~al.}{2011}]{walkowicz11}
{Walkowicz} L.~M.,  et~al., 2011, \mn@doi [\aj] {10.1088/0004-6256/141/2/50},
  \href {http://adsabs.harvard.edu/abs/2011AJ....141...50W} {141, 50}

\bibitem[\protect\citeauthoryear{{Weiss} \& {Marcy}}{{Weiss} \&
  {Marcy}}{2014}]{weiss14}
{Weiss} L.~M.,  {Marcy} G.~W.,  2014, \mn@doi [\apjl]
  {10.1088/2041-8205/783/1/L6}, \href
  {http://adsabs.harvard.edu/abs/2014ApJ...783L...6W} {783, L6}

\bibitem[\protect\citeauthoryear{{Winn}}{{Winn}}{2010}]{winn10}
{Winn} J.~N.,  2010, {Exoplanet Transits and Occultations}.
University of Arizona Press, pp 55--77

\bibitem[\protect\citeauthoryear{{Winters} et~al.,}{{Winters}
  et~al.}{2015}]{winters15}
{Winters} J.~G.,  et~al., 2015, \mn@doi [\aj] {10.1088/0004-6256/149/1/5},
  \href {https://ui.adsabs.harvard.edu/#abs/2015AJ....149....5W} {149, 5}

\bibitem[\protect\citeauthoryear{{Wolfgang}, {Rogers}  \& {Ford}}{{Wolfgang}
  et~al.}{2016}]{wolfgang16}
{Wolfgang} A.,  {Rogers} L.~A.,   {Ford} E.~B.,  2016, \mn@doi [\apj]
  {10.3847/0004-637X/825/1/19}, \href
  {http://adsabs.harvard.edu/abs/2016ApJ...825...19W} {825, 19}

\makeatother
\end{thebibliography}

\begin{deluxetable*}{cccccccccc}
\tabletypesize{\small}
\tablecaption{Stellar photometry for the 1599 TICs in our sample of low mass stars\label{table:starphot}}
\tablehead{TIC & $T$ & $G_{\text{BP}}$ & $G_{\text{RP}}$ & $J$ & $H$ & $K_{\text{S}}$}
\startdata
2733611 & 11.121 & 11.200 $\pm$ 0.001 & 13.106 $\pm$ 0.001 & 9.958 $\pm$ 0.020 & 9.312 $\pm$ 0.025 & 9.137 $\pm$ 0.023 \\ 
2758962 & 11.695 & 11.872 $\pm$ 0.002 & 13.888 $\pm$ 0.002 & 10.466 $\pm$ 0.023 & 9.842 $\pm$ 0.023 & 9.637 $\pm$ 0.021 \\ 
2761472 & 12.546 & 12.672 $\pm$ 0.003 & 15.983 $\pm$ 0.003 & 10.665 $\pm$ 0.023 & 10.102 $\pm$ 0.025 & 9.755 $\pm$ 0.020 \\ 
2761836 & 12.998 & 13.167 $\pm$ 0.002 & 15.986 $\pm$ 0.002 & 11.397 $\pm$ 0.024 & 10.759 $\pm$ 0.025 & 10.476 $\pm$ 0.025 \\ 
2762148 & 12.716 & 12.720 $\pm$ 0.001 & 15.495 $\pm$ 0.001 & 11.060 $\pm$ 0.023 & 10.561 $\pm$ 0.023 & 10.246 $\pm$ 0.019 \\ 
\enddata
\tablecomments{The first five rows are shown here to illustrate the table's content and format. The complete table in plain ascii format can be provided via an \href{mailto:cloutier@astro.utoronto.ca}{email request} and will be made available as a machine readable table in its final published version.}
\end{deluxetable*}
\begin{longrotatetable}
\begin{deluxetable*}{lccccccccc}
\tabletypesize{\small}
\tablecaption{Stellar parameters for the 1599 TICs in our sample of low mass stars\label{table:starparams}}
\tablehead{TIC & $\alpha$ (J2000) & $\delta$ (J2000) & $\varpi$ & $d$ & $M_{K_{\text{S}}}$ & $R_s$ & $T_{\text{eff}}$ & $M_s$ & $\log{g}$ \\ 
  & [deg] & [deg] & [mas] & [pc] &  & [R$_{\odot}$] & [K] & M$_{\odot}$ & [dex]}
\startdata
2733611 & 356.1299 & -15.5194 & 16.763 $\pm$ 0.053 & 59.654 $\pm$ 0.188 & 5.259 $\pm$ 0.024 & 0.5650 $\pm$ 0.0169 & 3880 $\pm$ 49 & 0.6115 $\pm$ 0.0053 & 4.720 $\pm$ 0.026 \\ 
2758962 & 356.2124 & -15.1250 & 13.530 $\pm$ 0.050 & 73.911 $\pm$ 0.272 & 5.293 $\pm$ 0.022 & 0.5590 $\pm$ 0.0166 & 3768 $\pm$ 49 & 0.6076 $\pm$ 0.0051 & 4.727 $\pm$ 0.026 \\ 
2761472 & 356.5801 & -11.9453 & 43.288 $\pm$ 0.103 & 23.101 $\pm$ 0.055 & 7.937 $\pm$ 0.021 & 0.2160 $\pm$ 0.0065 & 3028 $\pm$ 49 & 0.1799 $\pm$ 0.0021 & 5.024 $\pm$ 0.027 \\ 
2761836 & 356.5473 & -14.0693 & 15.645 $\pm$ 0.060 & 63.920 $\pm$ 0.246 & 6.447 $\pm$ 0.026 & 0.3804 $\pm$ 0.0116 & 3259 $\pm$ 49 & 0.4094 $\pm$ 0.0052 & 4.890 $\pm$ 0.027 \\ 
2762148 & 356.5982 & -15.9872 & 28.845 $\pm$ 0.083 & 34.668 $\pm$ 0.100 & 7.547 $\pm$ 0.020 & 0.2519 $\pm$ 0.0075 & 3262 $\pm$ 49 & 0.2249 $\pm$ 0.0027 & 4.988 $\pm$ 0.027 \\ 
\enddata
\tablecomments{The first five rows are shown here to illustrate the table's content and format. The complete table in plain ascii format can be provided via an \href{mailto:cloutier@astro.utoronto.ca}{email request} and will be made available as a machine readable table in its final published version.}
\end{deluxetable*}
\end{longrotatetable}
\begin{longrotatetable}
  \begin{deluxetable*}{clcl}
  \tabletypesize{\small}
\tablecaption{Descriptions of the free parameters controlling the performance of \pipeline{}\label{table:freeparams}}
\tablehead{Parameter & Definition & Default Value & Summary of Behavior}
\startdata
\multicolumn{4}{c}{\emph{Linear search parameters}} \\
\hline \\
$\Delta t$ & Linear search temporal bin width & 30 minutes &
Decreasing $\Delta t$ will improve sensitivity to ultra-short period \\
&&& planets while increasing the \pipeline{} runtime number of signals \\
&&& for confusion in the periodic search stage. \\ 
$D$ grid & Grid of transit durations considered during the & $\{1.2,2.4,$ &
Decreasing the minimum duration improves sensitivity to \\
& linear search stage & $4.8\}$ hours &
ultra-short period planets but makes \pipeline{} more \\
&&& susceptible to stochastic features with short timescales. \\ 
S/N$_{\text{thresh}}$ & Minimum linear search S/N of a transit- & 5 &
Decreasing S/N$_{\text{thresh}}$ will result in more light curve features \\
& like event &&
being flagged as false positives thus creating more signals for \\
&&& confusion within the periodic search stage.  \\
\multicolumn{4}{c}{\emph{Periodic search parameters}} \\
\hline \\
$f_{\text{P}}$ & Maximum relative difference between two  & 0.01 &
Increasing $f_{\text{P}}$ makes fewer period pairs consistent with  \\
& periods to be flagged as a multiple &&
being multiples thus increasing the sensitivity to resonant \\
&&& planet pairs while simultaneously increasing the number \\
&&& of single planets misidentifed as a resonant pair.  \\
\multicolumn{4}{c}{\emph{Automated light curve vetting parameters}} \\
\hline \\
$c_1$ & Minimum transit S/N & 8.4 & Increasing $c_1$ prevents the detection of some small planets \\
&&& but will also significantly reduce the number false positives \\
&&& due to residual systematics. \\
$c_2$ & Minimum number of MADs from the & 2.4 &
Similar behavior to $c_1$. \\
& out-of-transit flux that the difference in && \\
& median in and out-of-transit fluxes  && \\
& must exceed && \\
$c_3$ & Minimum fraction of in-transit points & 0.7 &
Increasing $c_3$ may result in more accurate determinations of \\
& below $Z+\sigma_Z$ &&
correct periods but will also cause some transits to be \\
&&& discarded if residual noise is also present during the transit.  \\
$c_4$ & Minimum fraction of in-transit points & 0.1 &
Increasing $c_4$ increases sensitivity to asymmetric transit \\
& prior to $T_0$ && 
shapes such as those from disintegrating planets. \\
$c_5$ & Minimum number of MADs for a flare & 8 &
Increasing $c_5$ makes flare detection smore robust but at the \\
& above the flux continuum &&
risk of missing some lower amplitude flares. \\
$c_6$ & Minimum number of successive points & 2 &
Increasing $c_6$ decreases sensitivity to flares of short relative \\
& within a flare duration & & to the light curve cadence. \\
$c_7$ & Assumed M dwarf flare duration & 30 minutes &
Increasing $c_7$ decreases the sensitivity to long duration flares \\
&&& while increasing the assumed fraction light curve fraction that \\
&&& is contaminated by flares. \\
$c_8$ & Number of transit durations from $T_0$ & 4 &
Decreasing $c_8$ decreases the probability of an transit being \\
& which cannot be affected by flares &&
contaminated by flares. \\
$c_9$ & Maximum time from the light curve & 4.8 hours &
Increasing $c_9$ descreases the probability of a transit being \\
& edges to not be rejected due to possible &&
affects by light curve systematics at its edges but also narrows \\
&  contamination at the edges. &&
the baseline over which transits can be found. \\
$c_{10}$ & Minimum autocorrelation of flux residuals & 0.6 &
Increasing $c_{10}$ improves the robustness of transit detections \\
&&& but decreases the detection sensitivity in light curves with \\
&&& imperfect systematics corrections. \\
\multicolumn{4}{c}{\emph{Eclipsing binary vetting parameters}} \\
\hline \\
$c_{\text{EB},1}$ & Maximum $r_p/R_s$ of a  transit-like event & 0.5 & - \\
$c_{\text{EB},2}$ & Maximum planet radius & 30 R$_{\oplus}$ & - \\
$c_{\text{EB},3}$ & Maximum transit duration, & $D(30\text{ R}_{\oplus},$ & - \\
& $D(r_p,P,a/R_s,Z,i)$ & $P,a/R_s,Z,i)$ & \\
$c_{\text{EB},4}$\tablenotemark{a} & Minimum occultation S/N of an EB (Eq.~\ref{eq:occ}) & 5 &
Decreasing $c_{\text{EB},4}$ makes a larger fraction of occultations \\
&&& consistent with being due to an EB rather than a transiting \\
&&& planet. \\
$c_{\text{EB},5}$\tablenotemark{a} & Minimum fraction of iterative occultation & 0.5 &
Increasing $c_{\text{EB},5}$ makes transit-like events more robust as \\
& searches consistent with an EB & &
the probability of being flagged as an EB is reduced. \\ 
$c_{\text{EB},6}$\tablenotemark{a} & Minimum ingress plus egress time fraction & 0.9 &
Increasing $c_{\text{EB},6}$ makes fewer transit-like events consistent \\
& of $D$ for a `V'-shaped transit && with having a `V'-shaped transit. \\
\enddata
\tablenotetext{a}{These eclising binary (EB) parameters are intended to identify favorable EBs rather than transiting planets. Planetary signals of interest are rejected if any of the EB criteria are satisfied.}
\end{deluxetable*}
\end{longrotatetable}

\begin{deluxetable*}{lc}
\tablecaption{Model parameter priors\label{table:priors}}
\tablehead{Parameter & Prior}
\startdata
\multicolumn{2}{c}{\emph{GP hyperparameters}\tablenotemark{a}} \\
Covariance amplitude, $\ln{a_{\text{GP}}}$ & $\mathcal{U}(-20,0)$ \\
Exponential timescale, $\ln{\lambda/}$days & $\mathcal{U}(-3,10)$ \\
Coherence, $\ln{\Gamma}$ & $\mathcal{U}(-5,5)$ \\
Periodic timescale, $\ln{P_{\text{GP}}/}$days & $\mathcal{U}(-3,10)$  \\
\multicolumn{2}{c}{\emph{Transit model parameters}} \\
Orbital period, $P$ [days] & $\mathcal{U}(0.9,1.11)\cdot P_{\text{opt}}$\tablenotemark{b} \\
Time of mid-transit, $T_0$ & $\mathcal{U}(-1.11,1.11)\cdot P_{\text{opt}} + T_{0,\text{opt}}$ \\
$[$BJD-2,457,000$]$ & \\
Scaled semimajor axis, $a/R_s$ & $\mathcal{U}(0.58,1.70)\cdot (a/R_s)_{\text{opt}}$ \\
Planet-star radius ratio, $r_p/R_s$ & $\mathcal{U}(0,1)$ \\
Orbital inclination, $i$ & $\mathcal{U}(-1,1)\cdot i((a/R_s)_{\text{opt}},b=1)$\tablenotemark{c} \\
\multicolumn{2}{c}{\emph{Single transit model parameters}} \\
Orbital period, $P$ [days] & $\mathcal{J}(1,100)\cdot P_{\text{inner}}$\tablenotemark{d} \\
Time of mid-transit, $T_0$ & $\mathcal{U}(-3,3)\cdot D$ \\
$[$BJD-2,457,000$]$ & \\
Scaled semimajor axis, $a/R_s$ & $\mathcal{J}(1,100)\cdot (a/R_s)_{\text{inner}}$ \\
Planet-star radius ratio, $r_p/R_s$ & $\mathcal{U}(0,1)$ \\
Orbital inclination, $i$ & $\mathcal{U}(-1,1)\cdot i((a/R_s)_i,b=1)$ \\
\enddata
\tablenotetext{a}{GP hyperparameter priors used during de-trending (i.e. with zero mean model) and during the simultaneous systematics plus transit modeling.}
\tablenotetext{b}{The designation `opt' is indicative of the 
optimized parameter values from the maximum likelihood model used for parameter initialization.}
\tablenotetext{c}{The function $i(a/R_s,b)=a \cos{i}/R_s$ returns the orbital inclination
  given $a/R_s$ and the impact parameter $b$ which is constrained to $|b|<1$ in our
  transit models.}
\tablenotetext{d}{The designation `inner' is indicative of the inner-most orbital period 
permissible for a single transit event over the \tess{} baseline.}
\end{deluxetable*}

\begin{longrotatetable}
\begin{deluxetable*}{cc|cc|cccccccc}
\tabletypesize{\small}
\tablecaption{\texttt{vespa} input parameters and inferred probabilities of transiting planet and astrophysical false positive models for our 24 objects of interest\label{table:vespa}}
\tablehead{\multicolumn{2}{c}{IDs} & \multicolumn{2}{c}{\texttt{vespa} Input} & \multicolumn{8}{c}{\texttt{vespa} Results} \\ OI & TOI & \texttt{maxrad} & \texttt{secthresh} & $\text{P}_{\text{EB}}$ & $\text{P}_{\text{EB}2}$ & $\text{P}_{\text{HEB}}$ & $\text{P}_{\text{HEB}2}$ & $\text{P}_{\text{BEB}}$ & $\text{P}_{\text{BEB}2}$ & FPP\tablenotemark{a} & Disposition\tablenotemark{b}}
\startdata
12421862.01 & 198.01 & 34.755 & 9.9$\times 10^{-4}$ & 2.0$\times 10^{-9}$ & 4.9$\times 10^{-5}$ & 1.1$\times 10^{-14}$ & 9.8$\times 10^{-8}$ & 5.8$\times 10^{-2}$ & 3.7$\times 10^{-2}$ & 9.6$\times 10^{-2}$ & PC \\ 
47484268.01 & 226.01 & 34.147 & 4.8$\times 10^{-3}$ & 1.3$\times 10^{-7}$ & 3.8$\times 10^{-3}$ & 1.4$\times 10^{-9}$ & 1.6$\times 10^{-4}$ & 3.9$\times 10^{-1}$ & 2.8$\times 10^{-1}$ & 6.8$\times 10^{-1}$ & pPC \\ 
49678165.01 & - & 75.094 & 7.7$\times 10^{-3}$ & 9.4$\times 10^{-27}$ & 1.4$\times 10^{-14}$ & 2.4$\times 10^{-26}$ & 3.5$\times 10^{-14}$ & 9.1$\times 10^{-3}$ & 7.1$\times 10^{-3}$ & 1.6$\times 10^{-2}$ & ST \\ 
92444219.01 & - & 35.723 & 3.8$\times 10^{-3}$ & 4.1$\times 10^{-15}$ & 6.7$\times 10^{-8}$ & 4.4$\times 10^{-15}$ & 1.6$\times 10^{-8}$ & 2.6$\times 10^{-1}$ & 1.6$\times 10^{-1}$ & 4.1$\times 10^{-1}$ & pST \\ 
100103200.01 & - & 47.832 & 7.0$\times 10^{-4}$ & 5.1$\times 10^{-19}$ & 4.3$\times 10^{-12}$ & 5.0$\times 10^{-33}$ & 2.9$\times 10^{-20}$ & 3.8$\times 10^{-3}$ & 2.0$\times 10^{-3}$ & 5.7$\times 10^{-3}$ & pPC\tablenotemark{c} \\ 
100103201.01 & - & 47.941 & 8.1$\times 10^{-4}$ & 2.2$\times 10^{-73}$ & 1.7$\times 10^{-67}$ & 5.1$\times 10^{-132}$ & 3.7$\times 10^{-91}$ & 6.2$\times 10^{-1}$ & 3.5$\times 10^{-1}$ & 9.7$\times 10^{-1}$ & BEB \\ 
100103201.02 & - & 47.941 & 8.1$\times 10^{-4}$ & 2.2$\times 10^{-73}$ & 1.7$\times 10^{-67}$ & 5.1$\times 10^{-132}$ & 3.7$\times 10^{-91}$ & 6.2$\times 10^{-1}$ & 3.5$\times 10^{-1}$ & 9.7$\times 10^{-1}$ & BEB \\ 
141708335.01 & - & 33.606 & 3.6$\times 10^{-3}$ & 6.7$\times 10^{-39}$ & 2.9$\times 10^{-31}$ & 3.4$\times 10^{-35}$ & 8.6$\times 10^{-26}$ & 6.3$\times 10^{-1}$ & 3.3$\times 10^{-1}$ & 9.5$\times 10^{-1}$ & BEB \\ 
206660104.01 & - & 38.276 & 8.4$\times 10^{-4}$ & 1.7$\times 10^{-65}$ & 8.6$\times 10^{-39}$ & 2.5$\times 10^{-87}$ & 1.2$\times 10^{-40}$ & 1.3$\times 10^{-2}$ & 9.3$\times 10^{-3}$ & 2.2$\times 10^{-2}$ & PC \\ 
231279823.01 & - & 36.346 & 5.3$\times 10^{-4}$ & 7.7$\times 10^{-43}$ & 7.9$\times 10^{-47}$ & 7.5$\times 10^{-55}$ & 4.0$\times 10^{-78}$ & 5.1$\times 10^{-1}$ & 4.9$\times 10^{-1}$ & $1.0\times 10^{0}$ & BEB \\ 
231702397.01 & 122.01 & 36.485 & 6.9$\times 10^{-3}$ & 7.5$\times 10^{-18}$ & 8.7$\times 10^{-6}$ & 1.8$\times 10^{-23}$ & 5.1$\times 10^{-9}$ & 2.3$\times 10^{-3}$ & 2.7$\times 10^{-3}$ & 5.0$\times 10^{-3}$ & PC \\ 
234994474.01 & 134.01 & 37.794 & 7.1$\times 10^{-4}$ & 1.2$\times 10^{-51}$ & 8.5$\times 10^{-34}$ & 3.9$\times 10^{-107}$ & 4.2$\times 10^{-61}$ & 9.6$\times 10^{-4}$ & 1.9$\times 10^{-4}$ & 1.2$\times 10^{-3}$ & PC \\ 
235037759.01 & - & 39.567 & 1.9$\times 10^{-2}$ & - & - & - & - & - & - & - & AFP\tablenotemark{d} \\ 
238027971.01 & - & 37.085 & 1.4$\times 10^{-3}$ & 2.5$\times 10^{-67}$ & 7.5$\times 10^{-92}$ & 5.6$\times 10^{-76}$ & 4.1$\times 10^{-79}$ & 3.4$\times 10^{-1}$ & 6.6$\times 10^{-1}$ & $1.0\times 10^{0}$ & BEB2 \\ 
260004324.01 & - & 35.803 & 1.1$\times 10^{-3}$ & 3.3$\times 10^{-44}$ & 4.1$\times 10^{-17}$ & 1.4$\times 10^{-61}$ & 8.8$\times 10^{-20}$ & 5.3$\times 10^{-2}$ & 2.8$\times 10^{-2}$ & 8.1$\times 10^{-2}$ & PC \\ 
262530407.01 & 177.01 & 40.509 & 8.0$\times 10^{-4}$ & 6.5$\times 10^{-38}$ & 3.2$\times 10^{-41}$ & 7.6$\times 10^{-45}$ & 5.5$\times 10^{-44}$ & 1.8$\times 10^{-7}$ & 5.7$\times 10^{-7}$ & 7.5$\times 10^{-7}$ & PC \\ 
278661431.01 & - & 48.894 & 5.1$\times 10^{-3}$ & 2.4$\times 10^{-31}$ & 1.4$\times 10^{-37}$ & 3.6$\times 10^{-48}$ & 6.9$\times 10^{-55}$ & 3.8$\times 10^{-1}$ & 4.8$\times 10^{-1}$ & 8.6$\times 10^{-1}$ & pPC \\ 
279574462.01 & - & 146.950 & 1.3$\times 10^{-2}$ & - & 1.7$\times 10^{-86}$ & 1.0$\times 10^{-304}$ & 1.9$\times 10^{-77}$ & 3.0$\times 10^{-1}$ & 7.0$\times 10^{-1}$ & $1.0\times 10^{0}$ & BEB2 \\ 
303586421.01 & - & 32.084 & 6.2$\times 10^{-3}$ & 7.8$\times 10^{-53}$ & 6.2$\times 10^{-47}$ & 3.1$\times 10^{-67}$ & 1.5$\times 10^{-31}$ & 3.3$\times 10^{-1}$ & 5.6$\times 10^{-1}$ & 9.0$\times 10^{-1}$ & BEB2 \\ 
305048087.01 & 237.01 & 34.445 & 8.7$\times 10^{-3}$ & 2.6$\times 10^{-30}$ & 1.4$\times 10^{-8}$ & 4.8$\times 10^{-27}$ & 8.0$\times 10^{-8}$ & 1.8$\times 10^{-2}$ & 1.4$\times 10^{-2}$ & 3.2$\times 10^{-2}$ & PC \\ 
307210830.01 & 175.01 & 38.817 & 7.7$\times 10^{-4}$ & 1.7$\times 10^{-5}$ & 8.0$\times 10^{-3}$ & 9.0$\times 10^{-10}$ & 7.5$\times 10^{-5}$ & 9.2$\times 10^{-3}$ & 4.3$\times 10^{-3}$ & 2.2$\times 10^{-2}$ & PC \\ 
415969908.01 & 233.01 & 32.000 & 2.3$\times 10^{-3}$ & 1.5$\times 10^{-11}$ & 1.1$\times 10^{-5}$ & 1.4$\times 10^{-18}$ & 2.0$\times 10^{-8}$ & 6.0$\times 10^{-3}$ & 5.5$\times 10^{-3}$ & 1.2$\times 10^{-2}$ & PC \\ 
415969908.02 & - & 32.000 & 2.3$\times 10^{-3}$ & 2.5$\times 10^{-19}$ & 2.4$\times 10^{-10}$ & 4.9$\times 10^{-28}$ & 2.7$\times 10^{-13}$ & 2.6$\times 10^{-3}$ & 2.0$\times 10^{-3}$ & 4.7$\times 10^{-3}$ & ST \\ 
441056702.01 & - & 35.795 & 1.9$\times 10^{-3}$ & 5.5$\times 10^{-19}$ & 5.3$\times 10^{-9}$ & 1.8$\times 10^{-37}$ & 6.0$\times 10^{-17}$ & 2.5$\times 10^{-2}$ & 1.7$\times 10^{-2}$ & 4.2$\times 10^{-2}$ & PC \\ 
\enddata
\tablecomments{P$_i$ represents the relative probability (i.e. the product of the model's prior and likelihood relative to the probabilties of all other models) of the $i^{\text{th}}$ astrophysical false positive scenario where $i$ is one of six possible scenarios: blended eclipsing binaries (BEB), undiluted eclipsing binaries (EB), and hiearchical eclipsing biaries (HEB), each with either one or twice its input orbital period.}
\tablenotetext{a}{The transiting planet false positive probability.}
\tablenotetext{b}{Possible dispositions of objects of interest are a planet candidate (PC), a putative planet candidate (pPC), a single transit event (ST), a putative single transit event (pST), an unclassifed astrophysical false positive (AFP), or any of the scenarios $i$. The putative dispositions have FPP $\in [0.1,0.9]$ whereas remaining candidates and astrophysical false positives have FPP $<0.1$ and $>0.9$ respectively.}
\tablenotetext{c}{OI 100103200.01 is assigned the pPC disposition despite having a FPP $<0.1$ because of its proximity to the comparably bright TIC 100103201 (see Fig.~\ref{fig:gaiafps}).}
\tablenotetext{d}{The MCMC for OI 235037759.01 failed to converge so we broadly classify it as an astrophysical false positive (AFP) based on the prevalence of nearby bright sources from \gaia{} (see panel `v' in Fig.~\ref{fig:gaiafps}). We are unable to classify the object as a particular type of astrophysical false positive.}
\end{deluxetable*}
\end{longrotatetable}
\begin{longrotatetable}
\begin{deluxetable*}{cccccccccc}
\tabletypesize{\small}
\tablecaption{Planetary parameters for our 16 vetted candidates\label{table:planets}}
\tablehead{OI & TOI & $P$ & $T_0$ & $a/R_s$ & $r_p/R_s$ & $i$ & $r_p$ & $S$ & Disposition\tablenotemark{a} \\ 
  &  & [days] & [BJD-2,457,000] &  &  & [deg] & [R$_{\oplus}$] & [S$_{\oplus}$] & }
\startdata
12421862.01 & 198.01 & 20.4282 $\pm$ 0.0042 & 1376.8028 $\pm$ 0.0023 & 68.05$^{+6.21}_{-9.86}$ & 0.033$^{+0.001}_{-0.002}$ & 90.24$^{+0.44}_{-0.40}$ & 1.59$^{+0.09}_{-0.10}$ & 2.7$\pm$ 0.2 & PC \\ 
47484268.01 & 226.01 & 20.2833 $\pm$ 0.0048 & 1378.7855 $\pm$ 0.0042 & 56.34$^{+9.13}_{-8.28}$ & 0.090$^{+0.004}_{-0.006}$ & 90.62$^{+0.32}_{-0.19}$ & 3.82$^{+0.20}_{-0.25}$ & 1.4$\pm$ 0.1 & pPC \\ 
49678165.01 & - & 35.64$^{+49.78}_{-14.61}$ & 1371.1945 $\pm$ 0.0038 & 74.13$^{+102.92}_{-26.55}$ & 0.10$^{+0.01}_{-0.00}$ & 90.00$^{+0.36}_{-0.36}$ & 3.02$^{+0.21}_{-0.19}$ & 0.3$^{+0.3}_{-0.2}$ & ST \\ 
92444219.01 & - & 39.52$^{+45.64}_{-16.90}$ & 1342.4729 $\pm$ 0.0037 & 49.71$^{+55.87}_{-19.54}$ & 0.07$^{+0.00}_{-0.00}$ & 89.99$^{+0.40}_{-0.41}$ & 3.07$^{+0.13}_{-0.14}$ & 0.5$^{+0.6}_{-0.3}$ & pST \\ 
100103200.01 & - & 18.4462 $\pm$ 0.0024 & 1380.0894 $\pm$ 0.0018 & 51.40$^{+9.45}_{-5.97}$ & 0.035$^{+0.002}_{-0.001}$ & 90.45$^{+0.33}_{-0.23}$ & 1.97$^{+0.11}_{-0.09}$ & 4.2$\pm$ 0.3 & pPC \\ 
206660104.01 & - & 13.4507 $\pm$ 0.0081 & 1379.7842 $\pm$ 0.0079 & 27.84$^{+1.82}_{-1.05}$ & 0.021$^{+0.001}_{-0.001}$ & 90.00$^{+0.44}_{-0.49}$ & 1.21$^{+0.08}_{-0.07}$ & 5.2$\pm$ 0.4 & PC \\ 
231702397.01 & 122.01 & 5.0789 $\pm$ 0.0010 & 1349.4338 $\pm$ 0.0031 & 31.23$^{+3.05}_{-4.64}$ & 0.077$^{+0.004}_{-0.005}$ & 90.55$^{+0.96}_{-1.12}$ & 2.80$^{+0.18}_{-0.19}$ & 7.5$\pm$ 0.7 & PC \\ 
234994474.01 & 134.01 & 1.4013 $\pm$ 0.0001 & 1345.6507 $\pm$ 0.0010 & 8.68$^{+0.73}_{-1.10}$ & 0.021$^{+0.001}_{-0.001}$ & 88.84$^{+2.67}_{-3.67}$ & 1.39$^{+0.07}_{-0.06}$ & 145.1$\pm$ 11.0 & PC \\ 
260004324.01 & - & 3.8157 $\pm$ 0.0007 & 1357.9369 $\pm$ 0.0036 & 17.89$^{+1.98}_{-2.73}$ & 0.021$^{+0.001}_{-0.002}$ & 91.12$^{+1.23}_{-2.14}$ & 1.15$^{+0.07}_{-0.09}$ & 24.6$\pm$ 1.9 & PC \\ 
262530407.01 & 177.01 & 2.8540 $\pm$ 0.0001 & 1364.7070 $\pm$ 0.0004 & 16.69$^{+1.86}_{-2.17}$ & 0.033$^{+0.001}_{-0.001}$ & 90.95$^{+1.45}_{-1.11}$ & 1.87$^{+0.08}_{-0.08}$ & 39.7$\pm$ 3.3 & PC \\ 
278661431.01 & - & 17.6317 $\pm$ 0.0048 & 1343.9299 $\pm$ 0.0025 & 46.41$^{+0.26}_{-0.38}$ & 0.094$^{+0.002}_{-0.002}$ & 89.98$^{+0.12}_{-0.08}$ & 2.82$^{+0.11}_{-0.10}$ & 1.1$\pm$ 0.1 & pPC \\ 
305048087.01 & 237.01 & 5.4310 $\pm$ 0.0014 & 1376.9753 $\pm$ 0.0037 & 35.04$^{+3.47}_{-5.12}$ & 0.073$^{+0.004}_{-0.006}$ & 90.33$^{+0.80}_{-1.14}$ & 1.67$^{+0.12}_{-0.14}$ & 3.6$\pm$ 0.4 & PC \\ 
307210830.01 & 175.01 & 3.6893 $\pm$ 0.0001 & 1374.6508 $\pm$ 0.0005 & 22.34$^{+1.64}_{-2.52}$ & 0.039$^{+0.001}_{-0.001}$ & 90.35$^{+0.88}_{-0.58}$ & 1.34$^{+0.05}_{-0.04}$ & 12.0$\pm$ 0.9 & PC \\ 
415969908.01 & 233.01 & 11.6658 $\pm$ 0.0056 & 1376.9247 $\pm$ 0.0040 & 45.23$^{+4.77}_{-7.04}$ & 0.046$^{+0.003}_{-0.003}$ & 90.11$^{+0.67}_{-0.73}$ & 1.90$^{+0.13}_{-0.13}$ & 3.8$\pm$ 0.3 & PC \\ 
415969908.02 & - & 52.74$^{+56.33}_{-20.31}$ & 1381.0703 $\pm$ 0.0042 & 98.07$^{+90.89}_{-33.09}$ & 0.05$^{+0.00}_{-0.00}$ & 90.01$^{+0.36}_{-0.37}$ & 2.02$^{+0.22}_{-0.20}$ & 0.5$^{+0.5}_{-0.3}$ & ST \\ 
441056702.01 & - & 6.3424 $\pm$ 0.0093 & 1371.8111 $\pm$ 0.0055 & 20.88$^{+3.80}_{-3.40}$ & 0.031$^{+0.003}_{-0.003}$ & 89.81$^{+1.67}_{-1.24}$ & 1.97$^{+0.19}_{-0.19}$ & 25.0$\pm$ 2.2 & PC \\ 
\enddata
\tablenotetext{a}{Possible dispositions of objects of interest are a planet candidate (PC), a putative planet candidate (pPC), a single transit event (ST), or a putative single transit event (pST). The putative dispositions have FPP $\in [0.1,0.9]$ whereas the remaining candidates have FPP $<0.1$.}
\end{deluxetable*}
\end{longrotatetable}
\begin{deluxetable*}{ccccccccccc}
\tabletypesize{\small}
\tablecaption{Metric values indicating the feasibility of a variety of follow-up programs for our 16 vetted candidates\label{table:followup}}
\tablehead{OI & TOI & $J$ & $P$ & $r_p$ & $m_p$\tablenotemark{a} & $K$ & $\Omega$\tablenotemark{b} & $T_{\text{eq}}$\tablenotemark{c} & TSM\tablenotemark{d} & ESM\tablenotemark{e} \\ 
  &  &  & [days] & [R$_{\oplus}$] & [M$_{\oplus}$] & [m/s] &  & [K] &  & }
\startdata
12421862.01 & 198.01 & 8.65 & 20.43 & 1.59 & 3.16 & 1.19 & 0.58 & 358 & 53.4 & 0.6 \\ 
47484268.01 & 226.01 & 10.85 & 20.28 & 3.82 & 14.03 & 5.93 & \textbf{1.40} & 302 & 69.0 & 0.9 \\ 
49678165.01 & - & 11.14 & 35.64 & 3.02 & 9.40 & 4.40 & 0.92 & 212 & 57.7 & 0.1 \\ 
92444219.01 & - & 10.49 & 39.52 & 3.07 & 9.68 & 3.35 & 0.90 & 238 & 51.1 & 0.1 \\ 
100103200.01 & - & 7.50 & 18.45 & 1.97 & 4.54 & 1.61 & \textbf{0.74} & 397 & \textbf{99.6} & 2.0 \\ 
206660104.01 & - & 8.36 & 13.45 & 1.21 & 1.92 & 0.76 & 0.51 & 419 & 5.9 & 0.7 \\ 
231702397.01 & 122.01 & 11.53 & 5.08 & 2.80 & 8.26 & 6.35 & \textbf{1.63} & 461 & 69.3 & 3.2 \\ 
234994474.01 & 134.01 & 7.94 & 1.40 & 1.39 & 2.52 & 1.98 & \textbf{1.24} & 965 & \textbf{14.2} & \textbf{9.8} \\ 
260004324.01 & - & 8.80 & 3.82 & 1.15 & 1.62 & 0.98 & 0.74 & 619 & 7.6 & 2.1 \\ 
262530407.01 & 177.01 & 8.17 & 2.85 & 1.87 & 4.16 & 2.74 & \textbf{1.32} & 698 & \textbf{119.8} & \textbf{10.1} \\ 
278661431.01 & - & 10.89 & 17.63 & 2.82 & 8.36 & 5.14 & 1.08 & 288 & \textbf{86.2} & 0.7 \\ 
305048087.01 & 237.01 & 11.74 & 5.43 & 1.67 & 3.44 & 4.06 & 0.95 & 384 & 66.2 & 1.3 \\ 
307210830.01 & 175.01 & 7.93 & 3.69 & 1.34 & 2.37 & 2.15 & \textbf{0.87} & 517 & \textbf{26.4} & 6.7 \\ 
415969908.01 & 233.01 & 9.94 & 11.67 & 1.90 & 4.26 & 2.21 & 0.84 & 387 & 55.5 & 1.1 \\ 
415969908.02 & - & 9.94 & 52.74 & 2.02 & 4.74 & 1.48 & 0.54 & 234 & 36.4 & 0.1 \\ 
441056702.01 & - & 9.89 & 6.34 & 1.97 & 4.53 & 2.17 & \textbf{1.06} & 622 & 39.8 & 2.4 \\ 
\enddata
\tablecomments{Bolded values are indicative of candidates that exceed threshold values of that parameter (see \citealt{cloutier18b} for $\Omega$ and \citealt{kempton18} for the TSM and ESM) and should be strongly considered for rapid confirmation and follow-up.}
\tablenotetext{a}{Planet masses are estimated from the planet radius using the deterministic version of the mass-radius relation from \cite{chen17}.}\tablenotetext{b}{$\Omega$ is a diagnostic metric that is indicative of the observing time required to characterize a planet's RV mass \citep{cloutier18b}. $\Omega= r_p/P^{1/3}$ where $r_p$ is given in Earth radii and $P$ in days.}
\tablenotetext{c}{Planetary equilibrium temperature is calculated assuming zero albedo and full heat redistribution via $T_{\text{eq}} = T_{\text{eff}} \sqrt{R_s/2a}$.}
\tablenotetext{d}{The transmission spectroscopy metric from \citep{kempton18}. See Sect.~\ref{sect:atmospheres} for the definition.}
\tablenotetext{e}{The emission spectroscopy metric from \citep{kempton18}. See Sect.~\ref{sect:atmospheres} for the definition.}
\end{deluxetable*}


\end{document}